\documentclass{aa}  
\usepackage{graphicx}
\usepackage{txfonts}
\usepackage{natbib}
\bibpunct{(}{)}{;}{a}{}{,} %% natbib format for A&A and ApJ

\begin{document}

   \title{A proto-pseudobulge in ESO 320-G030 fed by a massive molecular inflow
  driven by a nuclear bar}

   \titlerunning{A proto-pseudobulge in ESO 320-G030}
%   \subtitle{}

   \author{Eduardo Gonz\'alez-Alfonso
          \inst{1,2}
          \and
          Miguel Pereira-Santaella\inst{3}
          \and
          Jacqueline Fischer\inst{4}
          \and
          Santiago Garc\'{\i}a-Burillo\inst{5}
          \and
          Chentao Yang\inst{6}
          \and
          Almudena Alonso-Herrero\inst{7}
          \and
          Luis Colina\inst{3}
          \and
          Matthew L. N. Ashby\inst{2}
          \and
          Howard A. Smith\inst{2}
          \and
          Fernando Rico-Villas\inst{3}
          \and
          Jes\'us Mart\'{\i}n-Pintado\inst{3}
          \and
          Sara Cazzoli\inst{8}
          \and
          Kenneth P. Stewart\inst{4}
          }
   \institute{Universidad de Alcal\'a, Departamento de F\'{\i}sica
     y Matem\'aticas, Campus Universitario, E-28871 Alcal\'a de Henares,
     Madrid, Spain \\     
              \email{eduardo.gonzalez@uah.es}
         \and
             Center for Astrophysics $|$ Harvard \& Smithsonian,
             60 Garden Street, Cambridge, MA 02138, USA 
             \and
             Centro de Astrobiolog\'{\i}a (CSIC-INTA), Ctra. de Ajalvir,
             km. 4, E-28850, Torrej\'on de Ardoz, Madrid, Spain
             \and
             George Mason University, Department of Physics \& Astronomy,
             MS 3F3, 4400 University Drive, Fairfax, VA 22030, USA
             \and
             Observatorio Astron\'omico Nacional (OAN-IGN)-Observatorio de
             Madrid, Alfonso XII, 3, E-28014 Madrid, Spain
             \and
             European Southern Observatory, Alonso de C\'ordova 3107,
             Casilla 19001, Vitacura, Santiago, Chile
             \and
             Centro de Astrobiolog\'{\i}a (CAB, CSIC-INTA), ESAC Campus,
             E-28692, Villanueva de la Ca\~nada, Madrid, Spain
             \and
             IAA - Instituto de Astrof\'{\i}sica de Andaluc\'{\i}a (CSIC),
             Apdo. 3004, E-18008, Granada, Spain             
             }

%   \date{Received September 15, 1996; accepted March 16, 1997}

% \abstract{}{}{}{}{} 
% 5 {} token are mandatory
 
   \abstract{
  Galaxies with nuclear bars are believed to efficiently drive gas
  inward, generating a nuclear starburst and possibly an
  active galactic nucleus (AGN). 
  We confirm this scenario for the isolated, double-barred, luminous
  infrared galaxy ESO~320-G030 based on an analysis of
  {\it Herschel} and ALMA spectroscopic observations.
  {\it Herschel}/PACS and SPIRE
  observations of ESO~320-G030 show
  absorption or emission in 18 lines of H$_2$O, 
  which we combine with the ALMA 
  H$_2$O $4_{23}-3_{30}$ 448\,GHz line ($E_{\mathrm{upper}}\sim400$\,K) 
  and continuum images to study the 
  physical properties of the nuclear region.
  Radiative transfer models
  indicate that three nuclear components are required to account for the
  multi-transition 
  H$_2$O and continuum data. 
  An envelope, with radius $R\sim130-150$\,pc, dust temperature
  $T_{\mathrm{dust}}\approx50$\,K, and $N_{\mathrm{H2}}\sim2\times10^{23}$\,cm$^{-2}$,
  surrounds a nuclear disk with $R\sim40$\,pc
  that is optically thick in the far-infrared
  ($\tau_{100\,\mathrm{\mu m}}\sim1.5-3$,
  $N_{\mathrm{H2}}\sim2\times10^{24}$\,cm$^{-2}$).
  In addition, an extremely compact ($R\sim12$\,pc), warm
  ($\approx100$\,K), and buried
  ($\tau_{100\,\mathrm{\mu m}}>5$, $N_{\mathrm{H2}}\gtrsim5\times10^{24}$\,cm$^{-2}$)
  core component is
  required to account for the very high-lying H$_2$O absorption lines.
  The three nuclear components 
  account for $70$\% of the
  galaxy luminosity ($\mathrm{SFR}\sim16-18$\,M$_{\odot}$\,yr$^{-1}$).
  The nucleus is fed by a molecular inflow observed
  in CO 2-1 with ALMA, which is associated with the nuclear bar. With
  decreasing radius ($r=450-225$\,pc), the 
  mass inflow rate increases up to
  $\dot{M}_{\mathrm{inf}}\sim20$\,$M_{\odot}\,\mathrm{yr}^{-1}$,
  which is similar to the nuclear star formation rate (SFR),
  indicating that the starburst is
  sustained by the inflow. 
  At lower $r$, $\sim100-150$\,pc, the inflow is best
    probed by the far-infrared OH ground-state doublets, 
    with an estimated 
    $\dot{M}_{\mathrm{inf}}\sim30$\,$M_{\odot}\,\mathrm{yr}^{-1}$.
  The inferred short timescale of $\sim20$\,Myr for nuclear gas
    replenishment indicates quick secular evolution,
    and indicates that we are witnessing an intermediate stage
    ($<100$\,Myr) proto-pseudobulge
   fed by a massive inflow that is driven by a strong nuclear bar.
    We also
    apply the H$_2$O model to the {\it Herschel} far-infrared spectroscopic
    observations of H$_2^{18}$O, OH, $^{18}$OH,
OH$^+$, H$_2$O$^+$, H$_3$O$^+$, NH, NH$_2$, NH$_3$, CH, CH$^+$, $^{13}$CH$^+$,
HF, SH, and C$_3$, and we estimate their abundances. 
}

   \keywords{Galaxies: bulges --
               Galaxies: evolution  --
               Galaxies: individual: ESO 320~G030  --
               Galaxies: nuclei  --
               Infrared: galaxies  --
               Submillimeter: galaxies
               }

   \maketitle
%
%-------------------------------------------------------------------

\section{Introduction}

The funneling of large amounts of gas into galaxies' nuclear regions has
profound consequences for galaxy evolution because it triggers starbursts
and leads to buried galactic nuclei that are characterized
by high column densities and dust temperatures.
This is followed by a rapid growth of supermassive black holes (SMBHs),
which appear to gain most of their mass in bright quasar modes \citep{sol82}.
Eventually, feedback unbinds the local gas supply, terminating the inflow
and stalling further SMBH growth \citep[e.g.,][]{you08}.
While mergers are one obvious mechanism for generating central mass
concentrations, secular evolution via mass inflows caused by 
disk instabilities such as bars can generate
pseudobulges \citep{kor82,kor04,com81}; however, the
physical properties of such nuclei while they are still assembling gas
are not always well understood because of the high obscuration.

Taking advantage of the availability of
far-infrared (far-IR) and (sub)millimeter
(submm) wavelengths facilities
is the best way to overcome these difficulties. 
Spectroscopy of buried nuclei in the far-IR
with the {\it Infrared Space Observatory} and the {\it Herschel Space
  Observatory} has revealed high excitation of light hydrides, mostly
water vapor (H$_2$O) and hydroxyl (OH), with the far-IR ($<200$\,$\mu$m)
lines detected in absorption and lines at longer wavelengths observed
in emission \citep[e.g.,][and references therein]{gon04,gon08,gon12,gon17}.
The specific characteristic of these lines, as compared with the
rotational lines of other more commonly used tracers (CO, HCN, HCO$^+$, etc),
is that their rotational levels are excited through the intense
  far-IR radiation generated in buried galactic nuclei,
thus directly probing the
generation of the bulk of the luminosity in these environments.
The line ratios are thus sensitive to the strength of the far-IR
  radiation density responsible for the excitation, and the absolute line
  fluxes constrain the effective sizes of the involved regions, which
  will be similar to the physical sizes when the
  surface filling factor is $\sim1$. This provides an effective spatial
  resolution that is much better than the
  low spatial resolution of these powerful spectroscopic observations.
  These nuclei are also directly imaged through observations of
the continuum at (sub)mm wavelengths \citep[e.g.,][]{sak13} or
through the observation of vibrationally excited lines of HCN
\citep[e.g.,][]{aal15} and HC$_3$N \citep[e.g.,][]{ric20}.
Nevertheless, it is highly desirable to combine these far-IR
  observations of H$_2$O with (sub)mm
  interferometric observations of a transition
  of the same species, providing a more direct and complementary way
  to probe the size and morphology of highly obscured nuclear regions.

With ALMA, the first detections in space of the ortho-H$_{2}$O 
$4_{23}-3_{30}$ transition at 448 GHz (H$_2$O448), both in the local
Universe \citep{per17} and at high redshift \citep{yan20},
offer a new way to address the need to spatially resolve these regions.
Despite the large difference in the infrared luminosities of the
two reported detections in H$_2$O448
($L_{\mathrm{IR}}=1.7\times10^{11}$\,$L_{\odot}$ for the local luminous
infrared galaxy (LIRG) ESO~320-G030, also identified as IRAS~11506-3851,
and $\sim10^{13}$\,$L_{\odot}$ for the $z=3.6$ merger
G09v1.97), the fractional luminosity of the H$_2$O448 line is similar
($L_{\mathrm{H_2O448}}/L_{\mathrm{IR}}\approx(0.85-2)\times10^{-7}$); this is
surprising because in both sources the line is generated in a
small nuclear region that accounts for only a fraction of the
total galaxy luminosity.

The main characteristics that make this line a unique probe of buried
stages of galactic nuclei are \citep[see also][]{per17,yan20}:
$(i)$ Owing to the long wavelength of the line and low transition
  probabilities, it is a deeper probe than other H$_2$O lines at
  shorter wavelengths. 
$(ii)$ The high energy of the involved rotational levels 
($>400$\,K) guarantees the filtering out of relatively cold extended
regions, thus specifically tracing the warmest nuclear regions. 
The line in ESO~320-G030 indeed comes from a region that is even more
compact than the continuum emission at 448 GHz, but it is spatially
resolved with ALMA clearly probing the innermost rotating disk.
$(iii)$ Radiative transfer calculations confirm that 
the H$_2$O448 line is
pumped through absorption of far-IR photons in the high-lying H$_2$O
lines at 79 and 132 $\mu$m, and the 79\,$\mu$m line ($4_{23}-3_{12}$) is
indeed observed in absorption with {\it Herschel} toward buried galactic
nuclei including ESO~320-G030, thus tracing the far-IR
absorption detected with {\it Herschel}/PACS.
$(iv)$ While the H$_2$O448 line is strong, our models
indicate that it is not a maser (which 
would be difficult to interpret due to the uncertain amplification), and the
low $A-$Einstein coefficient ensures that the transition 
requires high columns to emit at the observed level.
$(v)$ The H$_2$O448 line can be modeled, in combination with other
  H$_2$O lines at shorter wavelengths,
to provide crucial parameters such as the nuclear IR luminosity,
  the columns of gas, the continuum opacities, dust temperatures, and
  the kinematics of the warm and luminous nuclear ISM. 

 The observation of multiple H$_2$O lines is required to obtain a
  complete description of buried nuclear regions, which can be
  thought of as an ensemble of components with differing characteristics.
  The most extreme
  nuclear components, characterized by optical depths at 100\,$\mu$m
  $\tau_{100}>>1$ and dust temperatures $T_{\mathrm{dust}}\gtrsim100$\,K,
  are best identified with far-IR absorption lines with level energies
  at $\gtrsim600$\,K \citep{gon12}. On the other hand, the opaque
  nuclei are surrounded by massive ISM components with moderate column
  densities and $T_{\mathrm{dust}}$, which are best traced by the H$_2$O
  lines at $240-400$\,$\mu$m lines with level energies below
  $\approx300$\,K.
  \cite{per17} presented a model of
the  nucleus of ESO~320-G030 based on
the H$_2$O448 ALMA emission line and the pumping
H$_2$O 79\,$\mu$m {\it Herschel} absorption line.
While the analysis of these two lines alone provided the average
  properties of a starburst nuclear disk in the galaxy,
  the extremely rich spectrum of H$_2$O in the far-IR and submm allows us
  to obtain a more complete description of the nuclear region.
  As shown below, up to 18 lines of H$_2$O have been detected
with {\it Herschel} in ESO~320-G030. In this paper, we fully exploit the
{\it Heschel}/ALMA synergy with the goal of inferring the
physical conditions in the nuclear region of ESO~320-G030 from the full set
of H$_2$O absorption and emission lines.

At a distance of 48\,Mpc \citep[$233$ pc\,arcsec$^{-1}$,][]{per17},
ESO~320-G030 is morphologically classified as class 0
\citep[i.e., an isolated galaxy with a symmetric disk and no sign
  of past or ongoing interaction;][]{arr08},
and with a regular velocity rotational
field \citep{bel13,bel16}. Nevertheless,
it is a double-bar system \citep{gre00}, with the nuclear bar
($\mathrm{PA}=75^{\circ}$, radius of $\sim0.8$\,kpc)
  nearly perpendicular to the primary bar \citep[$\sim9$\,kpc,][]{per16}.
Evidence of high nuclear star formation activity and obscuration has already
been derived from optical and near-IR observations \citep{alo06,rod11,piq16},
and the relatively deep 9.7\,$\mu$m silicate absorption \citep{per10a}.
While far-IR spectroscopy shows inverse P-Cygni profiles in the
ground-state OH doublets suggesting inflowing gas
\citep[Fig. 11 in][and Sect.\,\ref{inflowfarir} below]{gon17},
outflows from the nucleus have also been detected in H$\alpha$
\citep{arr14} and NaD \citep{caz14,caz16}
with moderate velocities ($\lesssim100$\,km\,s$^{-1}$), and in
CO 2-1 with higher velocities \citep{per16,per20}.
There is no clear evidence for the presence of an active galactic nucleus
(AGN), either from mid-IR indicators such as the undetected 
  [O\,{\sc iv}] and [Ne\,{\sc v}] tracers
  or the mid-IR slope of the continuum \citep{per10b,alo12}, the
  observed radio properties 
  \citep[in a survey of OH megamaser galaxies by][]{baa06}, or from the
  X-ray emission and the optical spectral classification \citep{per11}.
ESO~320-G030 can be thus considered a prototype of an isolated galaxy
with strong secular evolution driven by bars during a phase
of central gas assembly with feedback already in action.

This paper is structured as follows.
We present the observations in Sect.\,\ref{sec:obs}; the analysis of the H$_2$O
data and the continuum, including a 3D modeling approach, is described in
Sect.\,\ref{sec:analysis}; we discuss the formation of the buried nucleus
in ESO~320-G030 in light of the CO $2-1$ ALMA observations at higher
spatial scales in Sect.\,\ref{sec:coh2o}, including an estimate of the mass
inflow rate based on the CO data cube and on the far-IR profiles of
  the OH doublets. Our findings are discussed and summarized in
  Sect.\,\ref{sec:discussion}. We also present in Appendix~\ref{appa}
    all {\it Herschel}/PACS wavelength ranges observed in ESO~320-G030,
    and apply the H$_2$O-based model to all other observed absorption molecular
    features to estimate the molecular abundances.

%--------------------------------------------------------------------
\section{Observations}
\label{sec:obs}

   \begin{figure*}
   \centering
   \includegraphics[width=16.0cm]{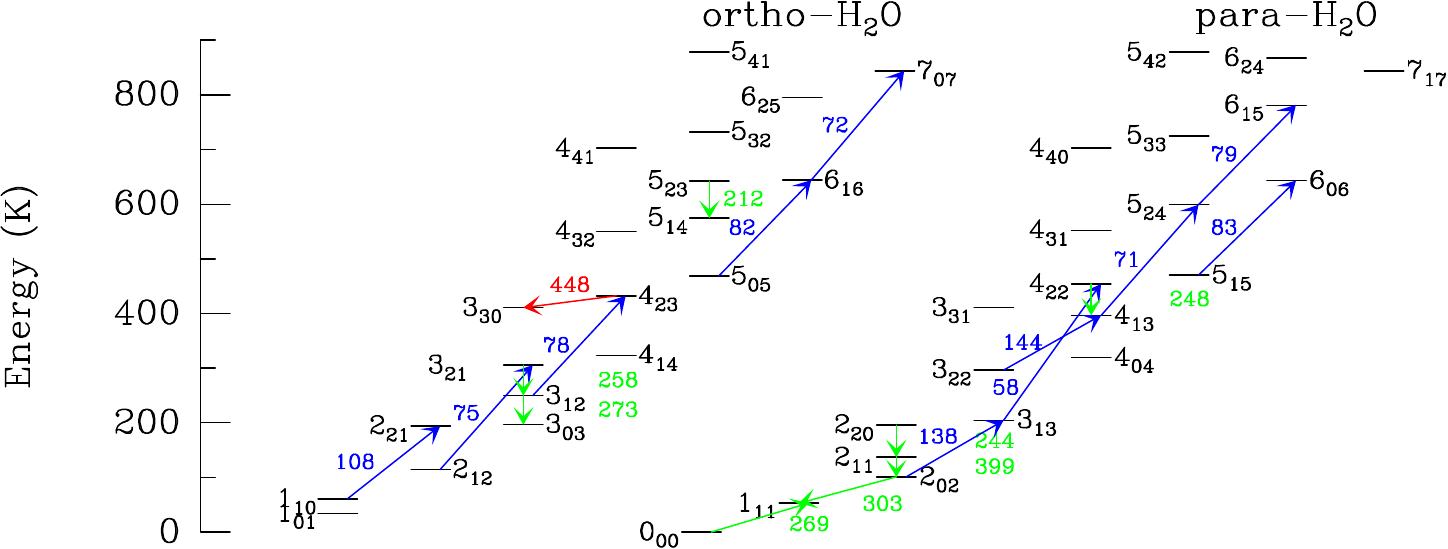}
   \caption{Energy level diagram of H$_2$O indicating the lines observed with
     {\it Herschel}/PACS (blue arrows and labels), with
     {\it Herschel}/SPIRE (green), and with ALMA (red). Labels denote
     the rounded wavelengths in $\mu$m as indicated in the second column
     of Table~\ref{tab:obs}, except for the line observed with ALMA which is
     denoted by its frequency in GHz. Upward (downward) arrows indicate lines
  detected in absorption (emission).
   }   
    \label{diag}
   \end{figure*}   

\subsection{{\it Herschel}/PACS data}

\begin{figure*}
   \centering
   \includegraphics[width=17.0cm]{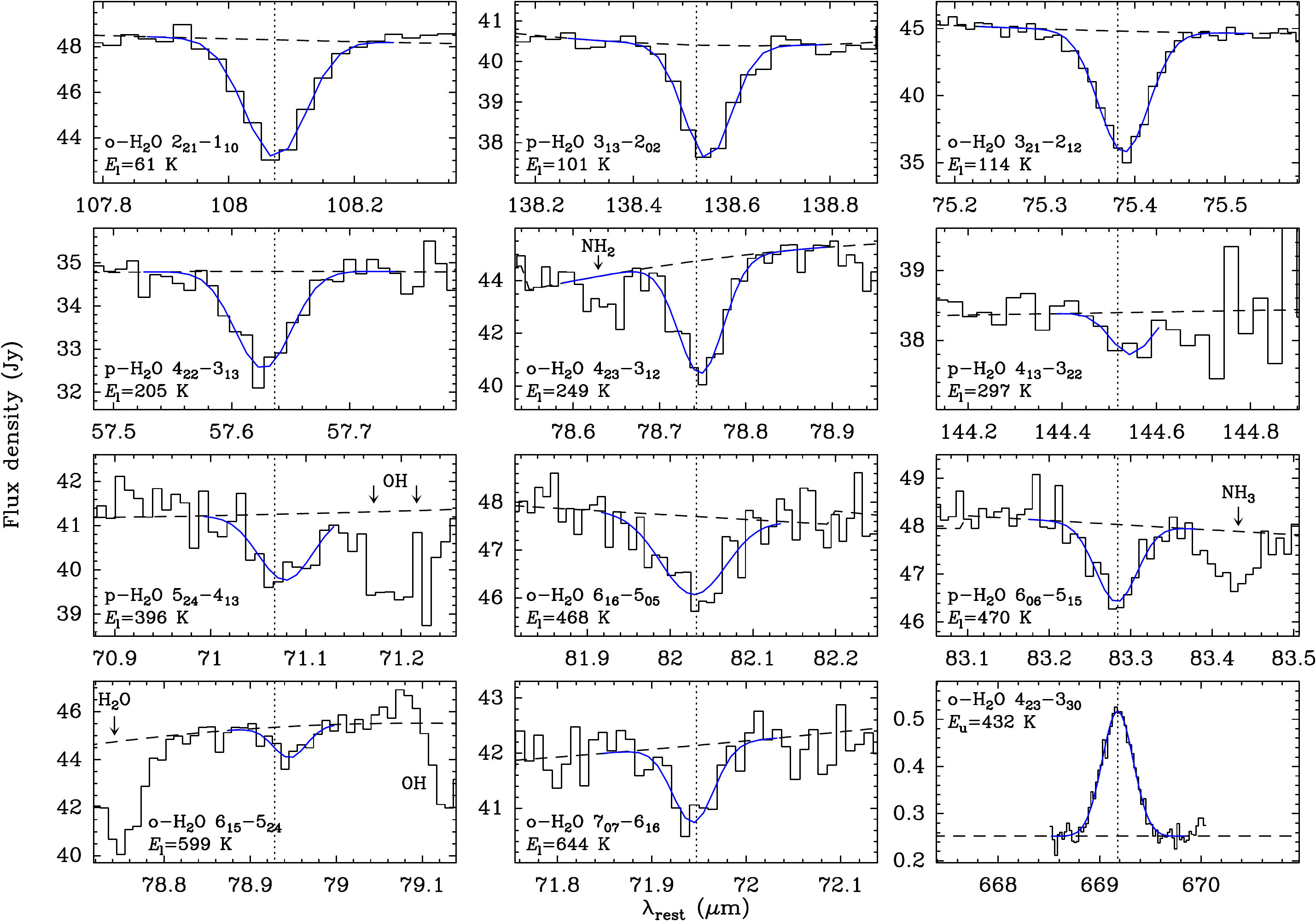}
   \caption{Spectra around the H$_2$O lines in ESO 320-G030 observed with
     {\it Herschel}/PACS and ALMA (lower-right panel), along with Gaussian
     fits to the lines (blue curves). In all panels,
     the plotted wavelength range corresponds to a velocity
     range of $\pm800$\,km\,s$^{-1}$.      
     The adopted baselines are shown with dashed lines.
     The vertical dotted lines indicate
     the expected central position of the lines by using $z=0.010266$,
     which is derived
     from the Gaussian fit to the H$_2$O448 line observed with ALMA. The
     {\it Herschel} lines are sorted by the lower-level energy ($E_l$) of the
     transition, which is also indicated in each panel.
     The species responsible for other lines in the spectra are also
       indicated (see also Appendix~\ref{appa}).
   }   
    \label{h2o-baselines}
    \end{figure*}

The {\it Herschel}/PACS \citep{pil10,pog10} spectra presented here were
obtained as part of the {\it Herschel} Open
Time (OT2) program HerMoLirg (PI: E. Gonz\'alez-Alfonso), which aimed to
observe a set of molecular lines including H$_2$O in a sample of local
(Ultra-)Luminous Infrared Galaxies ((U)LIRGs). The observed lines
  are indicated with blue arrows in the energy level diagram of
  Fig.~\ref{diag}. The
spectra were observed in high spectral sampling range-mode using the first and
second orders of the grating.  The velocity resolution of PACS in first order
ranges from $\approx320$ to 180\,km\,s$^{-1}$ over the wavelength
range from 105 to 190\,$\mu$m, and in second order from $\approx210$ to
$110$\,km\,s$^{-1}$ from 52 to 98\,$\mu$m.  The data reduction 
was performed using the PACS reduction and calibration pipeline (ipipe)
included in HIPE 14.0.1, with calibration tree version 72,
using an oversampling of four fully independent channels
(an upsample parameter of 1). The molecular
absorption lines are effectively 
point-like in ESO~320-G030, and we have thus used the point source calibrated
spectra ``c129'' produced by scaling the emission from the central
$\approx9''\times9''$ spatial pixel to the total emission from the central
$3\times3$ spaxels (``c9''), which is itself scaled according to the
point-source correction \citep[see also][]{gon17}. The absolute flux scale is  
robust to potential pointing jitter, with continuum flux reproducibility
of $\pm15$\%. The H$_2$O spectra observed with {\it Herschel}/PACS toward
ESO 320-G030 are shown in Fig.~\ref{h2o-baselines}, together with the
adopted baselines and Gaussian fits to the profiles. The resulting line
fluxes along with transition parameters are listed in
Table~\ref{tab:obs}. The individual H$_2$O lines are denoted
  according to their round-off wavelengths as indicated in the second column
  of Table~\ref{tab:obs}, except the line observed with ALMA that is denoted
according to its frequency in GHz (i.e., H$_2$O448).

All observed H$_2$O lines in the PACS wavelength range ($57-145$\,$\mu$m)
are detected in absorption, as also seen in several other buried galactic nuclei
\citep[e.g.,][]{gon04,gon08,gon12,fis14}.
Lower level energies cover a full range
of $61$ to $644$\,K, and are thus expected to probe regions with
significantly different dust temperatures ($T_{\mathrm{dust}}$). Specifically,
the highest-lying line H$_2$O72 is clearly detected, indicating the presence
of a very warm, optically thick component in the nucleus of ESO~320-G030
\citep{per17}.

\subsection{{\it Herschel}/SPIRE data}
\label{spiredata}

\begin{figure*}
   \centering
   \includegraphics[width=17.0cm]{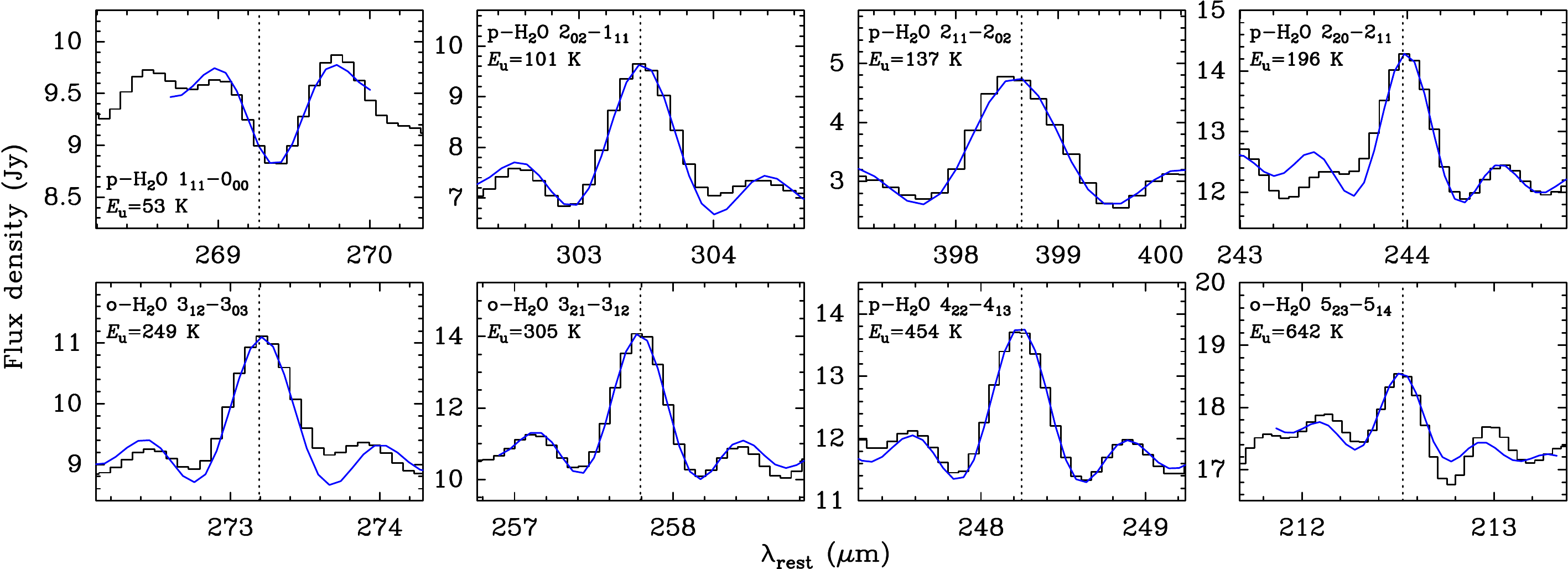}
   \caption{Spectra around the H$_2$O lines in ESO 320-G030 observed with
     {\it Herschel}/SPIRE, along with sinc fits to the lines
     (blue curves). In all panels, the plotted wavelength range
     corresponds to a velocity
     range of $\pm1200$\,km\,s$^{-1}$. The vertical dotted lines
     indicate the expected
     central position of the lines by using $z=0.01026$,
     as in Fig.~\ref{h2o-baselines}. The
     lines are sorted by the upper-level energy ($E_u$) of the
     transition, which is indicated in each panel.
   }   
    \label{h2o-baselines-spire}
    \end{figure*}

The {\it Herschel}/SPIRE \citep{gri10}
spectrum of ESO~320-G030 was obtained as part of the {\it Herschel}
Open Time Key Project {\it Hercules} (PI: P.~van der Werf). 
The SPIRE spectrometer observations cover the wavelength range 
$191-671$ $\mu$m with two spatial arrays covering two bands: SSW 
($191-318$ $\mu$m) and SLW ($294-671$ $\mu$m). 
The HIPE 15.0.1 unapodized spectra were downloaded from the archive,
with a spectral resolution of full-width half-maximum 
$\mathrm{FWHM\,(km \,s^{-1})} = 1.4472\,\lambda(\mu\mathrm{m})$. 
The observed lines
  are indicated with green arrows in Fig.~\ref{diag}, and the
  spectra are shown in Fig.~\ref{h2o-baselines-spire}.
Sinc functions on top of baselines of
order 1 were fitted around the spectra, and the resulting line fluxes
are also listed in Table~\ref{tab:obs}.

As observed in all other (U)LIRGs at low and high redshifts
\citep[e.g.,][]{gon10,yan13,yan16,liu17,lis11,per13,omo13}, the excited
submillimeter lines of H$_2$O (i.e., with the lower level of the transition
above ground-state) are observed in emission. Only the
ground-state H$_2$O\,$1_{11}-0_{00}$ (H$_2$O269) line is
observed either in emission and/or
absorption, depending on the source \citep{gon10,spi12,wei10,ran11}, with
complex intrinsic profiles in some galaxies 
as observed with the high-resolution {\it Herschel}/HIFI spectrometer
\citep{liu17}. In ESO~320-G030, the H$_2$O269 line is seen in absorption
but significantly redshifted relative to the other lines. This redshifted
absorption is also seen in other ground-state transitions, as the OH
doublets at 119 and 79\,$\mu$m, tracing an apparent
extended inflow \citep{gon17}. Cancellation of emission and absorption features
in the H$_2$O269 line within the SPIRE spectral resolution cannot be ruled out.
The ground-state H$_2$O$^+$\,$1_{11}-0_{00}\,3/2-1/2$ line at 268.85\,$\mu$m,
detected in strong absorption in M82 \citep{wei10}, is not detected in
ESO~320-G030.

The H$_2$O lines displayed in Fig.~\ref{h2o-baselines-spire} are the same as
those detected with {\it Herschel}/SPIRE toward Mrk~231 \citep{gon10}.
The lines in this wavelength range that trace the warmest dust are the
high-lying H$_2$O248 and H$_2$O212 transitions. In Mrk~231, the flux ratio of
these lines is $\mathrm{H_2O212/H_2O248}\approx0.8$, while this ratio in
ESO~320-G030 is significantly lower, $\approx0.4$. Since the H$_2$O212 line
requires warmer dust than the H$_2$O248 to be efficiently excited, the lower
ratio in ESO~320-G030 indicates lower $T_{\mathrm{dust}}$
than in Mrk~231 in the region sampled by these lines
(Section~\ref{excit}).


   \begin{figure*}
   \centering
   \includegraphics[width=17.0cm]{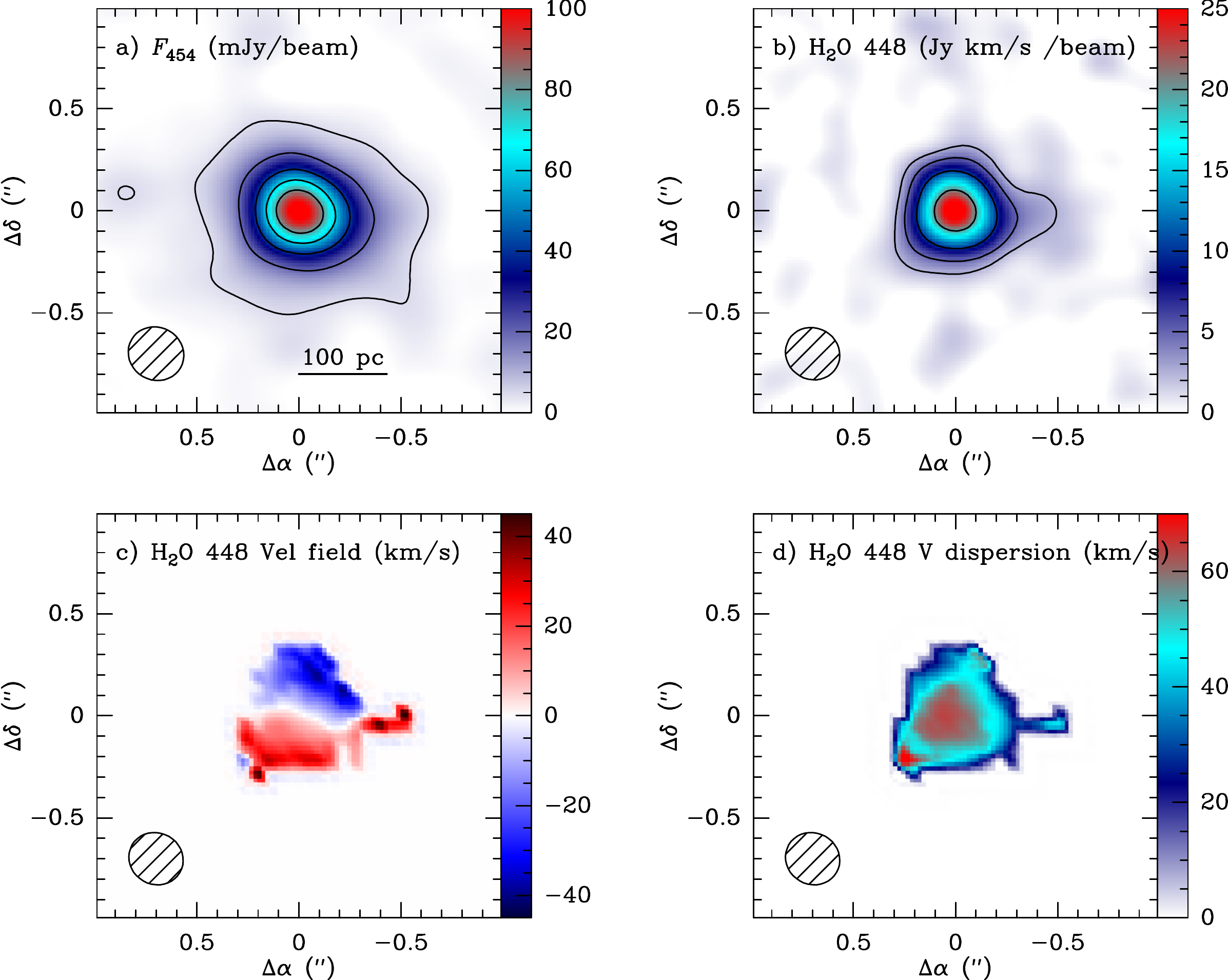}
   \caption{ALMA maps of the continuum at 454 GHz (660\,$\mu$m, panel a),
     and of the
     integrated intensity (moment 0, panel b), velocity field (moment 1,
     panel c), and velocity dispersion (moment 2, panel d) of the H$_2$O448
     line. North is up and east is left.
     The rms noise is $1.4$\,mJy\,beam$^{-1}$ in panel a and
     $0.8$\,Jy\,km\,s$^{-1}$\,beam$^{-1}$ in panel b.
     Contours are $4.5$ ($\approx3\sigma$), 20, 40, 60, and
     80\,mJy\,beam$^{-1}$ in panel a, and
     $2.5$ ($\approx3\sigma$), 5, 10, and 20\,Jy\,km\,s$^{-1}$\,beam$^{-1}$
     in panel b. The hatched ellipses indicate the synthesized beam.
   }   
    \label{almamaps}
   \end{figure*}
   
\subsection{ALMA data}

In the present study we use a new reduction of the band 8 ALMA data of
ESO~320-G030 based on the combination of extended
and compact array configurations. The observations with the extended
configuration (project \#2016.1.00263.S), with baselines ranging
from 15 to 920\,m, 42 antennas and a maximum recoverable scale of
$\sim2''$, were described in \cite{per17}.
The compact configuration has baselines from 15 to 160\,m, providing a
maximum recoverable scale of $\sim6''$.
The two data sets were calibrated using the standard ALMA reduction software
CASA \citep[version 5.4;][]{mul07}, and combined in the uv plane within the
LRSK frequency reference frame. For spectroscopic observations of the
H$_2$O448 line, a velocity resolution of $\approx10$\,km\,s$^{-1}$
($\approx16$\,MHz) was selected in the final data cubes, as well as
pixels with a size of $0.06''$. We used for the cleaning the Briggs weighting
with a robustness parameter of 0.5 \citep{bri95}, which provided
a beam with a full width at half maximum (FWHM) of $0.27\times0.25$\,arcsec$^2$
($63\times58$\,pc$^2$) and a position angle (PA) of $\approx60$\,deg.  
The resulting 1$\sigma$ sensitivity was of $\approx6$\,mJy\,beam$^{-1}$
for the 16\,MHz channels. The continuum was extracted from line-free channels
in the upper sideband at 454\,GHz. With a similar beam size and PA as for
the line observations, the achieved 1$\sigma$ sensitivity was of
$\approx1.4$\,mJy\,beam$^{-1}$.

The spectrum of the H$_2$O448 line extracted from a circular aperture 
of radius $0.48''$ is shown in the lower-right panel of
Fig.~\ref{h2o-baselines}, yielding a flux of $44.5\pm1.2$\,Jy\,km\,s$^{-1}$.
The maps of the 454\,GHz (660\,$\mu$m) continuum,
which is dominated by thermal dust emission \citep{per16,per17},
and of the
velocity-integrated intensity (moment 0), velocity field (moment 1), and
velocity dispersion (moment 2) of the H$_2$O448 line are shown in
Fig.~\ref{almamaps}. The 454\,GHz continuum flux extracted from an aperture
of radius $0.9''$ is $260.1\pm 1.7$\,mJy.
The fluxes measured in the H$_2$O448 line
and continuum are slightly higher than previously reported \citep{per17}
because of the inclusion of the compact array configuration.

   As noted in \cite{per17}, the 454\,GHz continuum, with a low-brightness
surface above $3\sigma$ level of $0.83$\,arcsec$^{2}$ (effective radius of
120\,pc), is significantly more spatially extended than the H$_2$O448 line,
which probes a nuclear disk (Fig.~\ref{almamaps}c).
The maps of both the continuum and H$_2$O448 line are elongated in
approximately the east-west direction, in contrast with the CO (2-1)
emission that traces much larger scales and probes a disk inclined
$i=43^{\circ}$ and with $\mathrm{PA}=133^{\circ}$ \citep{per16}. The continuum
at 454\,GHz and the H$_2$O448 line emission are however aproximately aligned
with the nuclear bar ($\mathrm{PA}=75^{\circ}$, see Sect.\,\ref{sec:inflowco}).

%__________________________________________________ One column table
   \begin{table*}
      \caption{H$_2$O lines in ESO 320-G030.}
         \label{tab:obs}
\begin{center}
          \begin{tabular}{lccccccc}   
            \hline
            \noalign{\smallskip}
            Transition  & Name & $\lambda_{\mathrm{rest}}$ & $E_{\mathrm{lower}}$ &
            $E_{\mathrm{upper}}$ & Flux & Obs ID   \\  
            &      & ($\mu$m)               &  (K) &
            (K)               & (Jy\,km\,s$^{-1}$) & \\
            \noalign{\smallskip}
            \hline
            \noalign{\smallskip}
 p-H$_2$O $4_{22}-3_{13}$  &  H$_2$O58   &  $ 57.636$  &  $204.7$  &  $454.3$  &  $  -700.2\pm74.5$  &  $1342248551$  \\
 p-H$_2$O $5_{24}-4_{13}$  &  H$_2$O71   &  $ 71.067$  &  $396.4$  &  $598.9$  &  $  -449.7\pm77.7$  &  $1342248549$  \\
 o-H$_2$O $7_{07}-6_{16}$  &  H$_2$O72   &  $ 71.947$  &  $643.5$  &  $843.5$  &  $  -317.7\pm55.7$  &  $1342248549$  \\
 o-H$_2$O $3_{21}-2_{12}$  &  H$_2$O75   &  $ 75.381$  &  $114.4$  &  $305.3$  &  $ -2512.2\pm54.1$  &  $1342248549$  \\
 o-H$_2$O $4_{23}-3_{12}$  &  H$_2$O78   &  $ 78.742$  &  $249.4$  &  $432.2$  &  $ -1076.5\pm52.5$  &  $1342248549$  \\
 p-H$_2$O $6_{15}-5_{24}$  &  H$_2$O79   &  $ 78.928$  &  $598.9$  &  $781.2$  &  $  -239.0\pm43.4$  &  $1342248549$  \\
 o-H$_2$O $6_{16}-5_{05}$  &  H$_2$O82   &  $ 82.031$  &  $468.1$  &  $643.5$  &  $  -592.6\pm64.9$  &  $1342248552$  \\
 p-H$_2$O $6_{06}-5_{15}$  &  H$_2$O83   &  $ 83.284$  &  $470.0$  &  $642.7$  &  $  -370.9\pm49.5$  &  $1342248552$  \\
 o-H$_2$O $2_{21}-1_{10}$  &  H$_2$O108  &  $108.073$  &  $ 61.0$  &  $194.1$  &  $ -1755.4\pm75.4$  &  $1342248550$  \\
 p-H$_2$O $3_{13}-2_{02}$  &  H$_2$O138  &  $138.528$  &  $100.8$  &  $204.7$  &  $  -736.7\pm31.2$  &  $1342248550$  \\
 p-H$_2$O $4_{13}-3_{22}$  &  H$_2$O144  &  $144.518$  &  $296.8$  &  $396.4$  &  $  -132.9\pm37.0$  &  $1342248549$  \\
 o-H$_2$O $5_{23}-5_{14}$  &  H$_2$O212  &  $212.526$  &  $574.7$  &  $642.4$  &  $   265.2\pm18.3$  &  $1342210861$  \\
 p-H$_2$O $2_{20}-2_{11}$  &  H$_2$O244  &  $243.974$  &  $136.9$  &  $195.9$  &  $   557.6\pm25.0$  &  $1342210861$  \\
 p-H$_2$O $4_{22}-4_{13}$  &  H$_2$O248  &  $248.247$  &  $396.4$  &  $454.3$  &  $   656.4\pm26.1$  &  $1342210861$  \\
 o-H$_2$O $3_{21}-3_{12}$  &  H$_2$O258  &  $257.795$  &  $249.4$  &  $305.3$  &  $  1013.5\pm26.9$  &  $1342210861$  \\
 p-H$_2$O $1_{11}-0_{00}$  &  H$_2$O269  &  $269.272$  &  $  0.0$  &  $ 53.4$  &  $  -237.4\pm30.2$  &  $1342210861$  \\
 o-H$_2$O $3_{12}-3_{03}$  &  H$_2$O273  &  $273.193$  &  $196.8$  &  $249.4$  &  $   700.1\pm29.3$  &  $1342210861$  \\
 p-H$_2$O $2_{02}-1_{11}$  &  H$_2$O303  &  $303.456$  &  $ 53.4$  &  $100.9$  &  $   883.0\pm42.8$  &  $1342210861$  \\
 p-H$_2$O $2_{11}-2_{02}$  &  H$_2$O399  &  $398.643$  &  $100.8$  &  $136.9$  &  $   878.4\pm26.9$  &  $1342210861$  \\
 o-H$_2$O $4_{23}-3_{30}$  &  H$_2$O448  &  $669.178$  &  $410.7$  &  $432.2$  &  $    44.5\pm 1.2$  &  ALMA\#2016.1.00263.S  \\
            \noalign{\smallskip}
            \hline
         \end{tabular} 
\end{center}
%\begin{list}{}{}
%\item[$^{\mathrm{a}}$]  
%\end{list}
   \end{table*}

\section{Analysis}
\label{sec:analysis}

As shown in Figs.~\ref{h2o-baselines} and \ref{h2o-baselines-spire},
a total of 20 H$_2$O lines in absorption or in emission,
with wavelengths ranging from 58 to 669\,$\mu$m, are detected in ESO~320-G030,
and an ALMA map of one high-lying line, the H$_2$O448 transition, is available,
as well as the map of the 454\,GHz continuum dominated by thermal dust emission.
This gives a unique opportunity to combine all these data, and exploit at the
maximum level the {\it Herschel}/ALMA synergy to infer the distribution
of luminosity sources, their spatial extent, dust temperatures, and
ISM column densities with unprecedented accuracy. To attain this goal, we
fit the data, including up to 3 continuum flux densities, to a linear
combination of spherically symmetric model components from a library
(Sect.\,\ref{fit}), which yields the solid angles, and hence the spatial scales
of the different components. Since H$_2$O is excited
primarily through absorption of dust-emitted photons, our fit also gives
specific predictions for the spectral energy distribution (SED) of the fitted
components, and the predicted combined SED is compared with the observed SED
(Sect.\,\ref{res}). In addition, the fit enables a Bayesian analysis that
yields the probability densities of the inferred physical parameters.
To check the reliability of these results, the components
inferred from the spherically symmetric models are combined into a physical
model in 3D, with predicted maps for the 454\,GHz continuum and the
H$_2$O448 line that are compared with the observed maps to further refine
our results (Sect.\,\ref{3d}).

\subsection{Fitting procedure}
\label{fit}

\subsubsection{Defining the data set}

We attempt to model the nuclear region of ESO~320-G030 from the
H$_2$O absorption and emission lines and the observed continuum flux
densities at some specific wavelengths. We include in the fit
all detected H$_2$O lines, which are observed with the {\it Herschel}
beam of $\sim9''$ (PACS) and $\sim20''$ (SPIRE). While the high-lying
absorption lines are indeed expected to be fully nuclear, this is not
necessarily true for the lowest-lying absorption and emission lines.
Nevertheless, the low-brightness emission observed in the 454\,GHz
continuum map (Fig.~\ref{almamaps}a) indicates the presence of a nuclear
but relatively extended ($\sim150$\,pc) component where the low-lying
absorption and emission can be formed. We will thus implicitly assume that
all H$_2$O lines with a non-ground-state lower level
($E_{\mathrm{lower}}>0$) are basically nuclear
and associated with the spatial scale of the 454\,GHz map, and results
below will show the plausibility of this assumption.

Nevertheless, we note that the
ground-state H$_2$O269 line, the only SPIRE line that is seen in absorption,
is significantly redshifted relative to the systemic velocity
(Fig.~\ref{h2o-baselines-spire}),
similar to the OH ground-state lines at 119 and 79\,$\mu$m
\citep{gon17}.
On the one hand, such absorption is expected to be produced by foreground gas
not necessarily forming part of the modeled nuclear gas.
On the other hand, an inner strong emission line is disfavored because there
is no hint of an emission feature in the blueshifted part of the line.
We therefore include the line in the fit 
with a high 1$\sigma$ uncertainty of 200\,Jy\,km\,s$^{-1}$, nearly the
value of the measured flux (Table~\ref{tab:obs}).
In addition, the very high-lying H$_2$O $8_{18}-7_{07}$ line at $63.32$\,$\mu$m,
lying close to the [O {\sc i}]\,63\,$\mu$m line, is detected in NGC~4418
\citep{gon12}, but is not detected in ESO~320-G030, with
$S<70$\,Jy\,km\,s$^{-1}$ (2\,$\sigma$). We also use below
this non-detection to further constrain the inferred physical parameters
of the core of the nucleus (Sect.\,\ref{bayesian}).

We consider in the fit
3 continuum flux densities, at 30, 428, and 660\,$\mu$m,
as constraints for fitting the SED.
The measured flux densities at 428\,$\mu$m (700\,GHz,
$1.00\pm0.05$\,Jy, Pereira-Santaella et al. in prep.) and 660\,$\mu$m
(454\,GHz, Fig.~\ref{almamaps}) are evidently
nuclear as they have been measured with ALMA. We also expect 
the 30\,$\mu$m
continuum as measured by {\it Spitzer} to be nuclear as well
and intrinsically
related to H$_2$O because H$_2$O probes the SED transition
from mid- to far-IR wavelengths \citep{gon12,fal15,fal17,ala18}.
No more continuum flux densities (e.g., in the far-IR) are included in the fit
because they may be contaminated by extended emission unrelated to H$_2$O.

\subsubsection{A library of model components}
\label{grid}

   \begin{figure*}
   \centering
   \includegraphics[width=15.0cm]{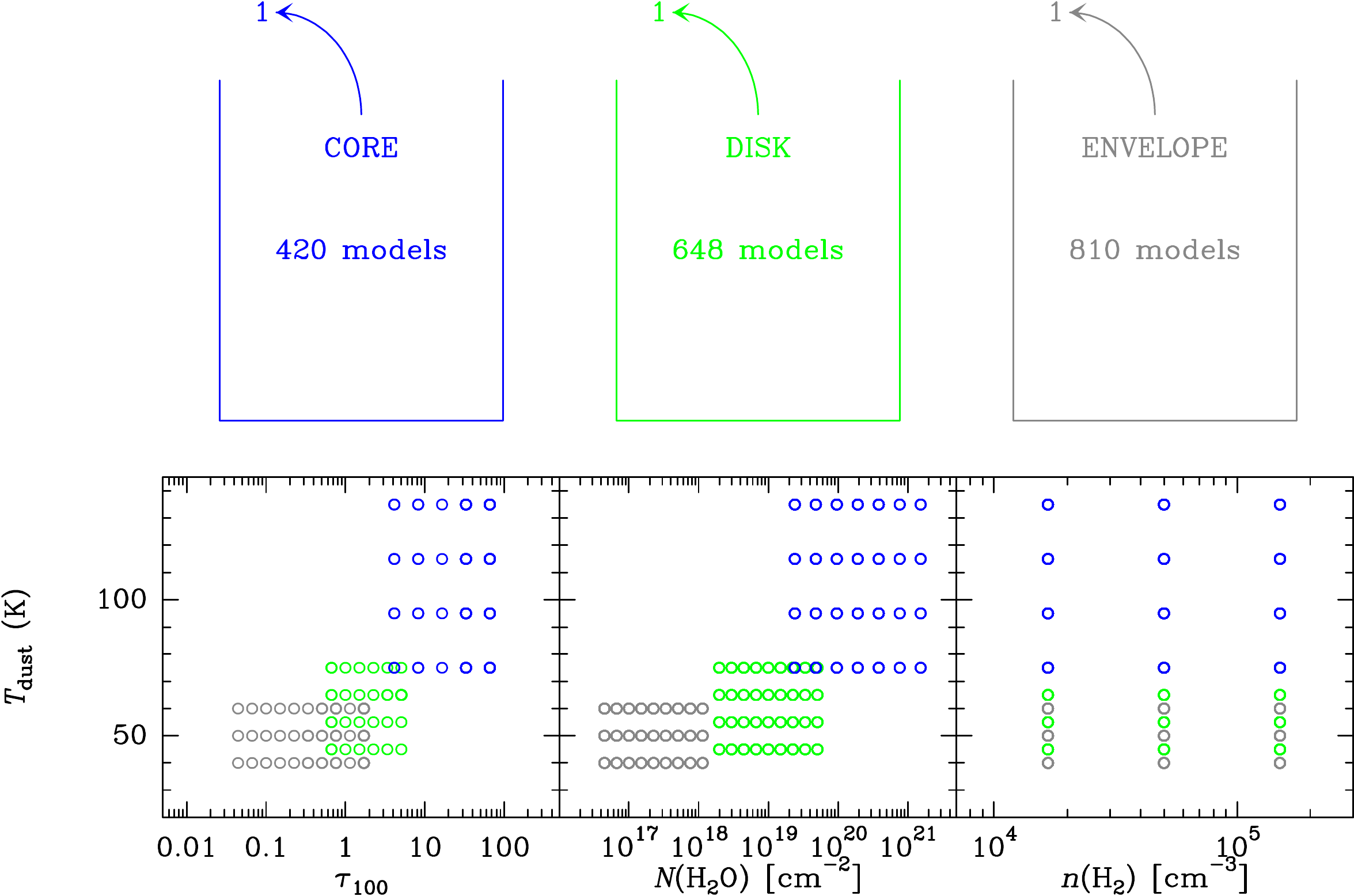}
   \caption{Spherically symmetric model components are clasified into
     three groups, the ``core'' (in blue), the ``disk'' (in green), and the
     ``envelope''  components (in gray), according to their physical
     parameters. The lower panels show the physical parameters, namely,
     $T_{\mathrm{dust}}$, $\tau_{100}$, $N_{\mathrm{H2O}}$, and $n_{\mathrm{H2}}$,
     covered by each group of components.
     Our models for ESO~320-G030 use one component from each group, yielding
     $\approx2.2\times10^8$ combinations. For each combination, $\chi^2$
     is minimized to give the solid angle subtended by each component.
   }   
    \label{figrid}
    \end{figure*}

A library of model components has been developed following the method
described in \cite{gon14}. In short, the model components
consist of spherically symmetric distributions of gas and dust,
for which the statistical equilibrium populations
of the H$_2$O rotational levels are calculated
through nonlocal, non-LTE radiative transfer calculations.
The fluxes and profiles of all involved lines and
the spectral energy distribution (SED) of the dust continuum from the
mid-IR to the mm are subsequently computed.
We assume for each component uniform physical properties:
$T_{\mathrm{dust}}$, the continuum optical depth at 100\,$\mu$m along a radial
path $\tau_{100}$, the column density of H$_2$O along a radial path
$N_{\mathrm{H2O}}$, the H$_2$ density $n_{\mathrm{H2}}$, the velocity dispersion
$\Delta V$, and the gas temperature $T_{\mathrm{gas}}$. The gas and dust
are assumed to be mixed. The physical parameters that are modified 
from model to model are $T_{\mathrm{dust}}$, $\tau_{100}$, $N_{\mathrm{H2O}}$,
and $n_{\mathrm{H2}}$, and we keep fixed $\Delta V=100$\,km\,s$^{-1}$ and
$T_{\mathrm{gas}}=150$\,K. As shown in \cite{gon14}, the excitation depends
on $N_{\mathrm{H2O}}/\Delta V$ and line fluxes are then proportional to
$\Delta V$, so that results can be easily scaled to any other value
of $\Delta V$.

While the excitation of H$_2$O is dominated by
radiative pumping, and thus our data are much more sensitive to the
parameters defining the radiation field ($T_{\mathrm{dust}}$ and $\tau_{100}$)
than to the collisional parameters ($T_{\mathrm{gas}}$ and $n_{\mathrm{H2}}$),
collisional excitation can still have an impact in populating the low-lying
(excited) levels from which the pumping cycle operates \citep{gon14}.
A significant role of H$_2$O excitation through collisions is not a
  priori expected in ESO~320-G030 given the lack of emission in the
  H$_2$O269 ground-state line (contrary to the case of NGC~1068), but we
  aim to further check this point by looking for any trend in the line
  ratios that would favor some role of collisions. To do this, we vary
$n_{\mathrm{H2}}$ keeping $T_{\mathrm{gas}}=150$\,K as a constant 
fiducial value characterizing warm (shocked) molecular gas,
so that any significant impact of collisions would be reflected in a trend
favoring high values of $n_{\mathrm{H2}}$. That we do not find  
such a trend below (Section~\ref{excit}) indicates that our results are
insensitive to our choice of $T_{\mathrm{gas}}$.

\subsubsection{Groups of model components}
\label{groups}

The model components are classified into 3 groups according to their physical
parameters (Fig.~\ref{figrid}). 
The ``core'' models are all optically thick
in the far-IR ($\tau_{100}\ge4$) and very warm ($T_{\mathrm{dust}}\ge75$\,K).
The ``disk'' models have lower $\tau_{100}$ but are still (nearly) optically
thick ($\tau_{100}\ge0.7$) with $T_{\mathrm{dust}}=45-75$\,K.
The ``envelope'' models mainly cover optically thin conditions
but can reach optically thick values ($\tau_{100}\lesssim2$),
and have moderate $T_{\mathrm{dust}}=40-60$\,K.
Each of these 3 groups covers a regular grid
in the free parameters ($T_{\mathrm{dust}}$, $\tau_{100}$,
$N_{\mathrm{H2O}}$, $n_{\mathrm{H2}}$). Models were generated and added
  to each group as needed to obtain reliable likelihood distributions of
  the above parameters, as shown below.  While $N_{\mathrm{H2O}}$ is varied by
more than 1\,dex within each group with multiplicative factors of $1.5-2$,
the grid for $n_{\mathrm{H2}}$ is coarser with only 3 values, representing
typical densities of buried galactic nuclei ($(1.7-15)\times10^4$\,cm$^{-3}$,
see Fig.~\ref{figrid}).

\subsubsection{Minimizing $\chi_{\mathrm{red}}^2$}
\label{min}

As shown below, a reasonable model fit to the present data set requires the
combination of $N_c=3$ components, one from each group (Fig.~\ref{figrid}).
We then consider all possible combinations, in a number of $2.2\times10^8$,
that are obtained by taking 1 component of each group.
For each combination, and since each component $j$ yields line fluxes and
continuum flux densities that are proportional to the solid angle
$\Delta\Omega_j=\pi\,R_j^2/D^2$, where $R_j$ is the the effective radius,
the reduced $\chi^2$ ($\chi_{\mathrm{red}}^2$) is minimized to give
$\Delta\Omega_j$:
\begin{equation}
  \chi_{\mathrm{red}}^2=\frac{1}{N_L-N_c} \sum_{i=1}^{N_L} \frac{1}{\sigma_i^2}
  \left[ \left(\sum_{j=1}^{N_c} s_{ji}^{\mathrm{comp}}\Delta\Omega_j\right)-
    S_{i}^{\mathrm{obs}}\right]^2,
  \label{chi}
\end{equation}
where $N_L=24$ is the number of H$_2$O lines and continuum flux densities
that are fitted, $S_{i}^{\mathrm{obs}}$ are the observed fluxes, $\sigma_i$ are
their uncertainties, and $s_{ji}^{\mathrm{comp}}$ is the predicted flux per unit
solid angle for line $i$ by model component $j$.
To obtain $\sigma_i$, we sum in quadrature the errors in
Table~\ref{tab:obs} and the systematic uncertanties of 15\% and 10\% for the
{\it Herschel} and ALMA measurements, respectively. 
The minimization is performed by
a standard procedure and yields both $R_j$ for each component of the
combination and then the minimum $\chi_{\mathrm{red}}^2$.

\subsubsection{Likelihood distributions}
\label{bayesian}

Our model for ESO~320-G030 has a total of $N_f=12$ free physical parameters,
($T_{\mathrm{dust}}$, $\tau_{100}$, $N_{\mathrm{H2O}}$, $n_{\mathrm{H2}}$) for each of
the 3 model components. In our approach, the sizes $R_j$ for each combination
are treated as derived rather than free parameters, as they are uniquely
determined from the $\chi_{\mathrm{red}}^2$ minimization above.
We follow \cite{war03} and \cite{kam11} in calculating the likelihood
distributions of the free physical parameters, which are collected into
vector $\mathbf{a}$. A given set of values $\mathbf{a}$ yields modeled line
fluxes or continuum flux densities that are inserted into the vector
$\mathbf{S(a)}$ of $N_L=24$ components.
%, where $N_u=1$ denotes the number of undetected lines.
We also denote as vectors $\mathbf{S^{\mathrm{obs}}}$ and $\boldsymbol{\sigma}$
the values and uncertainties measured for these quantities.
For a given set of physical parameters $\mathbf{a}$, the probability
density for measuring a set of values $\mathbf{S^{\mathrm{obs}}}$ is
\begin{eqnarray}
  P(\mathbf{S^{\mathrm{obs}}}|\mathbf{a},\boldsymbol{\sigma})
&  = & \prod_{i=1}^{N_d} \frac{1}{\sqrt{2\pi}\sigma_i} \exp \left\{
    -\frac{1}{2} \left[ \frac{S^{\mathrm{obs}}_i-S_i(\mathbf{a})}{\sigma_i}\right]^2
    \right\} \nonumber \\
& \times & \prod_{i=1}^{N_u} \frac{1}{2}
\left[1+\mathrm{erf}\left(\frac{\sigma_i-|S_i(\mathbf{a})|}{\sqrt{2}\sigma_i}
        \right)\right],
\label{prob}
\end{eqnarray}
where $N_d=23$ corresponds to the line and continuum detections and
  $N_u=1$ to the undetected H$_2$O $8_{18}-7_{07}$ line at 63.32\,$\mu$m, which
  is treated according to Appendix B by \cite{per15}.

The likelihood of a particular set of parameters $\mathbf{a}$, for a set
of measurements $\mathbf{S^{\mathrm{obs}}}$, is given by the Bayes's theorem:
\begin{equation}
  P(\mathbf{a}|\mathbf{S^{\mathrm{obs}}},\boldsymbol{\sigma})
  =\frac{P(\mathbf{a})\,
    P(\mathbf{S^{\mathrm{obs}}}|\mathbf{a},\boldsymbol{\sigma})}{
    \int d\mathbf{a} \, P(\mathbf{a}) \,
P(\mathbf{S^{\mathrm{obs}}}|\mathbf{a},\boldsymbol{\sigma})},
\label{prob2}
\end{equation}
where $P(\mathbf{a})$ is the prior probability density function.
The posterior distribution of eq.~(\ref{prob2}) is marginalized over to
  obtain the likelihood distribution of a specific parameter $a_i$,
  and of any function of parameters $f(\mathbf{a})$
  \cite[Eqs. 5 and 6 in][]{war03}.

Besides calculating the probability densities of the $N_f=12$ free parameters,
we also determine for each component the likelihood distributions for
the sizes $R_j$, the H$_2$O abundances relative to H nuclei
\citep[$X_{\mathrm{H2O}}=N_{\mathrm{H2O}}/(1.3\times10^{24}\tau_{100})$,][]{gon14},
the infrared luminosities $L_{\mathrm{IR}}$, and the fractions of the H$_2$O448
flux and 454\,GHz continuum flux density arising from each component
($f[F(\mathrm{H_2O448GHz})]$ and $f[F(\mathrm{454GHz})]$, respectively).

We started running calculations with the prior probability density function
$P(\mathbf{a})=1$ for all sets of parameters. We found in this case that some
solutions for the disk, characterized by extremely high $N_{\mathrm{H2O}}$
and low $T_{\mathrm{dust}}<50$\,K, yielded significant likelihood.
A similar situation was also found by
\cite{war03} in their bayesian analysis of the $^{12}$CO emission in M82,
where solutions with unphysically large CO column densities and low volume
densities were rejected.
We have then put a single constraint on the H$_2$O
abundance as derived above from $N_{\mathrm{H2O}}$ and $\tau_{100}$, which
implicitly assumes a gas-to-dust ratio of 100 by mass.
Models that have accounted for the H$_2$O absorption and emission in buried
galactic nuclei have shown that high H$_2$O abundances are inferred in the very
warm nuclear cores with $T_{\mathrm{dust}}\gtrsim90$\,K
\citep{gon12,fal15,fal17,ala18}. However, in the more extended regions
surrounding these cores where $T_{\mathrm{dust}}$ is moderate, $X_{\mathrm{H2O}}$
decreases to values $<10^{-5}$. To avoid unphysical solutions
of extremely high $X_{\mathrm{H2O}}$ in moderately warm environments, we use the
prior $P(\mathbf{a})=0$ whenever a model component with $T_{\mathrm{dust}}<60$\,K
and $X_{\mathrm{H2O}}>3\times10^{-5}$ is found, and $P(\mathbf{a})=1$ otherwise.

\subsection{Results}
\label{res}

   \begin{figure*}
   \centering
   \includegraphics[width=18.0cm]{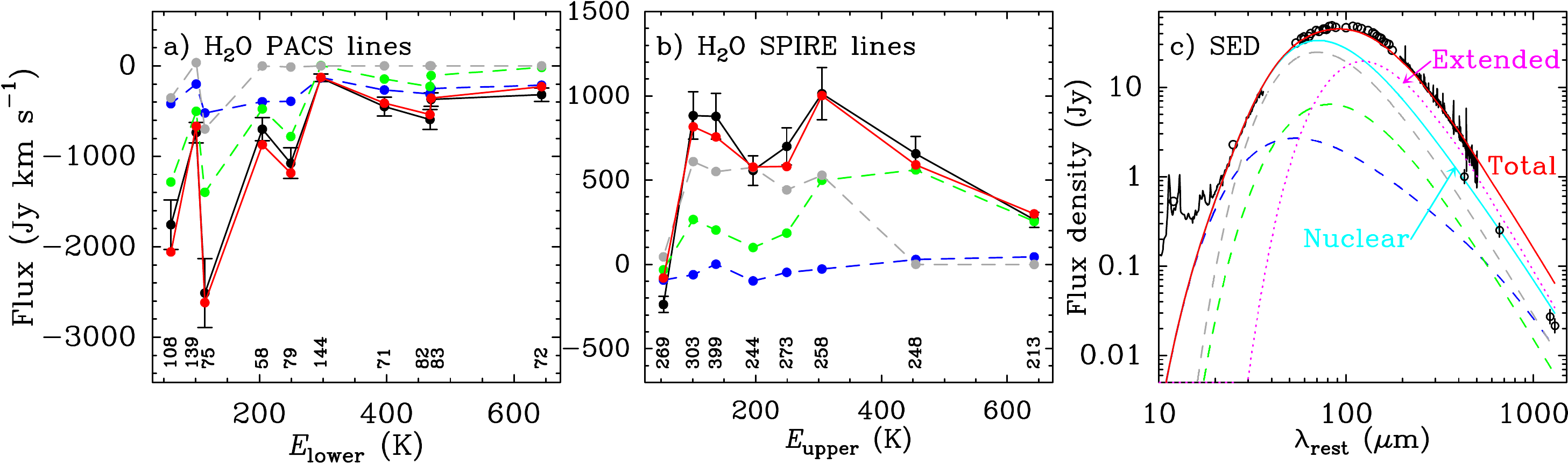}
   \caption{Our fiducial model fit (with parameters listed in
     Table~\ref{tab:res}) to the H$_2$O PACS lines (panel a),
     H$_2$O SPIRE lines (panel b), and the SED (panel c). Dashed blue, green,
     and gray lines indicate results for the three nuclear components: the core,
     the disk, and the envelope, respectively. In panels a-b,
     the combined (total) absorption or emission of the three components is
     shown in red, and the small numbers at the bottom indicate the approximate
     wavelength of the line. In panel c,
     circles at $<200$\,$\mu$m
       show both {\it IRAS} data and {\it Herschel}/PACS spectrophotometric
       data (see Appendix~\ref{appa}), with uncertainties
       better than 15\%, and circles with error bars at $>400$\,$\mu$m 
       are ALMA data for the nuclear region modeled in this work;
       we also show the {\it Spitzer}/IRS and the
       {\it Herschel}/SPIRE spectra.
     The continuum of the combined three nuclear
     components related to H$_2$O is shown in light-blue, and a nonnuclear
     (extended) component (in magenta, with $T_{\mathrm{dust}}=28$\,K) is required
     to reproduce the full SED at long wavelengths. The red line indicates
     the total (nuclear+extended) modeled SED.
   }   
    \label{bestfit}
   \end{figure*}   
   \begin{figure*}
   \centering
   \includegraphics[width=16.0cm]{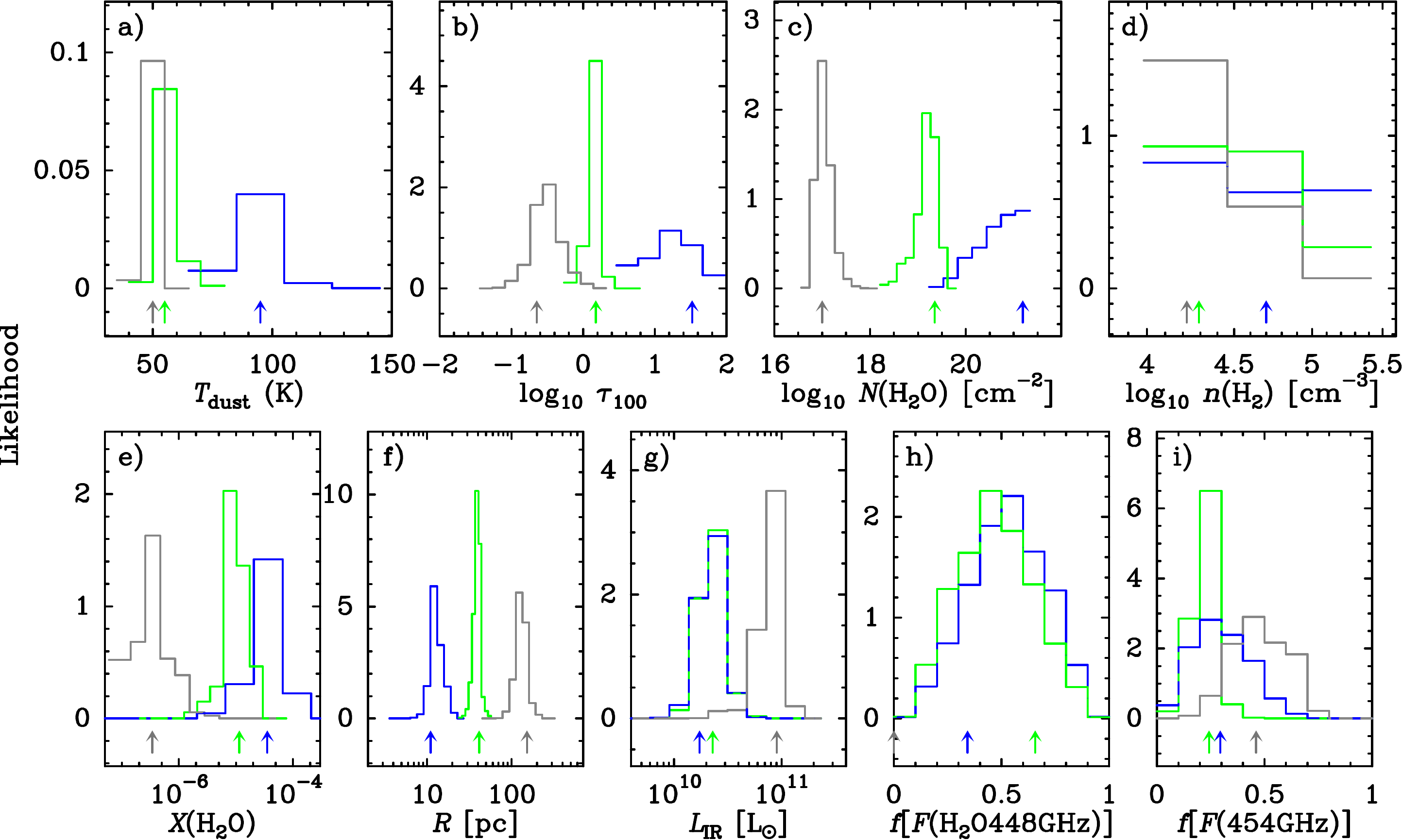}
   \caption{Bayesian analysis showing the probability densities of the
     physical parameters associated with the core (blue histograms), 
     disk (green), and envelope (gray). Panels a-d indicate results
     for the free physical parameters ($T_{\mathrm{dust}}$, $\tau_{100}$,
     $N(\mathrm{H_2O})$, and $n(\mathrm{H_2})$) and panels e-i show results
     for the derived parameters ($X(\mathrm{H_2O})$, $R$, $L_{\mathrm{IR}}$,
     and the fractions $f$ of the 448\,GHz continuum and of the H$_2$O448
     emission that arise from each component). The small arrows at the
     bottom of each panel indicate the values of the fiducial model
     in Fig.~\ref{bestfit}. In panel h, the contribution $f$ to the
     H$_2$O448 line from the envelope is not shown because it is negligible
     in all models. The median and 90\% confidence intervals are
     listed in Table~\ref{tab:res}.
   }   
    \label{bayes}
    \end{figure*}

%__________________________________________________ Two columns table
   \begin{table*}
 \caption{Model results from H$_2$O multitransition analysis of ESO 320-G030}
         \label{tab:res}
\begin{center}
          \begin{tabular}{l|ccc|ccc|ccc}   
            \hline
            \noalign{\smallskip}
  & \multicolumn{3}{c|}{CORE} & \multicolumn{3}{c|}{DISK} & \multicolumn{3}{c}{ENVELOPE}  \\  
Parameter    &  Median & Range$^{\mathrm{a}}$ & Fiducial$^{\mathrm{b}}$ & Median & Range$^{\mathrm{a}}$ & Fiducial$^{\mathrm{b}}$ & Median & Range$^{\mathrm{a}}$ & Fiducial$^{\mathrm{b}}$   \\
            \noalign{\smallskip}
            \hline
            \noalign{\smallskip}
 $T_{\mathrm{dust}}$\,(K)                       & $ 97.3$ & $ 85.7,121.0$ & 95.0 & $ 54.1$ & $ 42.9, 59.8$ & 55.0 & $ 49.7$ & $ 45.0, 54.5$ & 50.0 \\
 $\tau_{100}$                                   & $ 21.8$ & $  4.0, 74.7$ & 32.0 & $  1.5$ & $  0.9,  1.8$ & 1.5 & $  0.3$ & $  0.1,  0.7$ & 0.22 \\
 log$_{10}$\,$N_{\mathrm{H2O}}$\,(cm$^{-2}$)    & $ 20.9$ & $\gtrsim20.0$ & 21.2 & $ 19.2$ & $ 18.9, 19.7$ & 19.3 & $ 17.0$ & $ 16.8, 17.5$ & 17.0 \\
 log$_{10}$\,$n_{\mathrm{H2}}$\,(cm$^{-3}$)     & $  4.7$ & $  4.1,  5.3$ & 4.7 & $  4.6$ & $  4.0,  5.3$ & 4.2 & $  4.4$ & $  4.0,  5.1$ & 4.2 \\
 $R$\,(pc)                                      & $ 11.8$ & $  8.9, 15.5$$^{\mathrm{c}}$ & 11.0 & $ 41.7$ & $ 35.0, 54.3$ & 41.4 & $129.7$ & $ 91.7,159.6$ & 130 \\
 log$_{10}$\,$L_{\mathrm{IR}}$\,(L$_{\odot}$)   & $ 10.4$ & $ 10.1, 10.6$ & 10.2 & $ 10.4$ & $ 10.1, 10.5$ & 10.4 & $ 10.9$ & $ 10.5, 11.0$ & 10.9 \\
 log$_{10}$\,$M_{\mathrm{gas}}$\,(M$_{\odot}$)  & $  8.1$ & $  7.5,  8.7$ &  8.2    & $  8.0$ & $  7.7,  8.2$ & 8.1 & $  8.4$ & $  8.1,  8.6$ & 8.2     \\
 $f[F(\mathrm{H_2O}448)]$                       & $  0.5$ & $  0.2,  0.7$ & 0.34 & $  0.5$ & $  0.3,  0.8$ & 0.66 & $  0$ & $  0.0,  0.1$ & 0 \\
 $f[F(454\,\mathrm{GHz})]$                      & $  0.3$ & $  0.1,  0.5$ & 0.29 & $  0.2$ & $  0.1,  0.4$ & 0.24 & $  0.5$ & $  0.2,  0.7$ & 0.47 \\
 log$_{10}$\,$X_{\mathrm{H2O}}$                 & $ -4.5$ & $ -5.1, -3.9$ & $-4.36$ & $ -5.0$ & $ -5.3, -4.4$ & $-4.85$ & $ -6.5$ & $ -7.0, -5.8$ & $-6.38$ \\
            \noalign{\smallskip}
            \hline
         \end{tabular} 
\end{center}
\begin{list}{}{}
\item[$^{\mathrm{a}}$] 90\% confidence intervals
\item[$^{\mathrm{b}}$] Values for the fiducial model, selected for detailed
  comparison with data (see Sect.\,\ref{res})
\item[$^{\mathrm{c}}$] Assuming $\Delta V=100$\,km\,s$^{-1}$ for the core component
\end{list}
   \end{table*}
%__________________________________________________ 

   The values of $\chi_{\mathrm{red}}^2$ for the best-fit $10^3$ combinations
   are in the range $1.0-1.4$, indicating that three components
     provide a good fit and more are not needed. On the other hand, 
     the minimum value of $\chi_{\mathrm{red}}^2$ significantly increases to 2.3
     when only 2 components are used.
   Based on the superior comparison between the observed and
model-predicted maps of the H$_2$O448 and 454\,GHz continuum emission
(see Sect.\,\ref{3d}), we have selected a specific model combination, with
$\chi_{\mathrm{red}}^2=1.098$, as the fiducial model for detailed comparison
with the data.
Results for the fiducial model are compared with {\it Herschel}
data and with the observed SED of ESO~320-G030 in Fig.~\ref{bestfit}, and
Fig.~\ref{bayes} displays the probability distributions of the free (upper row)
and derived (lower row) parameters. The modeled and observed profiles of
the {\it Herschel}/PACS and ALMA lines, and of the {\it Herschel}/SPIRE lines
are compared in Figs.~\ref{fitpacsalma} and \ref{fitspire}, respectively.
Median likelihood estimators and 90\% confidence intervals, together with
the values of the parameters of the fiducial model, are listed in
Table~\ref{tab:res}. We also evaluate the degeneracy among the
  free parameters by showing in Appendix~\ref{appb} their marginalized
  2D posterior distributions.
   
\subsubsection{The core, disk, and envelope components}
\label{fiducial}

   As stated in Sect.\,\ref{min}, we require 3 components to attain a
   reasonable fit to the whole data set,
   which can be now justified in the light of
Figs.~\ref{bestfit}a-b and \ref{bayes}. To fit the high-lying absorption lines
($E_{\mathrm{lower}}\gtrsim300$\,K) observed with {\it Herschel}/PACS, a very warm
($T_{\mathrm{dust}}\gtrsim80$\,K) and optically thick at 100\,$\mu$m
``core component'' is required.
Its very small effective size ($R\sim12$\,pc, Fig.~\ref{bayes}f) suggests
  a torus around an AGN, such as that of NGC~1068 \citep[see][]{gar19}
  but with a much higher column density and mass (Sect.\,\ref{masses}); it could
  also represent super star clusters in a very early stage of evolution
  \citep{ric20} spread over the nuclear region.
  Because of the compactness of this component, it cannot be solely
  responsible for the measured fluxes of the rest of H$_2$O absorption lines. 
Therefore, the inclusion of a ``disk component'' is required, with a more
moderate $T_{\mathrm{dust}}\gtrsim55$\,K but still optically thick in the far-IR
($\tau_{100}\approx1.5$, Fig.~\ref{bayes}a-b).
Its size, $R\sim40$\,pc (Fig.~\ref{bayes}f), is similar to the size
  of the disk observed in the H$_2$O448 ALMA line (Fig.~\ref{almamaps}c);
  this line is indeed predicted
to be formed in both the core and disk components (Fig.~\ref{bayes}h).
The disk mainly accounts for most of the observed flux in the absorption
lines with $E_{\mathrm{lower}}<300$\,K and for the high-lying lines observed
with SPIRE in emission (H$_2$O248 and 212), contributing in addition
significantly to many of the remaining lines.
However, the disk cannot fully account for the low-lying
($E_{\mathrm{lower}}\lesssim300$\,K) emission lines (Fig.~\ref{bestfit}b),
and an extended, optically
thin component ($\tau_{100}<1$) is additionally required. This ``envelope
component'', which is also moderately warm ($T_{\mathrm{dust}}\approx50$\,K),
is predicted to have a radius of $\sim130-150$\,pc (Table~\ref{tab:res}),
similar to the extent of the low-brightness
surface seen in the 454\,GHz map (Fig.~\ref{almamaps}a). This consistency in
sizes supports our assumption that most of the H$_2$O low-lying
emission observed with {\it Herschel}/SPIRE is of nuclear origin,
although some extra-nuclear contribution to the lowest-lying
  H$_2$O lines is not ruled out.
The optical depths at 100\,$\mu$m, sizes, and H$_2$O column densities of the
three components have distributions with little overlap
(Fig.~\ref{bayes}b-c-f), which supports the reliability of our three
model components approach.

\subsubsection{H$_2$O excitation, column densities, and abundances}
\label{excit}

While our model grid only explores results for a fixed
$T_{\mathrm{gas}}=150$\,K and 3 (expectedly representative) values of
$n_{\mathrm{H2}}$, Fig.~\ref{bayes}d indicates that the excitation of H$_2$O is
dominated by radiative pumping: The flat distribution in densities
for the core indicates that results for this component are insensitive to
$n_{\mathrm{H2}}$; for the envelope, results strongly favor low $n_{\mathrm{H2}}$,
and low or moderate densities of several\,$\times10^4$\,cm$^{-3}$ are favored
for the disk. We expect that $T_{\mathrm{gas}}$ varies across the
  different components, and that the derived
  densities would be higher than suggested by
Fig.~\ref{bayes}d if $T_{\mathrm{gas}}$ were lower than assumed.
Our models, however, do not require the use of a varying $T_{\mathrm{gas}}$
  because collisional excitation does not appear to play a key role in the
excitation of H$_2$O.

The column densities $N_{\mathrm{H2O}}$ of the envelope and disk components are
well defined and very different, $\sim10^{17}$ and $\sim10^{19}$\,cm$^{-2}$
respectively (Fig.~\ref{bayes}c). Estimating the H column densities from the
continuum optical depth at 100\,$\mu$m, the resulting abundances
  $X_{\mathrm{H2O}}$ are $\sim3\times10^{-7}$ and $\sim10^{-5}$ for the envelope
  and the disk, respectively.

   \begin{figure*}
   \centering
   \includegraphics[width=15.0cm]{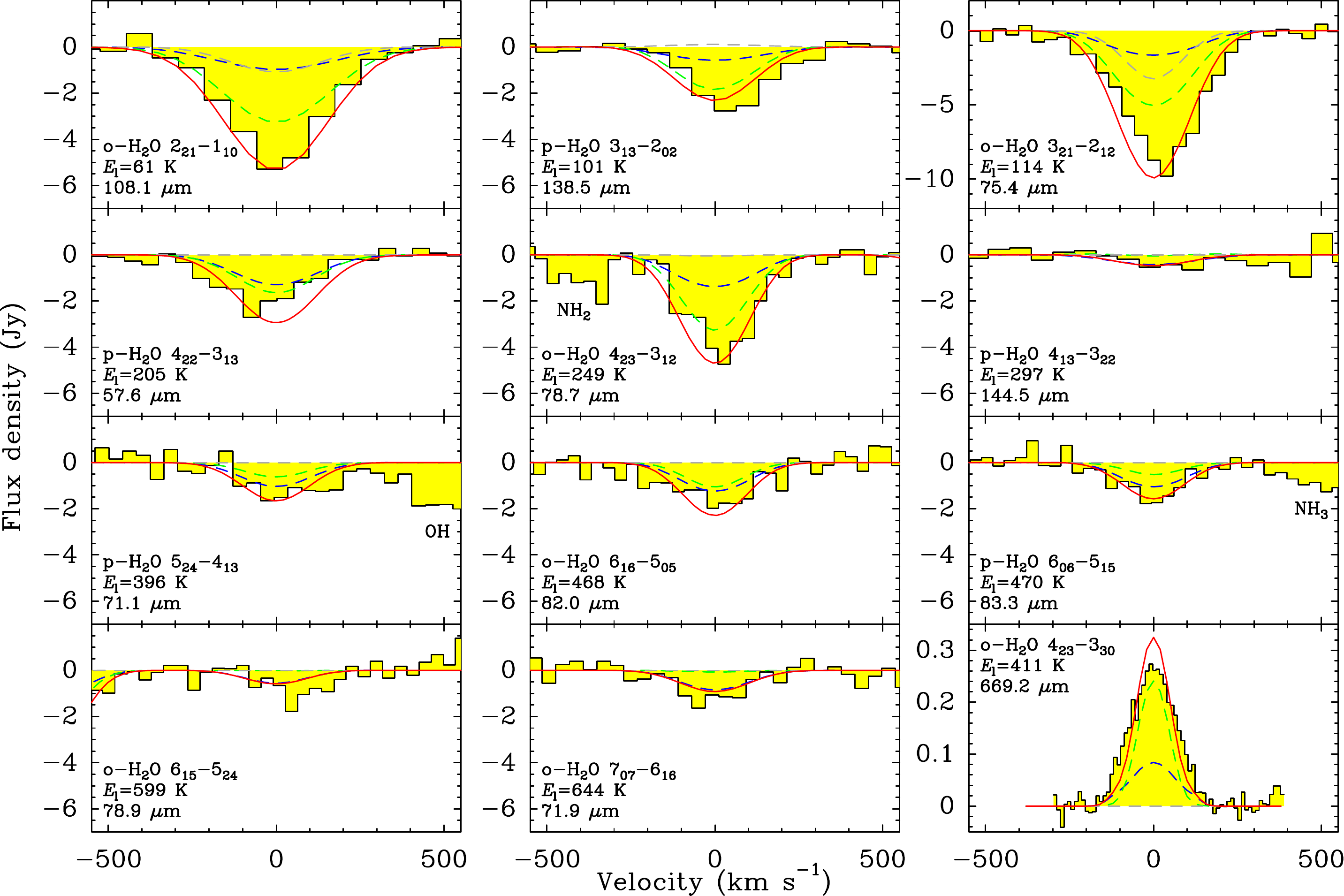}
   \caption{Fiducial model fit to the H$_2$O PACS, and ALMA lines.
     Black histograms show the observed continuum-subtracted spectra, and
     dashed curves show the contribution by the core component (blue), the
     inner disk (green), and the outer component (gray). The total predicted
     absorption or emission is shown in red.
     Spectral features due to NH$_2$, OH,
     and NH$_3$ lying in the plotted wavelength ranges are also indicated
     (see Appendix~\ref{appa}).
   }   
    \label{fitpacsalma}
    \end{figure*}
   \begin{figure*}
   \centering
   \includegraphics[width=15.0cm]{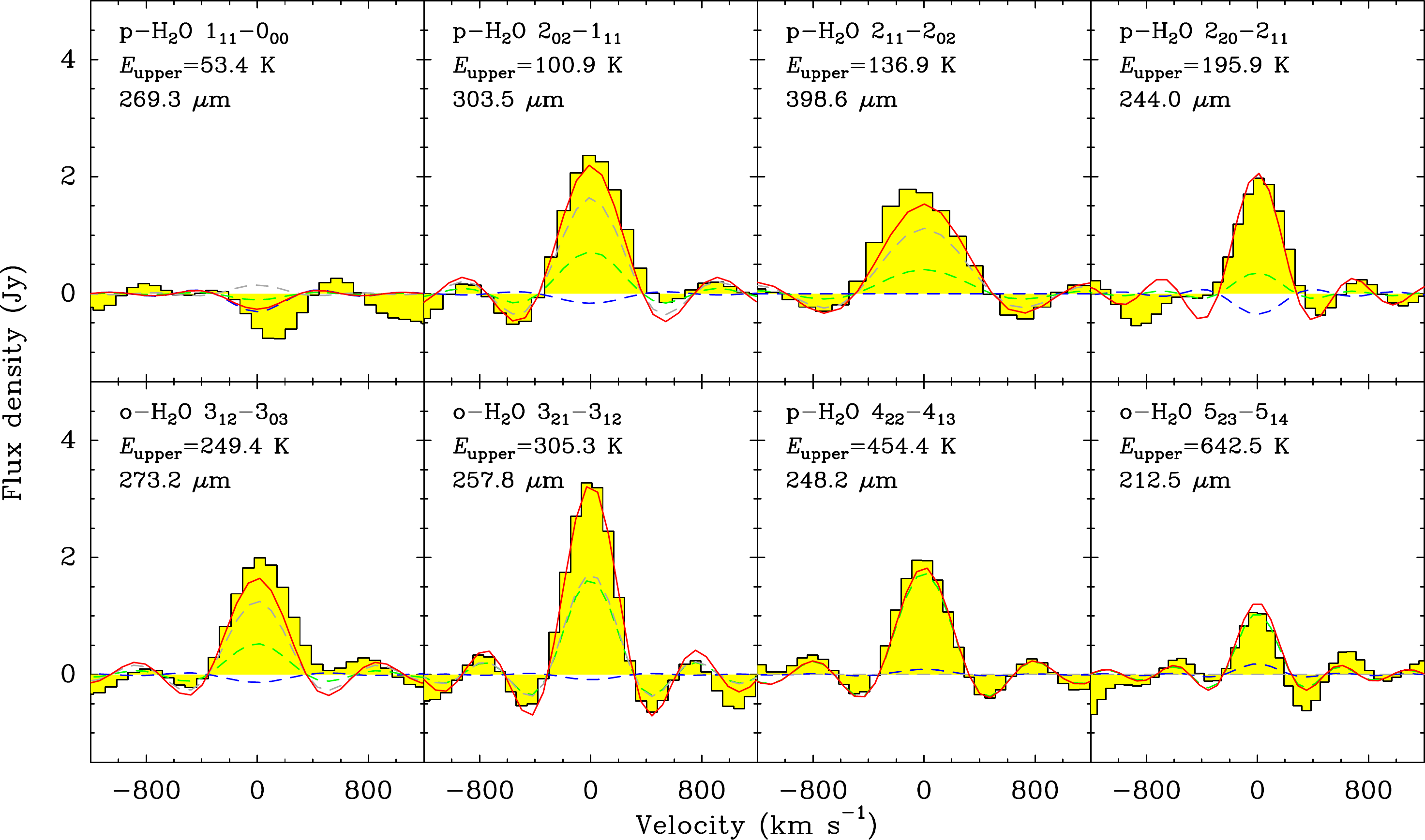}
   \caption{Fiducial model fit to the H$_2$O SPIRE lines.
     Black histograms show the observed continuum-subtracted spectra, and
     dashed curves show the contribution by the core component (blue), the
     inner disk (green), and the outer component (gray). The total predicted
     absorption or emission is shown in red.
   }   
    \label{fitspire}
   \end{figure*}
   
\begin{figure}
   \centering
   \includegraphics[width=8.8cm]{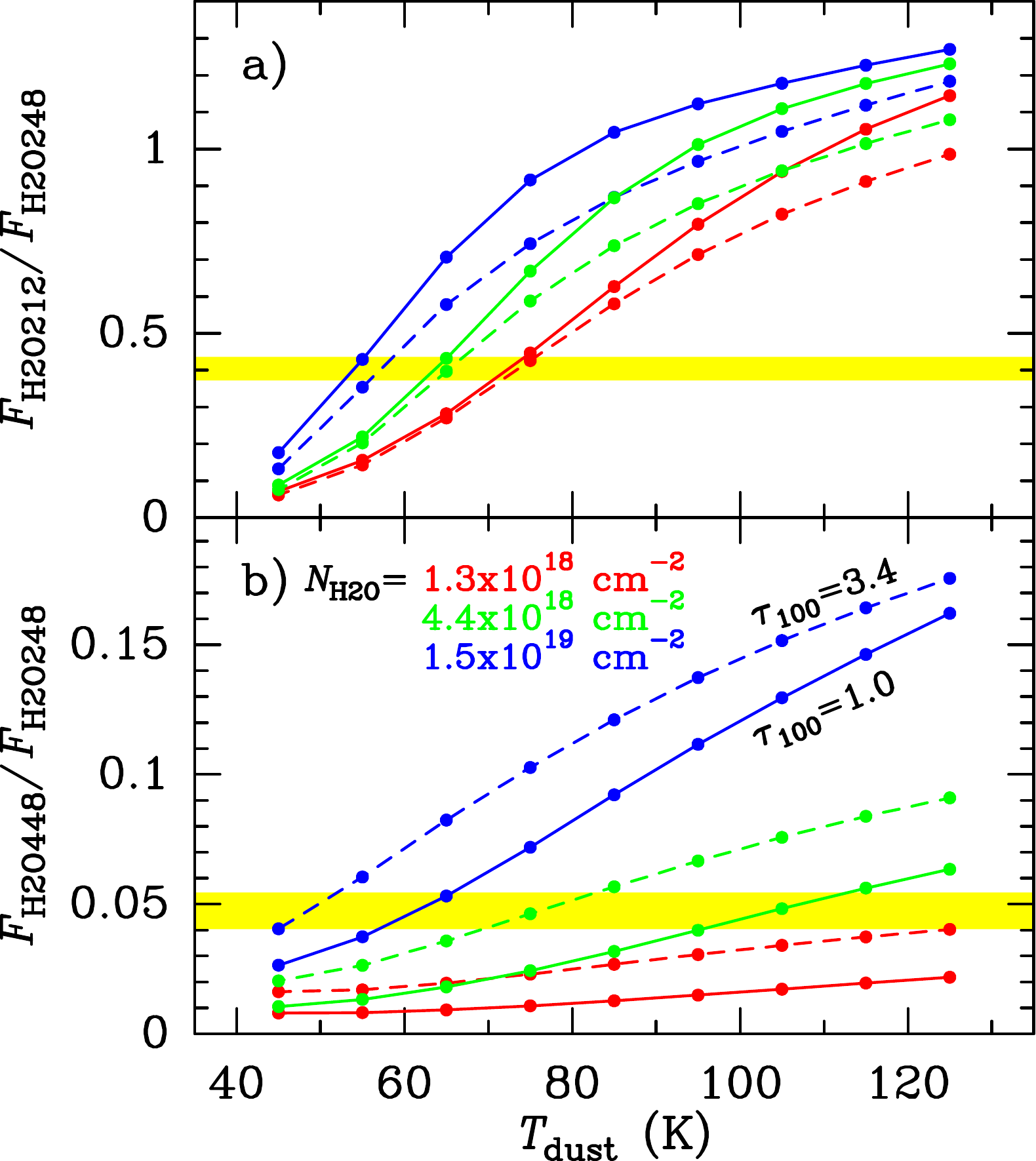}
   \caption{Modeled H$_2$O line ratios as a function of $T_{\mathrm{dust}}$
     (colored lines), compared with the measured ratios (appropriate
     for the disk, in yellow).
     The colors indicate the H$_2$O column densities as indicated in panel b,
     and solid and dashed lines correspond to $\tau_{100}=1.0$ and 3.4,
     respectively.
     In panel b, the measured $F_{\mathrm{H2O448}}/F_{\mathrm{H2O248}}$ has been
     corrected by assuming that 70\% of $F_{\mathrm{H2O448}}$ arises from the disk.
     While the observed $F_{\mathrm{H2O212}}/F_{\mathrm{H2O248}}\approx0.4$ ratio in
     panel a can be explained with a range of $T_{\mathrm{dust}}$ and
     $N_{\mathrm{H2O}}$ (increasing $T_{\mathrm{dust}}$ with decreasing
     $N_{\mathrm{H2O}}$), the measured
     $F_{\mathrm{H2O448}}/F_{\mathrm{H2O248}}$ breaks the degeneracy favoring
     the highest $N_{\mathrm{H2O}}$ and moderate $T_{\mathrm{dust}}\lesssim65$\,K.
   }   
    \label{h2oratios}
   \end{figure}

The disk component, which has high $N_{\mathrm{H2O}}$ and is optically thick in
the far-IR, nevertheless has a moderate $T_{\mathrm{dust}}\approx55$\,K. 
Since the H$_2$O248 and H$_2$O212 emission lines are expected to arise
  predominantly from the nuclear disk (Figs.~\ref{bestfit}b and \ref{fitspire}),
  we use their ratio in Fig.~\ref{h2oratios}a to better demonstrate the
  origin of the physical conditions inferred for this component.
  The measured H$_2$O212-to-H$_2$O248
  flux ratio of $\approx0.4$ is by itself consistent with a range of
  $T_{\mathrm{dust}}\approx55-75$\,K depending on $N_{\mathrm{H2O}}$, with
  $T_{\mathrm{dust}}$ decreasing with increasing $N_{\mathrm{H2O}}$. This
  degeneracy is broken when considering the H$_2$O448 line observed
  with ALMA. The measured H$_2$O448-to-H$_2$O248 flux ratio in
  Fig.~\ref{h2oratios}b has been corrected to account for only the fraction
  of the H$_2$O448 flux, $\approx70$\%, arising from the disk (see
  Section~\ref{h2o3d}). Even with this correction, the resulting ratio of
  $\approx0.048$ is so high that it cannot be explained with the lowest
  $N_{\mathrm{H2O}}=1.3\times10^{18}$\,cm$^{-2}$ considered in
  Fig.~\ref{h2oratios}b, but requires higher columns. The highest
  $N_{\mathrm{H2O}}=1.5\times10^{19}$\,cm$^{-2}$ and $T_{\mathrm{dust}}\approx55$\,K
  are mostly consistent with both ratios displayed in Fig.~\ref{h2oratios},
  although the increase in $\tau_{100}$, as favored in Appendix~\ref{appa},
  would also enable warmer $T_{\mathrm{dust}}\sim65$\,K and lower
  $N_{\mathrm{H2O}}\sim5\times10^{18}$\,cm$^{-2}$. The inferred extremely high
  $N_{\mathrm{H2O}}$ in the disk is consistent with the strong absorption and
  emission in the H$_2^{18}$O and $^{18}$OH lines, which still require a low
  $\mathrm{^{16}O/^{18}O}\sim100$ abundance ratio (Appendix~\ref{appa}).

In the core component, only a lower limit for $N_{\mathrm{H2O}}$ of
$\sim10^{20}$\,cm$^{-2}$ is obtained.
Primarily responsible for the very high-lying excitation observed with
{\it Herschel}/PACS H$_2$O lines in absorption (specifically the H$_2$O72 line
with $E_{\mathrm{lower}}=644$\,K), all lines -including the submillimeter
H$_2$O448 transition- are saturated in this component.
The values of $\tau_{100}$ and $X_{\mathrm{H2O}}$ are also rather uncertain
for the core given its extremely buried conditions.

\subsubsection{The fit to the SED and the nuclear SFR}
\label{sed}

The SED predicted by our fiducial model, shown in Fig.~\ref{bestfit}c,
is rather representative of all best-fit combinations. In the transition
from the mid- to far-IR wavelengths ($30-50$\,$\mu$m), the SED is dominated
by the optically thin, extended envelope, but the flux densities in the
(sub)millimeter from the three components are expected to be comparable.
The three nuclear components combined, however, account for a
luminosity of $L_{\mathrm{IR}}=(1.23\pm0.17)\times10^{11}$\,L$_{\odot}$
  (light-blue curve in Fig.~\ref{bestfit}c for the fiducial model),
  that is, $\sim70$\% of the total galaxy
luminosity. To fit the whole SED from $20$ to
$550$\,$\mu$m as observed with {\it Spitzer} and {\it Herschel}/PACS and SPIRE,
an additional extra-nuclear component has been included in
Fig.~\ref{bestfit}c (magenta curve, labeled extended),
with $T_{\mathrm{dust}}=28$\,K and $L_{\mathrm{IR}}\approx4\times10^{10}$\,L$_{\odot}$.
ALMA and {\it Herschel}/SPIRE measure continuum 
flux densities at 428 $\mu$m of $\approx1$ and $\approx2.5$\,Jy, respectively,
indeed indicating the presence of a continuum component missed by ALMA,
also at longer wavelengths ($>1$\,mm, Fig.~\ref{bestfit}c).
This component is expected to be of much larger extent than the nuclear
components traced by H$_2$O, and associated with star formation in the rest
of the galaxy.
There is indeed prominent Pa-$\alpha$ emission well outside the
nuclear region \citep[][see also Sect.\,\ref{sec:inflowco}]{alo06}.

Using the \cite{ken12} star formation rate (SFR) calibration of the total
IR luminosity, which is based on the works by \cite{mur11} and \cite{hao11},
the total and nuclear SFR are $\approx25$ and
$\approx18$\,M$_{\odot}$\,yr$^{-1}$, respectively.
These values also assume that the
very optically thick and compact core
component is powered by star formation; if we assume that its luminosity
is driven by an extremely buried AGN, the nuclear SFR is derived
  from the IR luminosities of only the disk and envelope to give
$\approx16$\,M$_{\odot}$\,yr$^{-1}$.
Our inferred nuclear SFR is $\sim40$\% higher than the
values previously estimated
\citep[$11-13$\,M$_{\odot}$\,yr$^{-1}$;][]{rod11,per16}.

The distribution of infrared luminosities $L_{\mathrm{IR}}$ of the three nuclear
components, shown in Fig.~\ref{bayes}g, indicate rather surprisingly similar
values for the compact core and the more extended disk. This could suggest
that the disk is to some extent heated by (and reemitting) the radiation
coming out from the core. However, the disk cannot surround the core on the
front side (as seen from the Earth) because the former is optically thick
in the far-IR continuum and hence the core would not be detected in the
far-IR H$_2$O lines. If the disk extends only on the sides of the core, it will
only intercept a fraction of the luminosity emitted by the latter, thus limiting
the nonlocal heating effect at spatial scales of $\sim40$\,pc.

\subsubsection{The H$_2$O448 line and the 454\,GHz continuum emission}
\label{h2o448}

The relative contributions of the three nuclear components to
the 454\,GHz total flux density of $\approx250$\,mJy
($f[F(\mathrm{454GHz})]$) are uncertain. While the contribution
by the disk is expected to be around $25$\%, the contributions by
the core and the envelope show broad distributions (Fig.~\ref{bayes}i).
This uncertainty is due to the distributions in sizes for the envelope
and core, and to the relatively broad distribution found for the optical depth
$\tau_{100}$ of the core (Fig.~\ref{bayes}b).
The fiducial model has continuum optical depths at 454\,GHz of $1.5$, $0.07$,
and $0.01$ for the core, disk, and envelope, respectively.

The relative contributions of the core and disk components to the
H$_2$O448 flux are even more uncertain (Fig.~\ref{bayes}h).
(The envelope makes a negligible contribution to this line in any case.)
Since the H$_2$O448 line is seen in emission, it is potentially sensitive to the
volume of the source (rather than to the surface, as is the case for
absorption lines) except when saturated, and its flux also depends on the
details of the extinction within the core at 448\,GHz. 
Nevertheless, this ambiguity is solved below (Sect.\,\ref{3d})
because we have the maps of both the H$_2$O448 line and 454\,GHz continuum
emission, which can be compared with predictions from the model combinations.

\subsubsection{Gas masses}
\label{masses}

We calculate the gas mass of each component traced by H$_2$O as
\begin{equation}
  M_{\mathrm{gas}}=\pi R^2 \tau_{100} \left(\frac{N_{\mathrm{H}}}{\tau_{100}}\right)
  \,\mu\,m_{\mathrm{H}},
\end{equation}
where $N_{\mathrm{H}}/\tau_{100}=1.3\times10^{24}$\,cm$^{-2}$ is the gas column
per unit optical depth at 100\,$\mu$m \citep{gon14}, and $\mu=1.4$ accounts
for He. The computed values are also listed in Table~\ref{tab:res}.
  The mass associated with the core component has a large uncertainty
  because its $\tau_{100}$ is not well constrained.
  Our 3D approach in Sect.\,\ref{3d}
  indicates that its mass likely does not exceed $10^8$\,M$_{\odot}$. The
  combined gas mass of the 3 nuclear components is
  $4.5_{-0.6}^{+1.5}\times10^8$\,M$_{\odot}$.

The CO $2-1$ emission from \cite{per16} has also been used to estimate
  gas masses. Using the CO emission within the 3$\sigma$ contour
  of the 454\,GHz emission displayed in Fig.~\ref{almamaps}a, thus
  covering accurately the three components traced by H$_2$O,
  and assuming the same
  brightness for the $2-1$ and $1-0$ lines with a ULIRG conversion factor of
  $\alpha_{\mathrm{CO}}=0.78$\,$M_{\odot}/(\mathrm{K\,km\,s^{-1}\,pc^2})$,
  the gas mass is $3.4\times10^8$\,M$_{\odot}$. This value is comparable to
  the mass derived from the
  H$_2$O model. The CO $2-1$ emission is however
  much more extended than the 454\,GHz contiuum (see Sect.\,\ref{sec:inflowco});
  the gas masses within radii of $1''$, $2''$, and $3''$ in the plane of
  the galaxy (233, 466, and 700\,pc, respectively) are $4.8\times10^8$,  
  $6.8\times10^8$, and $8.3\times10^8$\,M$_{\odot}$, respectively.

The three components probed by H$_2$O lie within a radius of
$r_{\mathrm{H_2O}}=0.9''$ ($\approx200$\,pc) from the galaxy center.
The dynamical mass within this radius
can be estimated from the rotation curve shown by \cite{per16}, which gives
$M_{\mathrm{dyn}}\sim2.1\times10^9$\,M$_{\odot}$
(M. Pereira-Santaella et al., in prep).
Using the combined gas mass as derived above from the H$_2$O model,
the gas fraction is
$f_g=M_{\mathrm{gas}}/M_{\mathrm{dyn}}\sim20$\%. At the current rate of nuclear
star formation (Sect.\,\ref{sed}), the nuclear starburst has an
age of $\sim100$\,Myr. This should be considered an upper limit owing to
the plausible presence of a stellar population prior to the current burst.

\subsubsection{Summary and limitations of the model}
\label{summary}

   \begin{figure*}
   \centering
   \includegraphics[width=16.0cm]{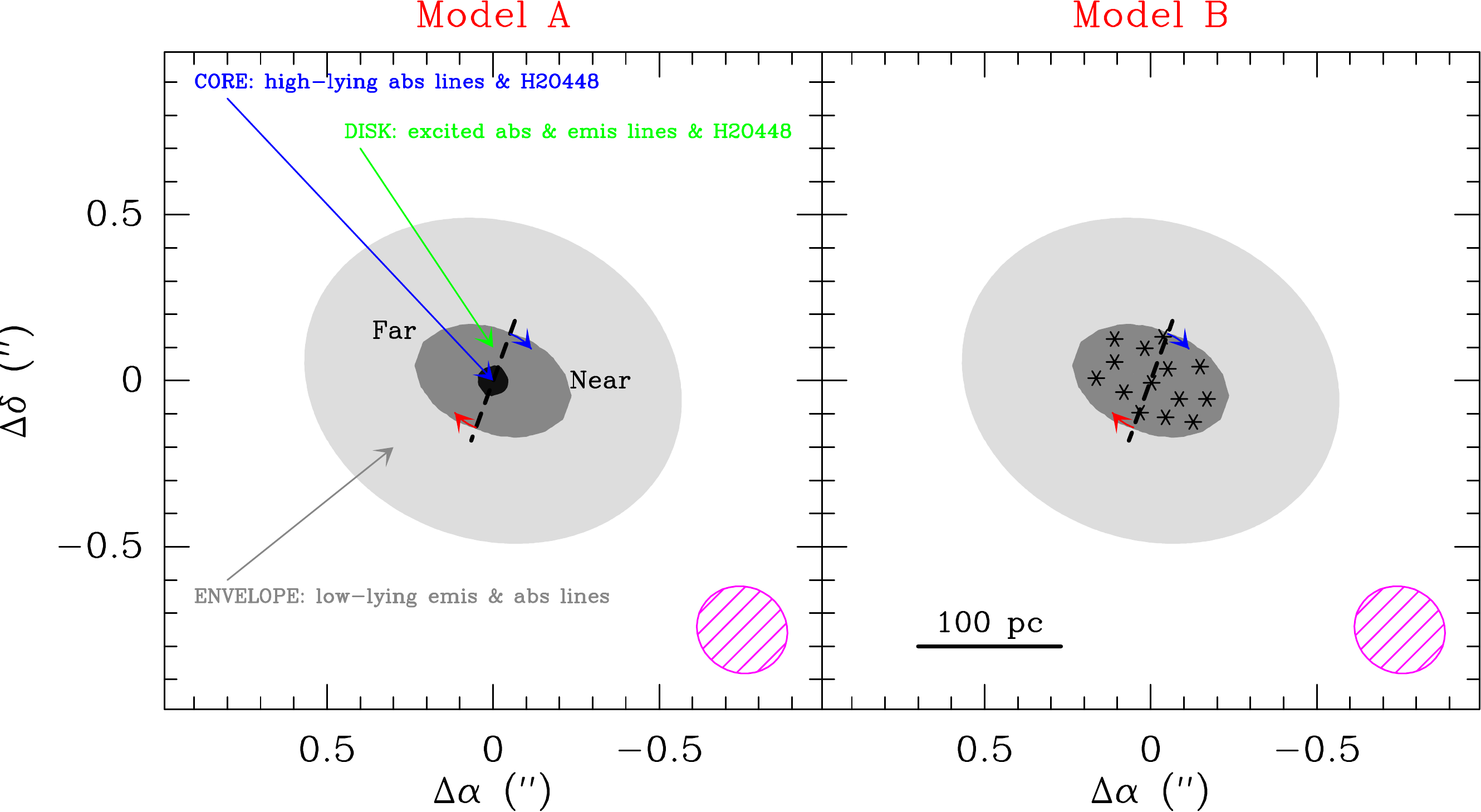}
   \caption{Two possible sketches of the nuclear region of ESO~320-G030,
     based on the model fit to the 20 detected and one undetected H$_2$O lines,
     and three continuum
     flux densities. We have in both sketches three components: the compact
     core, the disk, and the envelope. In model A, the core is a physically
     coherent component located at the center of the galaxy, while in model
     B the core is composed of discrete spots widespread over the disk.
     The dashed line indicates the major kinematic axis, but the source
     appears to be elongated along the minor axis that nearly coincidates
     with the direction of the inner bar. The arrows indicate the clockwise
     direction of rotation, with far and near sides of the disk also
     indicated. The envelope is assumed to be fully surrounding the disk.
   }   
    \label{figsketch}
    \end{figure*}

Figure \ref{figsketch} summarizes visually two possible scenarios of the
model source based on our three component fitting of the nuclear region
of ESO~320-G030. The most extended component, the envelope, has a luminosity
of $\sim8\times10^{10}$\,L$_{\odot}$ and an effective radius of $\sim130$\,pc,
is optically thin in the far-IR,
and only contributes significantly to the absorption or emission of the
lowest-lying far-IR and submillimeter lines. Its contribution to the H$_2$O448
line is negligible, as this line is exclusively formed in environments that
are optically thick in the far-IR, the disk and the core. 
The disk has a luminosity of $\sim2\times10^{10}$\,L$_{\odot}$ and an effective
radius of $\sim40$\,pc, and contributes significantly to the excited lines
of H$_2$O both in absorption and in emission. Our sketch in
Fig.~\ref{figsketch} shows the envelope and the disk as ellipses with their
major axis coincident with the minor kinematic axis to account
for the apparent elongation of the source in that direction, which
nearly coincides with the direction of the nuclear bar.
The different components can indeed be inclined
and shaped arbitrarily provided that the solid angle as derived from our
models remains unchanged (ignoring possible significant changes in
level populations as a consequence of the different geometry). 
The disk component is (partially) resolved by the ALMA beam of
$\approx0.26''$. Finally, we identify from the very high-lying absorption
lines of H$_2$O an additional, very compact component with an effective radius
of $\approx12$\,pc, very warm ($T_{\mathrm{dust}}\sim100$\,K), and with a
luminosity similar to that of the disk despite its small size.
It is extremely buried with H$_2$ columns probably above
$\sim10^{25}$\,cm$^{-2}$, resembling the
buried galactic nuclei (BGNs) detected in HCN vibrational emission
\citep[e.g.,][]{sak10,aal15,mar16}. This core is however unresolved by the ALMA
beam, and our fit to the H$_2$O fluxes cannot distinguish between
a physically coherent region at the center of the galaxy, as depicted in
model A, or a discrete set of star-forming cores
spread out over the disk volume (model B) or even beyond. Nevertheless, we
can discriminate between both models by comparing the observed spatial
distribution of the 454\,GHz continuum and H$_2$O448 emission with the
predicted distributions involved by the two scenarios in Fig.~\ref{figsketch},
as shown below.

\subsection{A 3D approach}
\label{3d}

A 3D model approach is here used with three main purposes:
first, to check the reliability of our model fits, and in particular
of the calculated sizes of the three components. This is performed
by inspecting whether any of our best-fit model combinations, obtained from
spherically symmetric models, can predict spatial distributions
for the 454\,GHz continuum and H$_2$O448 line emission that are consistent
with the ALMA maps. The comparison will provide a way to refine
the overall model and discriminate among
combinations with low $\chi_{\mathrm{red}}^2$, as
the contributions by the several model components to the H$_2$O448 line
and to the 454\,GHz continuum emission are poorly determined
(Fig.~\ref{bayes}h-i). Second,
we also aim to discriminate between scenarios A and B in Fig.~\ref{figsketch}.
Finally, analysis of the velocity field will establish the dynamical mass
as a function of inclination, favoring a given geometric disposition relative
to the plane of the galaxy that may shed light on the gas motions responsible
for the formation of the nuclear structure.

\subsubsection{Description of the 3D model}
\label{descrip3d}

   \begin{figure*}
   \centering
   \includegraphics[width=16.0cm]{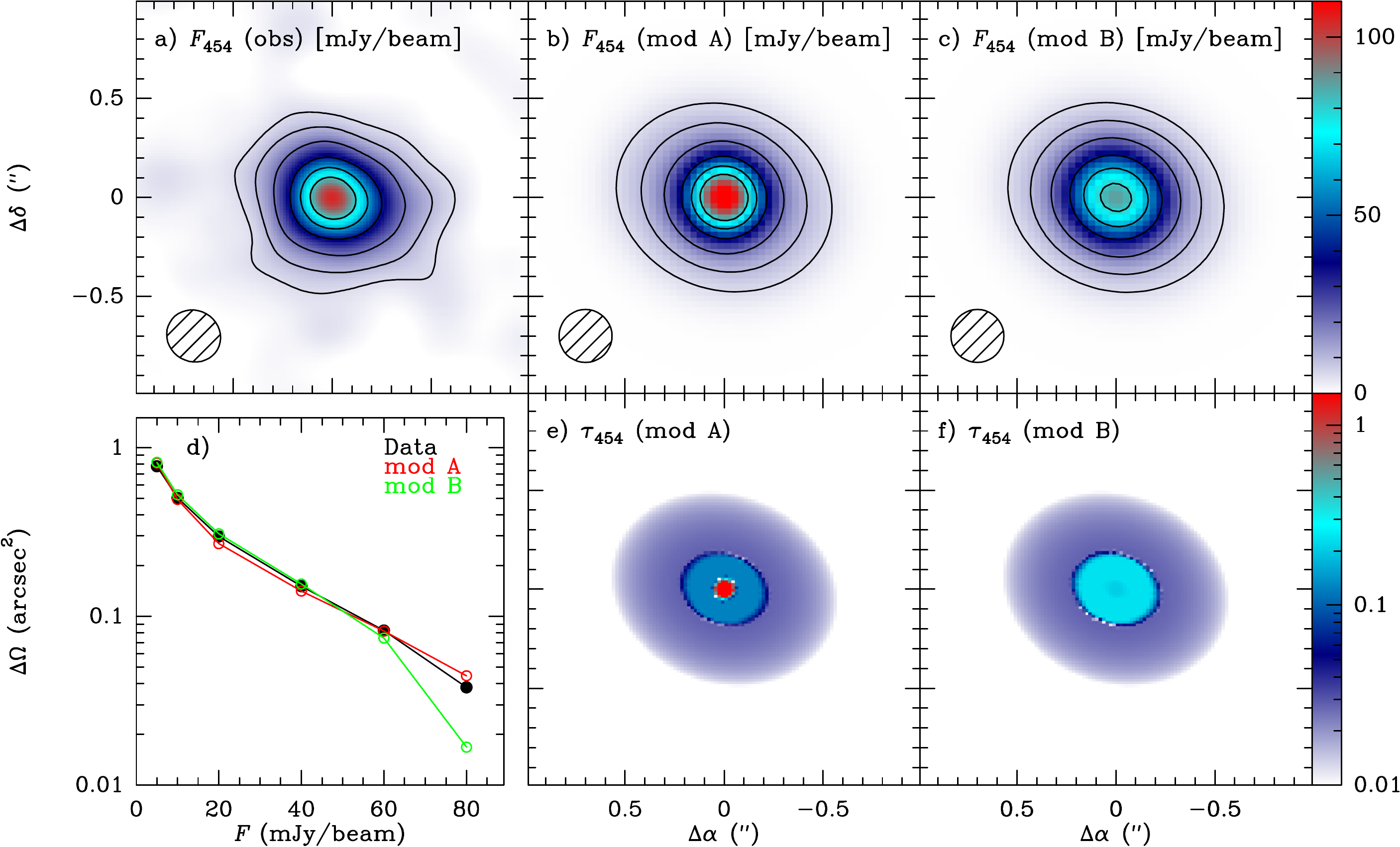}
   \caption{Comparison between the observed 454\,GHz continuum map (panel a)
     and two 3D models (A and B) based on our fiducial model. In model A
     (panel b, with $\tau_{454}$ in panel e), the core component is assumed to
     be a real physical component concentrated at the center of the galaxy,
     and in model B (panel c, with $\tau_{454}$ in panel f), the core component
     is assumed to be widespread in the inner disk. Panel d compares the
     solid angle subtended by the plotted contours (5, 10, 20, 40, 60, and
     80 mJy/beam) in the observed map (black line and symbols) and in models
     A and B (red and green, respectively). Model A fits the observed map 
     better than model B.
   }   
    \label{cont4543d}
    \end{figure*}

   \begin{figure*}
   \centering
   \includegraphics[width=18.0cm]{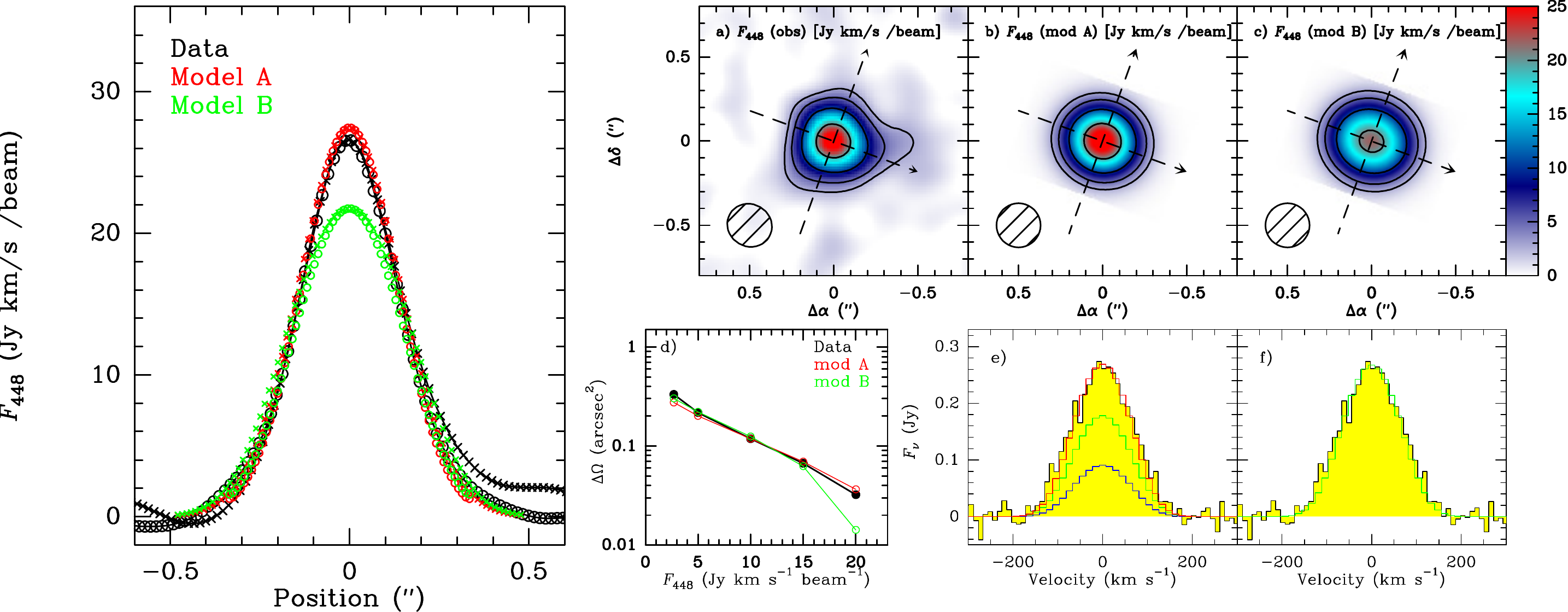}
   \caption{Comparison between the observed H$_2$O448 map (panel a)
     and two 3D models (A and B) based on our fiducial model. In model A
     (panel b, with the predicted spectrum in panel e), the core component
     is assumed to
     be a real physical component concentrated at the center of the galaxy,
     and in model B (panel c, with the predicted spectrum in panel f), the
     core component is assumed to be widespread in the inner disk.
     Panel d compares the
     solid angle subtended by the plotted contours (2.7, 5, 10, 15, and 20
     Jy\,km\,s$^{-1}$/beam) in the observed map (black line and symbols)
     and in models A and B (red and green, respectively). In the left hand
     panel, strips in the direction of the two axes indicated in panels a-c
     compare the data to both models. Model A fits the observed map 
     better than model B.
   }   
    \label{h2o4483d}
   \end{figure*}

Our model simulates arbitrarily complex source geometries and velocity fields
by means of small cubes defined within a large
cube of side 480\,pc. The small cubes have sides of $2-3$\,pc, which determines
the resolution of the simulations.
While calculations for the equilibrium
$T_{\mathrm{dust}}$ can be performed with a
Monte Carlo approach, we simply use in the present calculations the values
of $T_{\mathrm{dust}}$ and optical depths (i.e., the brightness), and sizes
(i.e., the solid angle) of the three components as
inferred from our fiducial model to generate beam-convolved maps at 454\,GHz
that can be directly compared with the observed maps. Likewise, we use 
the brightness and solid angle of the H$_2$O448 line for the core and disk
components, as derived from the fiducial model, to generate beam-convolved
maps for the H$_2$O448 line that are compared with the observed spatial
distribution.

We use the geometry depicted in Fig.~\ref{figsketch}.
The disk and envelope are assumed to lie in the plane of the galaxy,
and are observed with an inclination angle
of $i=43^{\circ}$ \citep{per16}. The actual sizes for the fiducial model
in Table~\ref{tab:res} are then increased to match the required solid angles.
The kinematic major axis observed in the nuclear region, however, has a PA
of $160^{\circ}$, significantly
higher than the value of $133^{\circ}$ derived from the large-scale CO $2-1$
observations \citep{per16}.
To approximately account for the elongated shapes along the minor
kinematic axis observed in the lowest contours 
of the 454\,GHz continuum and H$_2$O448 line images (Fig.~\ref{almamaps}),
the disk and envelope are modeled as
ellipses with aspect ratio $b/a=0.6$. The envelope is assumed to cover
the disk on the front and back sides, with an effective radius fixed at
130\,pc.

The unresolved core component in model A is simulated as a spherical
source. In model B, no core is included and the brightness of the disk
in both the 454\,GHz continuum and the H$_2$O448 line are increased
to match the combined flux of both components.

As pointed out above, there is a wide range in both $f[F(\mathrm{H_2O}448)]$
and $f[F(454\,\mathrm{GHz})]$ (i.e., the relative contributions of the
different components to the H$_2$O448 line and 454\,GHz continuum emission)
among our best-fit solutions (Table~\ref{tab:res} and Fig.~\ref{bayes}h-i).
Our fiducial model was selected because it generates maps for both the
454\,GHz continuum and H$_2$O448 line that compare well with the observed
maps, as shown in the next sections.

\subsubsection{The 454\,GHz continuum}
\label{cont3d}

The 3D simulation of the 454\,GHz continuum for the fiducial model is
compared with the observed map in Fig.~\ref{cont4543d}. Maps of the
continuum optical depth at 454\,GHz for models A and B are displayed
in panels e and f, respectively, and the corresponding intensity maps 
are shown in panels b and c. The solid angle subtended by each isocontour
in panels a-c is shown in panel d.

Our fiducial model reproduces the overall distribution of intensities
rather satisfactorily. Specifically, the envelope
is required to account for the observed extended emission of the continuum.
It is also evident from Fig.~\ref{cont4543d} that model A more closely
resembles the observed map than model B, indicating the presence of an
intensity peak of the continuum at the center that we associate with the
very high-lying H$_2$O absorption lines, that is, the core.
However, model A slightly
overpredicts the intensity continuum from the center. While the model
predicts fluxes of $72$ and $66$\,mJy from the core and the disk, 
a better match to the map would be obtained with $\sim60$ and $\sim78$\,mJy,
respectively.

\subsubsection{The H$_2$O448 line emission}
\label{h2o3d}

Only the core and the disk are included in the simulations for the
H$_2$O448 line emission, as the optically thin envelope does not
  obscure or contribute to this intrinsically weak line.
The simulated velocity-integrated intensity maps
of the line for models A and B are compared with the observed map
in Fig.~\ref{h2o4483d}a-c, and the solid angles subtended by the isocontours
in these panels are compared in panel d. The emergent line profiles from the
whole region are compared with the observed profile in panels e and f, and the
observed and modeled intensities along the major and minor axes are compared
in the left-hand panel.

From the comparison of the maps, we conclude that an effective disk radius
of $\approx40$\,pc matches rather well the observed map. In addition,
model A matches the observed intensity distribution slightly better than
model B, although higher angular resolution is required to verify this
point. In our fiducial model, the core accounts for $13.8$\,Jy\,km\,s$^{-1}$
($f[F(\mathrm{H_2O}448)]=0.31$, Fig.~\ref{bayes}h) so that the disk
dominates the H$_2$O448 line emission.

\subsubsection{The velocity field}
\label{vel3d}

    \begin{figure*}
   \centering
   \includegraphics[width=17.0cm]{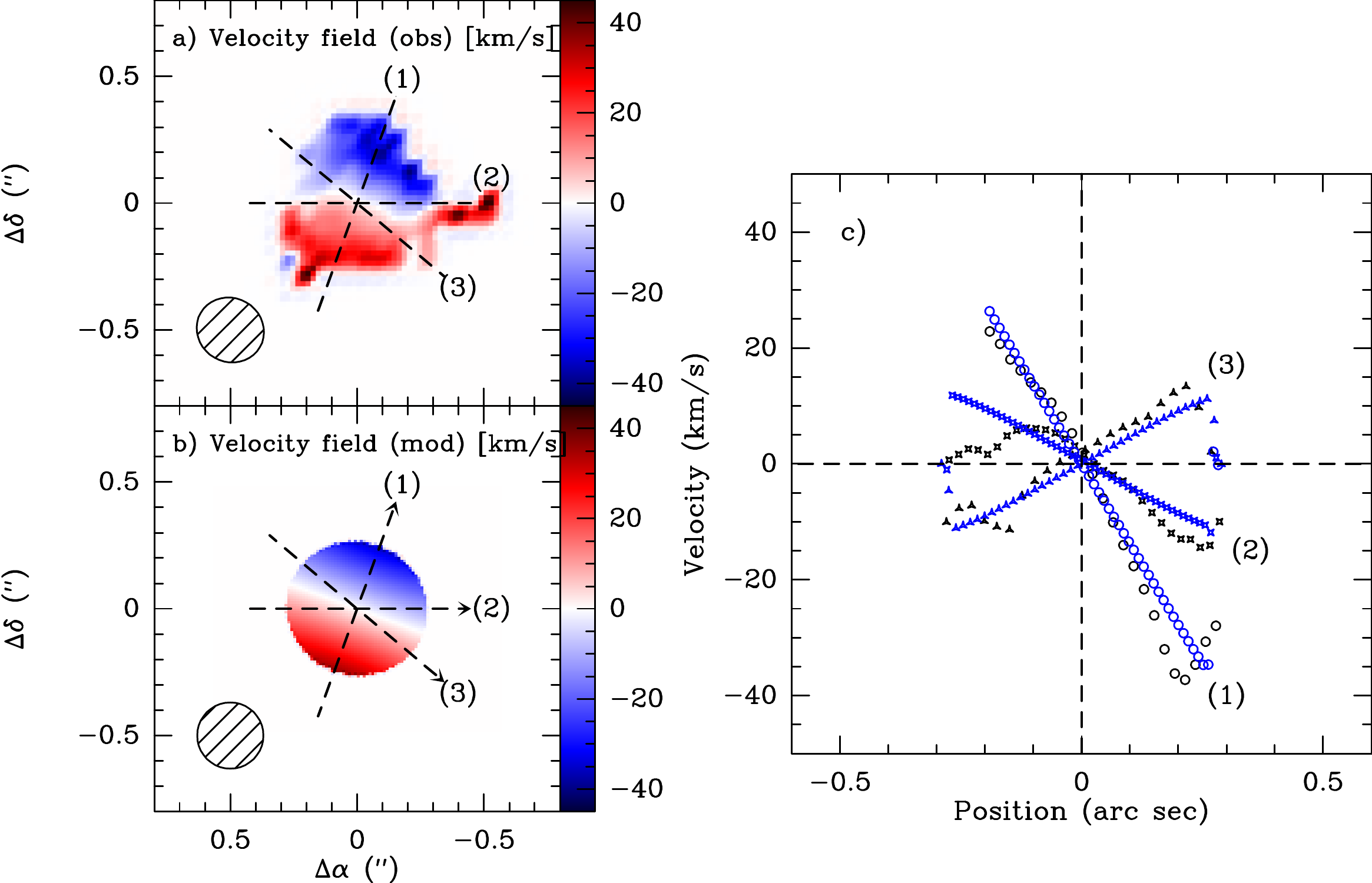}
   \caption{Velocity field associated with the H$_2$O448 map (panel a)
     is compared with our 3D model A (panel b). The velocity along the three
     axes indicated in these panels is shown in panel c, where black
     and blue symbols correspond to data and model, respectively.
   }   
    \label{h2o448vfield3d}
   \end{figure*}
  
The 3D simulations also provide a good match to the observed H$_2$O448
line shape (Fig.~\ref{h2o4483d}e-f). Line broadening is here simulated
by both microturbulence, with the same $\Delta V=100$\,km\,s$^{-1}$ as
adopted for the 1D models, and a rotating velocity field that further
broadens the line, of the form:
\begin{eqnarray}
  V_{\mathrm{rot}} (r<R_{\mathrm{core}}) & = & 100 \times r/R_{\mathrm{core}}  \\
  V_{\mathrm{rot}} (r>R_{\mathrm{core}}) & = & 100  \,\, \mathrm{km\,s^{-1}},
  \label{eq:vrot}
\end{eqnarray}
where $R_{\mathrm{core}}$ is the radius of the core component. We have not
attempted more complex velocity fields given our limited spatial resolution
and significant beam smearing. Figure~\ref{h2o448vfield3d} compares the
observed and modeled maps of the line-of-sight velocity. Although our
adopted velocity field approximately accounts for the observed rotation,
the observed field is quite distorted, as is also seen from the strips
along the three axes in panel c. The apparent S-shape of the zero
  velocity contour may indicate an elongated disk \citep{fra92}
  or warping, but also the presence of 
  inflowing gas motions along the minor kinematic axis of the
  disk. A massive inflow is indeed observed on larger spatial scales, as
  described below (Sect.\,\ref{sec:coh2o}).

The rotational velocity of $\approx100$\,km\,s$^{-1}$ of the disk gives
a dynamical mass $M_{\mathrm{dyn}}$ that is inconsistent with the high
concentration of gas in the nuclear region.
Considering both the rotation and dispersion motions as in
  \cite{bel13}, $M_{\mathrm{dyn}}=232\,r\,(V_{\mathrm{rot}}^2+1.35\sigma^2)$
  (where the velocities are in km\,s$^{-1}$ and $r$ in pc)
  gives $1.4\times10^8$\,$M_{\odot}$ at $r=40$\,pc, while the combined
  gas mass (i.e., not including the stellar mass) of the core and disk
 components is $\sim2\times10^8$\,$M_{\odot}$ (Sect.\,\ref{masses}).
This discrepancy can be attributed to a lower inclination of the nuclear
disk relative to that of the host galaxy; indeed, the kinematic major axis
of the nuclear disk is significantly rotated relative to that of the
host, which may suggest some degree of kinematic decoupling.
Alternatively, $V_{\mathrm{rot}}$ could underestimate $M_{\mathrm{dyn}}$
if the nuclear gas is not rotationally supported, but supported by
radiation pressure and turbulence.
 
\subsubsection{Additional remarks}

 While the model with 3 components accounts for the main properties
  of H$_2$O448 and continuum emission as observed at $0.25''$ (60\,pc)
  resolution, it is obviously very schematic with sharp edges and transitions
  from one component to the next. In reality, we may expect a 
  smoother transition between the different components, with the envelope
  representing the optically thin extension of the nuclear disk, and
  the core a cusp of gas column density and $T_{\mathrm{dust}}$ located
  at the center of the galaxy. On the other hand, the majority of the
  H$_2$O absorption lines observed with {\it Herschel}/PACS have rest
  wavelengths $\lesssim110$\,$\mu$m, with only 2 absorption lines
  observed at longer wavelengths.
  This means that the continuum optical depth of the disk at
  $120-200$\,$\mu$m is better probed by species with lines observed in
  this wavelength range. Our model for
  the remaining molecular species in Appendix~\ref{appa} indeed indicates
  that $\tau_{100}$ (disk) is probably somewhat higher ($1.5-3$) than in our
  fiducial model. Finally, we note that the sizes estimated for
    the different components depend on the assumed velocity dispersion of
    100\,km\,s$^{-1}$. While these sizes and $\Delta V$ are well constrained
    for the disk and envelope given the spatial resolution of our ALMA data
    and the spectral resolution of the H2O448 line, the size of the compact
    core is not so well constrained as it would increase with lower
    $\Delta V$.
    Higher spatial resolution observations would be required to better
    constrain the size and kinematics of this component.
  
%-------------------------------------------------------------

\section{A massive molecular inflow feeding the nucleus of ESO~320-G030}
\label{sec:coh2o}

\subsection{The inflow seen in CO 2-1}
\label{sec:inflowco}

We have so far analyzed the nuclear (inner $\sim200$\,pc) region of
ESO~320-G030 by combining the {\it Herschel} and ALMA H$_2$O
lines and the far-IR and submm continuum, but can we trace the formation of such
an extreme nuclear region from the observed kinematics at larger spatial
scales? \cite{per16} reported the CO\,$2-1$ map observed with ALMA
on spatial scales of $10''$ and with a resolution similar to that of the
H$_2$O448 observations, $\approx0.25''$, thus providing an ideal tool to
search for hints of inflowing gas.
\cite{per16} fit the large-scale velocity field in ESO~320-G030 by
excluding the nuclear region where complex CO profiles were found; here
we focus on the CO\,$2-1$ in this nuclear region.

   \begin{figure*}
   \centering
   \includegraphics[width=18cm]{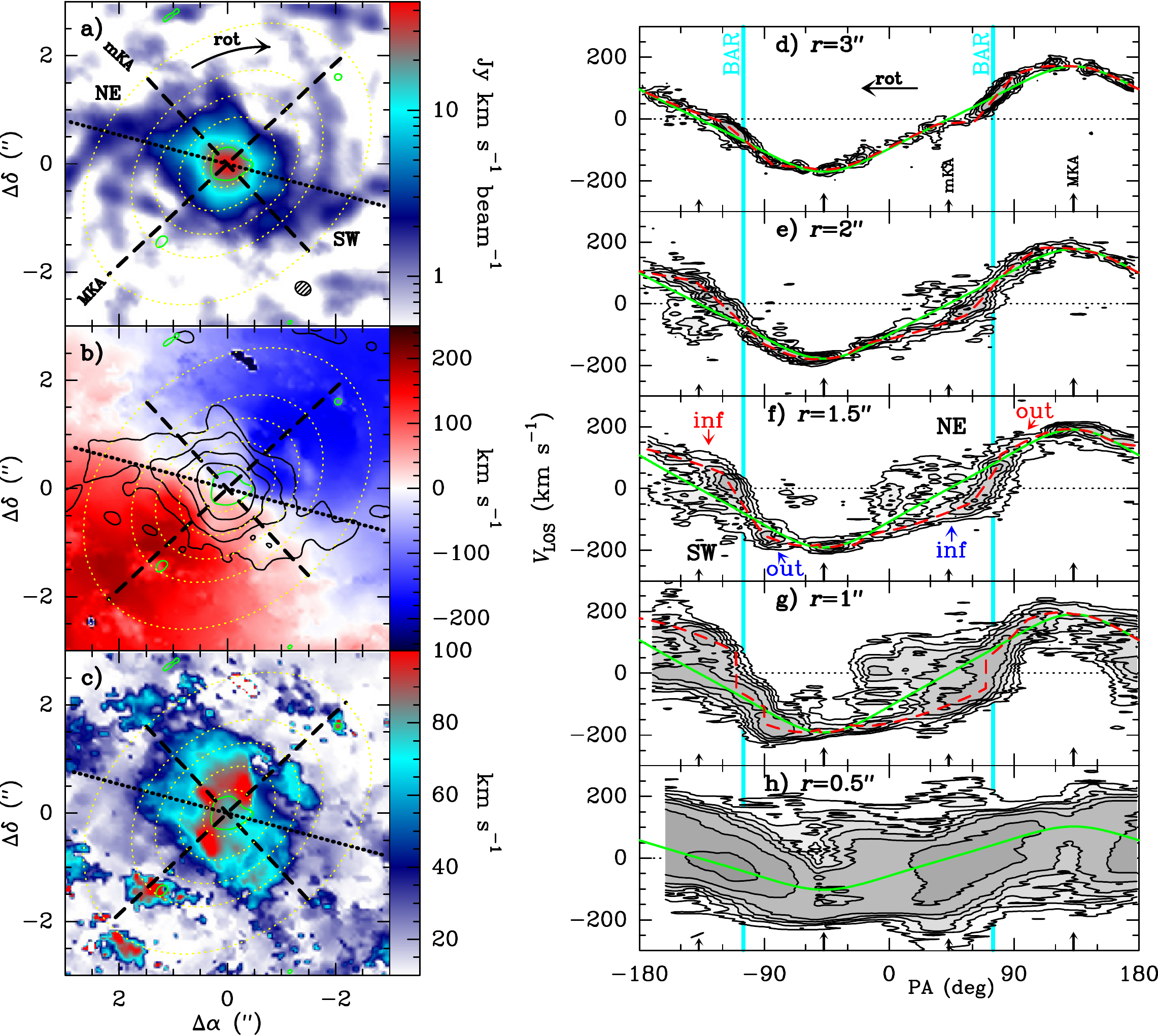}
   \caption{CO (2-1) emission observed with ALMA in the central region
     of ESO~320-G030. Panels a-c show with colors the integrated
     intensity (moment 0), velocity field (moment 1), and velocity
     dispersion (moment 2). The hatched ellipse in panel a indicates the
     ALMA beam. The dotted black line indicates the approximate
     direction of the nuclear bar
     ($\mathrm{PA}=75^{\circ}$, see Fig.~\ref{overplot}a), and the
     dashed lines are the kinematic major and minor axes (MKA and mKA).
     The small green contour at the
     center is the lowest H$_2$O448 contour in Fig.~\ref{almamaps}b.
     The yellow dotted curves indicate circles in the plane
     of the galaxy with radii $r=3''$, $2''$, $1.5''$, $1''$, and $0.5''$.
     d-h) Position-velocity diagrams along the above circles. The green
     curves show the purely rotational velocity field fitted by
     \cite{per16} to a region of $10''$ in size that excludes the nuclear
     region, with $V_{\mathrm{rot}}=250$, 260, 280, 280, and 150\,km\,s$^{-1}$
     in panels d, e, f, g, and h, respectively.
     We have here modified this velocity field in the nuclear
     region, including azimuthal variations of the rotational 
     and radial velocity components of the gas, as depicted with the red
     dashed curves, with values for $V_{\mathrm{rot}}$ and $V_{\mathrm{rad}}$
     displayed in Fig.~\ref{vrotradco}a-b.
     The PA of the stellar bar is indicated in cyan, and the PA of the MKA
     and mKA are indicated with long and short vertical
     arrows, respectively. The clockwise direction of rotation is indicated
     with arrows in panels a and d.
     In panel f, labels ``inf'' and ``out'' indicate regions with
     a radial velocity component negative (inflow) and positive (outflow),
     respectively, colored according to the velocity shift.
     Contour levels in panels d-h are 4.5, 9.0, 13.5, 19, 27, 54, 
     and 108\,mJy\,beam$^{-1}$.
   }   
    \label{co}
    \end{figure*}
   \begin{figure}
   \centering
   \includegraphics[width=8.0cm]{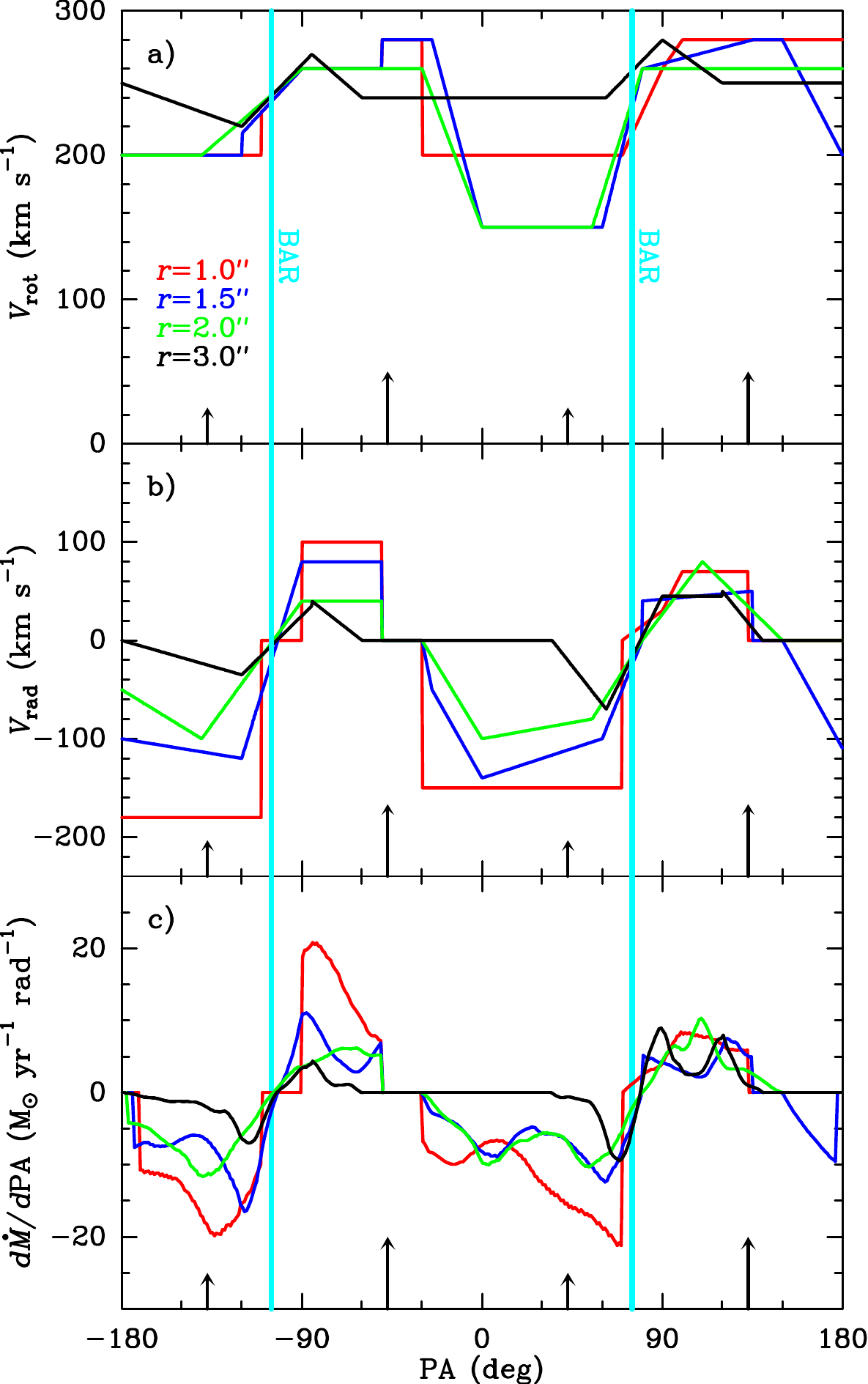}
   \caption{a) Rotational and b) radial velocity components of the gas
     along the four outer circles ($r=1''-3''$) depicted in Fig.~\ref{co}a-c,
     as derived from the CO (2-1) PV diagrams of Fig.~\ref{co}d-g.
     c) The rate at which gas mass
     is crossing the quoted circles per unit interval of PA.
     The PA of the bar
     is indicated with cyan vertical lines, and the vertical long and short
     arrows indicate the PA of the MKA and mKA, respectively.
     Values of $V_{\mathrm{rad}}$ and
     $d\dot{M}/d\mathrm{PA}$ positive (negative)
     indicate gas flowing in the outward (inward) direction. 
   }   
    \label{vrotradco}
   \end{figure}

\subsubsection{The velocity field}
   
\begin{figure*}
   \centering
   \includegraphics[width=17.5cm]{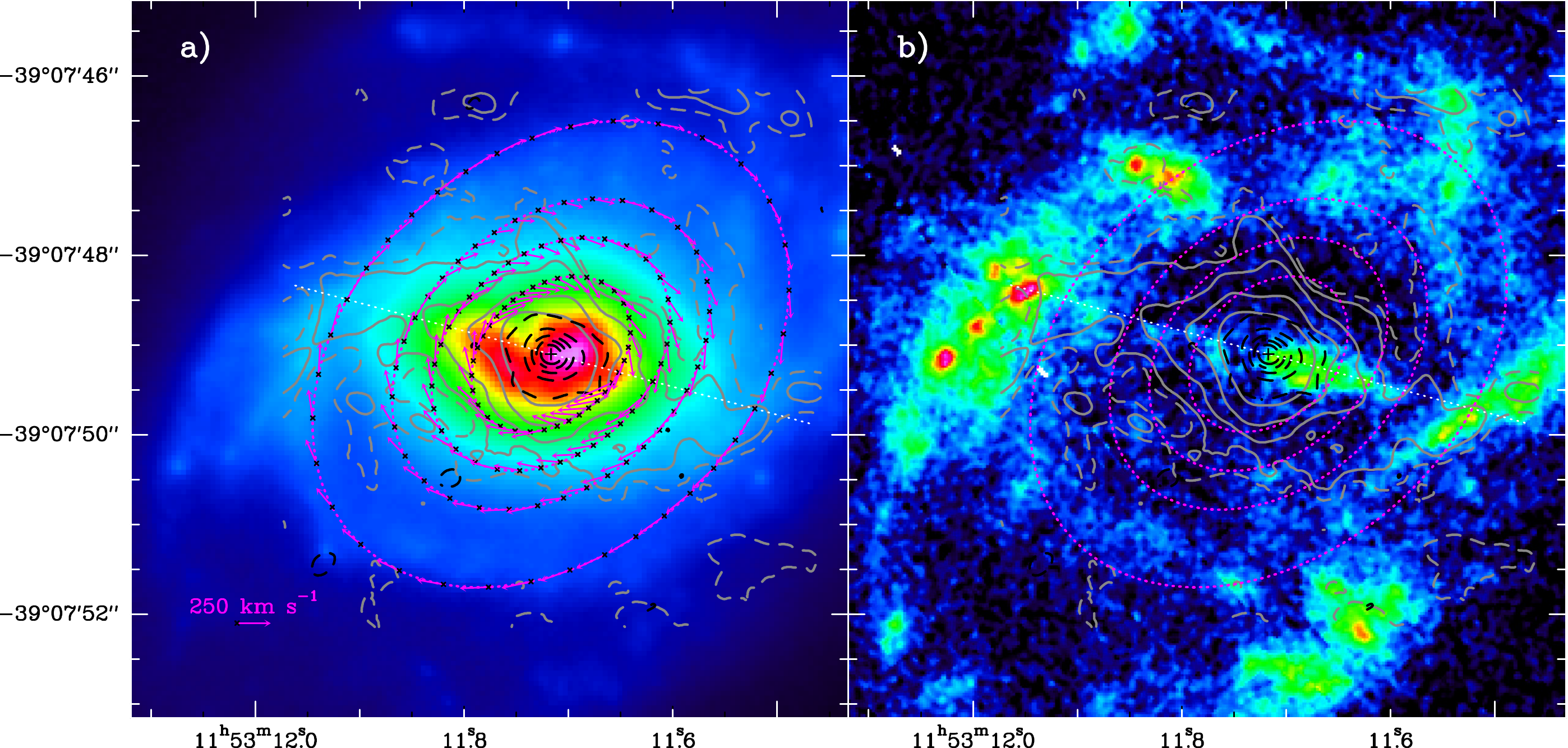}
   \caption{Images of the central region of ESO 320-G030 in
     VLT/NACO K-band (A. Crespo G\'omez et al., in prep.; panel {\bf a})
     and in the {\it HST}/NICMOS2 F190N-F187N 
     \citep[continuum-subtracted Pa-$\alpha$, from][reprocessed using
       the latest NICMOS pipeline by S\'anchez-Garc\'{\i}a et al. in prep.;
       panel {\bf b}]{alo06}. 
     The CO (2-1) emission observed with ALMA (gray contours) and
     the 454\,GHz continuum (dashed black contours) 
     are overlaid in both panels.     
     The direction of the nuclear gas bar is indicated by the dotted
     white line ($\mathrm{PA}=75^{\circ}$). The four outer circles of
     Fig.~\ref{co}a-c ($r=1''-3''$) where the velocity field
     is estimated are also indicated, with the arrows in panel a
     showing the gas velocity
     vectors projected on the plane of sky. The cross marks the
     position of the peak emission in CO, H$_2$O448, and 454\,GHz continuum.
     For consistency with the ALMA astrometry, the VLT and {\it HST} images
     were aligned to the Gaia catalog using several stars in the field.
     The spatial resolutions are $\sim0.25''$ and $\sim0.15''$ for the
     VLT/NACO and {\it HST}/NICMOS images, respectively.
   }   
    \label{overplot}
    \end{figure*}

Figure~\ref{co}a-c shows the CO color-coded maps of integrated intensity
(moment 0, between $-300$ and $+300$\,km\,s$^{-1}$), the velocity
field (moment 1) and velocity dispersion (moment 2), respectively.
The dotted line in these panels indicates the approximate direction
of the nuclear bar ($\mathrm{PA}=75^{\circ}$), which is well
  traced by the VLT/NACO K-band image
shown in Fig.~\ref{overplot}a.
The green contour at the center in Fig.~\ref{co}a-c is the lowest
(most extended) contour of the H$_2$O448 line in Fig.~\ref{almamaps}b,
emphasizing the compactness of the core+disk nuclear structure probed by
the H$_2$O line as compared with the large-scale CO emission.
The yellow dotted curves depict
circles in the plane of the galaxy with radii $r=3''$, $2''$, $1.5''$, $1''$,
and $0.5''$. PV diagrams along these circles are shown in panels d-h,
where the green curves indicate the velocity field fitted by \cite{per16},
that is, a uniform rotational velocity with no radial component.
It is clearly seen in panels d-g that,
around the PA of the bar and mostly overshooting it (i.e., at lower PA,
in the clockwise direction of rotation),
the bulk of the gas shows significant departures from these ordered
circular motions.

We have then modified this regular velocity field to account
for the main kinematic departures from the green curves:
\begin{eqnarray}
  V_{\mathrm{LOS}} & = & V_{\mathrm{rot}}\,\cos(\theta-\theta_0)\,\sin i
  \nonumber \\
  & + & V_{\mathrm{rad}}\,\sin(\theta-\theta_0)\,\sin i,
  \label{vfieldco}
\end{eqnarray}
where $\theta$, increasing in the clockwise (rotation) direction,
measures the angular position in the plane of the galaxy,
$\theta_0=133^{\circ}$ is the position angle of the major
kinematic axis (MKA), $i=43^{\circ}$ is the inclination angle, $V_{\mathrm{rot}}$
is the rotational velocity, and $V_{\mathrm{rad}}$ is the radial component
of the velocity.
Equation~(\ref{vfieldco}) takes into account that the NE region is the
far-side of the disk
\citep[Fig.~\ref{figsketch}; see also Fig.~9 in][]{caz14},
and hence any inflowing component ($V_{\mathrm{rad}}<0$)
in that region ($\sin(\theta-\theta_0)>0$) will be blueshifted
(see Fig.~\ref{co}f).
Similarly, the SW region corresponds to the near side
of the disk, and any inflow component here would be redshifted.
The modified velocity field is displayed with dashed red curves in
panels d-h, and the curves of $V_{\mathrm{rot}}$ and $V_{\mathrm{rad}}$
along these circles are shown in Fig.~\ref{vrotradco}a-b.

The integrated intensity map
in Fig.~\ref{co}a shows relatively strong emission not only along the bar
but also at lower PA, resulting in an elongated shape along the minor
kinematic axis (mKA). This is the region where the velocity dispersion is above
60\,km\,s$^{-1}$ (panel c) and where the velocity field shows
strong disturbances (panel b). Two trailing spiral arms at pitch angle of
$\approx90^{\circ}$ arise from each side of the bar.

The overall kinematics shown in Fig.~\ref{co}d-g clearly illustrate
that the bulk of the gas in the NE region of the disk, ahead 
of the bar major axis in the forward (rotation) direction
($0<\mathrm{PA}<75^{\circ}$, i.e., at around the mKA),
is blueshifted, and the gas on the opposite SW region of the
disk ($\mathrm{PA}<-105^{\circ}$) is redshifted (as indicated in panel f).
This effect is already seen at $r=3''$
(700\,pc), and becomes increasingly pronounced toward the center. If the gas
in these regions remains in the plane of the galaxy, the observed velocity
shifts are ascribed to an inflow ($V_{\mathrm{rad}}<0$), as
$\cos(\theta-\theta_0)=0$ along the mKA.
In addition, we also find clear evidence of outflowing gas
($V_{\mathrm{rad}}>0$) at PA higher than that of the bar (i.e., for gas that has
not still arrived at the bar). The outflowing gas is clearly seen at
$\mathrm{PA}\approx-90^{\circ}$ (see Fig.~\ref{co}f); it is blueshifted
(redshifted) on the western (eastern) side of the disk.
Nevertheless, the magnitude of this velocity is significantly lower
than the inflow velocity ahead of the bar, except at $r=3''$ where both are
similar (Fig.~\ref{vrotradco}b).

Our fit to the PV diagrams in Fig.~\ref{co}d-g, with results for
$V_{\mathrm{rot}}$ and $V_{\mathrm{rad}}$ in Fig.~\ref{vrotradco}a-b, respectively,
is based on the observed slope and values of $V_{\mathrm{LOS}}$.
Around the mKA, where gas on both sides of the disk shows inflowing
velocities, $V_{\mathrm{rad}}<0$ is well constrained from the values
of $V_{\mathrm{LOS}}$. Inflow velocities as large as
$80-180$\,km\,s$^{-1}$ are obtained at $r=2''-1''$ ($460-230$\,pc).
The slope of $V_{\mathrm{LOS}}$ in these PA regions
indicates that, at $r=1''-2''$, $V_{\mathrm{rot}}$ sharply decreases to
$150-200$\,km\,s$^{-1}$. At PA around $105^{\circ}$ and  $-75^{\circ}$,
where outflowing gas is detected, we have some degeneracy between 
$V_{\mathrm{rot}}$ and $V_{\mathrm{rad}}$, which is approximately solved from
the slope of $V_{\mathrm{LOS}}$. At $r=3''$, we find some evidence of
increasing $V_{\mathrm{rot}}$ at the trailing edge of the bar.
At $r\le1''$, the CO lines become very broad with FWZI of
$\approx400$\,km\,s$^{-1}$; the high turbulence masks both the rotation field
and any possible inflow in these innermost regions, although hints of
a velocity pattern similar to that found at higher $r$ are seen
on the NE side of the disk. The inflow in this region is better
probed by the OH lines (Sect.\,\ref{inflowfarir}).

Besides the above velocity field that applies to the bulk of gas at
different radii and PA, Fig.~\ref{co} also shows a low intensity
component that is fully decoupled from the overall pattern
but is also symmetric relative to the center.
It is traced by the lowest contour(s) in the velocity-position maps of
panels e-g, showing very-high velocity dispersion.
This component is already seen at $r=2''$ around $\mathrm{PA}\approx30^{\circ}$ 
and, symmetrically, around $\mathrm{PA}\approx-150^{\circ}$ (panel e),
with line-of-sight velocities that extend from the velocity of the rotating gas
at that position to a similar velocity but with opposite sign. As
$r$ decreases to $1''$, the component becomes more extended in PA.
The overall direction of this component is similar to that of the
outflowing clumps observed in CO 2-1 \citep{per16}, and to the direction
of the bipolar outflow seen in NaD as well \citep{caz14}.
It is thus possible that this CO component represents the low-velocity
counterpart of the CO outflow, with a relatively high opening angle that
enables both negative and positive line-of-sight velocities at a given position.
Nevertheless, a more plausible interpretation suggested by the limiting
velocities and also by the location of this component ahead of the bar
major axis,
is that it represents the kinematic effect of the strong shock produced
by the gas overshooting the bar.
The fraction of gas mass sampled within a velocity range of
$\pm50$\,km\,s$^{-1}$ around the red curves in Fig.~\ref{co}d-g 
ranges from 43\% at $r=1''$ to 87\% at $r=3''$.

\subsubsection{The gas flow}

The Pa-$\alpha$ image in Fig.~\ref{overplot}b \citep[from][]{alo06}
  shows ring-like emission with a radius of $\sim4''\sim1$\,kpc,
  at the expected location of the inner Lindblad resonance (ILR) of the
  primary bar where the gas tends to pile up
  and star formation is likely to proceed. \cite{fri93} argued that, in order
  to avoid chaos around the principal resonances, a double-bar system evolves
  with the corotation radius $R_{\mathrm{cor}}$ of the nuclear bar coincident with
  the ILR of the large-scale bar \citep[see also][]{hun08}, and we indeed
  observe CO emission along the nuclear gas bar approaching the Pa-$\alpha$
  ring (Fig.~\ref{overplot}b). Using
  $R_{\mathrm{cor}}\sim(1.2-1.4)\times R_{\mathrm{bar}}$
  \citep{ath92}, appropriate for fast rotating bars,
  also gives a similar $R_{\mathrm{cor}}\sim1$\,kpc.
The nuclear bar pattern speed is expected to be
$\Omega_s= V_{\mathrm{rot}}(R_{\mathrm{cor}})/R_{\mathrm{cor}}\sim250$\,km\,s$^{-1}$\,kpc$^{-1}$, where we have used the observed velocity field fitted by
\cite{per16}. Such a high value of $\Omega_s$ indicates that the nuclear
bar is decoupled from the primary bar; simulations indeed indicate that
the decoupling requires both the presence of the primary bar ILR and the
anti-bar $x_2$ orbit family \citep{fri93}.
The properties of the nuclear bar of ESO~320-G030 (length and
$\Omega_s$) are similar to those of NGC~2782 \citep{hun08}.
At $r=1''-3''$ ($230-700$\,pc), the
velocity of the bar is $60-180$\,km\,s$^{-1}$. Therefore, the gas
on the trailing edge outruns the bar, but the gas on the leading edge
has a small rotational velocity ($50-150$\,km\,s$^{-1}$) in the frame of
the rotating bar, comparable to or even lower than the inflow velocities
in the same region.

   \begin{figure}
   \centering
   \includegraphics[width=8.5cm]{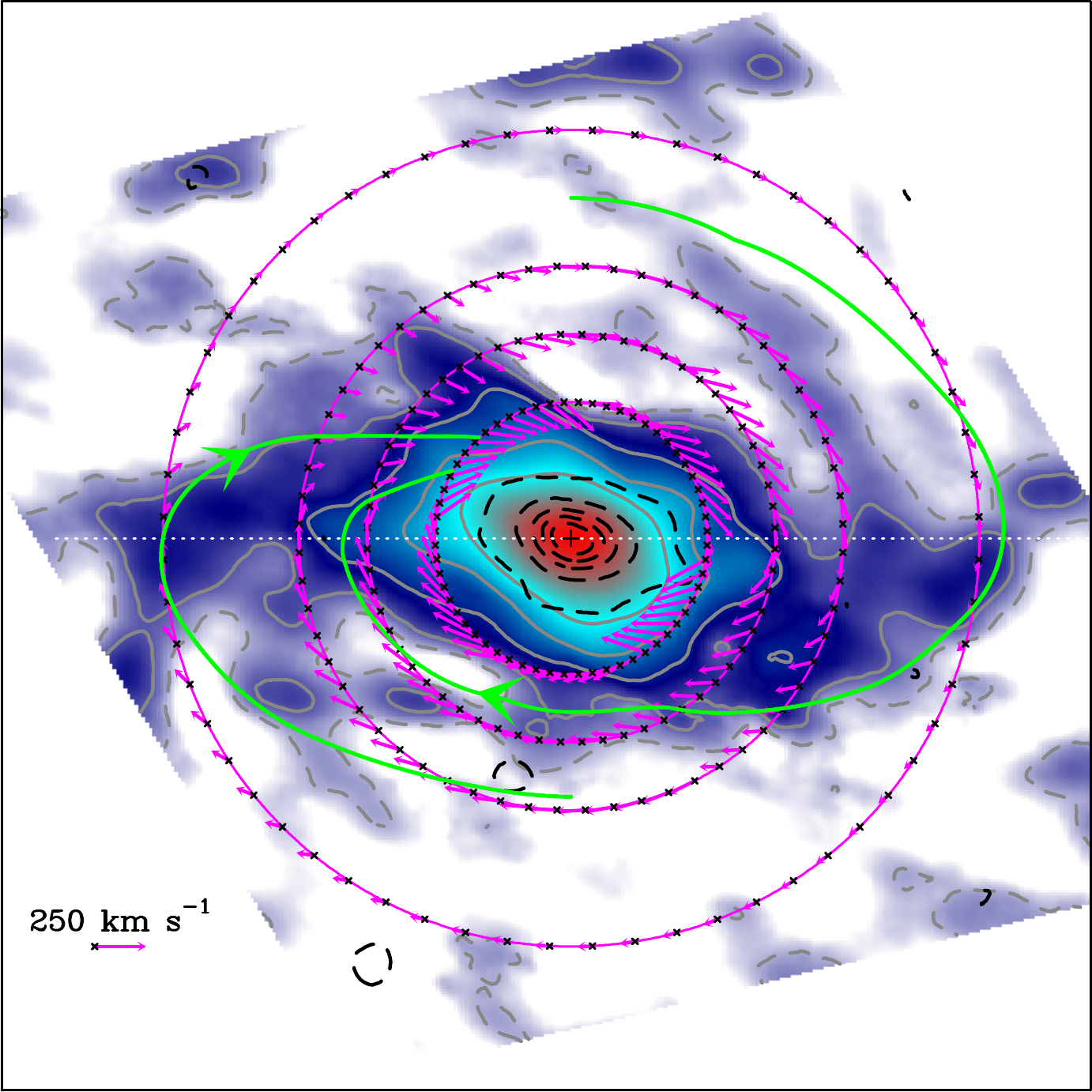}
   \caption{Deprojected images of the CO $2-1$ emission (colored scale and
     gray contours) and of the 454\,GHz continuum (dashed black contours), which
     have been also rotated to have the bar (dotted white line) horizontal.
     The magenta arrows show the inferred velocity vectors along the
     $r=1''-3''$ circles in the frame of the rotating bar, after correcting
     for an assumed nuclear bar pattern speed of
     $\Omega_s= 250$\,km\,s$^{-1}$\,kpc$^{-1}$. The green
     lines are the result of integrating the velocity vector in this
     rotating frame (with linear
     radial interpolation of the velocity field), with departing points at
     $\pm90^{\circ}$ from the bar and $r=1.9''-2.5''$, up to the point
     where they intersect the $r=1''$ circle.
   }   
    \label{barframe}
    \end{figure}

The K-band image of ESO~320-G030, displayed in Fig.~\ref{overplot}a,
probes the nuclear bar rather well, with still a V-shaped apparent
absorption at $\mathrm{PA}=20^{\circ}-60^{\circ}$
probably caused by the outflow observed around that direction
\citep{per16,caz14}. 
The velocity vectors of the molecular gas along the $r=1''-3''$ circles,
projected on the plane of sky, are overlaid on this image.
Most of the CO emission along the gas bar is spatially shifted in the
clockwise (rotation) direction relative to the stellar bar.
The inflowing gas ($V_{\mathrm{rad}}<0$) is also seen ahead of the bar major axis,
and the outflowing gas is observed on the opposite sides, so that
both mark the intersections of the gas flow, which is elongated along
the bar, with the circles. Nevertheless, owing to the asymmetry of the
negative and positive values of $V_{\mathrm{rad}}$ (Fig.~\ref{vrotradco}b),
the lines of gas flow are not
expected to be closed, but will spiral onto the nuclear region.

Since the gas orbits are expected to be approximately stationary in
the rotating bar frame, we show in Fig.~\ref{barframe} the deprojected
images of CO $2-1$ and 454\,GHz continuum, together with the
inferred velocity vectors in the frame of the bar after correcting
for the assumed $\Omega_s= 250$\,km\,s$^{-1}$\,kpc$^{-1}$. The whole
image is rotated such that the bar lies in the horizontal direction.
In this frame, the velocity vectors are nearly parallel to the isocontours
of CO emission at the leading edge of the gas bar, and are perpendicular
to the bar where the gas crosses it. Therefore, our inferred
velocity field approximately accounts for the morphology of the
leading edge of the gas bar, where the gas flows parallel to the bar and
in the inward direction. This point is better seen with the two green curves
of Fig.~\ref{barframe}, which are generated by integrating over time the
velocity vector.
The departing points are selected so as the lines get close to the $3''$
circle when crossing the bar. The curves should be considered with caution,
as the velocity field is only determined at four radial positions and a
linear interpolation is performed at all other radii. Nevertheless, they
seem to delineate rather well the leading edge of the CO gas bar. This
connection between kinematics and morphology gives support to the model,
and illustrates the very efficient bar mechanism to drive a massive
inflow. While these green lines cannot be considered realistic
gas ``orbits,'' due to complex events such as shocks at the bar position,
they represent prominent (dominant) lines of gas flow associated with the
velocity component in red in Fig.~\ref{co}d-g, and as such they have an
associated timescale. The elapsed time along the calculated lines is 13 and
24\,Myr, corresponding to $\sim0.5-1$ turns of the bar.

\subsubsection{The mass inflow rate}
\label{sec:mdotinf}

We estimate the instantaneous mass inflow rate at a radius $r$ as the
net gas mass crossing in the inward direction the circles depicted in
Fig.~\ref{co}a-c per unit time:
\begin{eqnarray}
  \dot{M}_{\mathrm{inf}}(r) & = & -\oint dl_c \,\frac{d M_{\mathrm{rad}}}{dl_c}\,
  \frac{V_{\mathrm{rad}}}{\Delta r_B} \nonumber \\
  & = & -\frac{\alpha_{\mathrm{CO}}\,r\,}{A_{\mathrm{B}}}
  \int_0^{2\pi} d\theta\, L'_{\mathrm{CO}}(r,\theta)\,V_{\mathrm{rad}}(r,\theta)\,
  f_r(\theta)\,f_s(\theta),
  \label{eq:mdotinf}
\end{eqnarray}
where $dl_c$ is an arc element in the plane of the galaxy, $d M_{\mathrm{rad}}$
measures the gas mass with $V_{\mathrm{rad}}\ne0$, $A_{\mathrm{B}}$ is the beam
area at the source distance, and $\Delta r_B=A_{\mathrm{B}}^{1/2}/f_r$ is the
radial interval sampled by the beam in the galaxy plane.
This equation integrates over the circles the gas mass flowing with
$V_{\mathrm{rad}}\ne0$ divided by $\Delta t=\Delta r_B/V_{\mathrm{rad}}$,
where $V_{\mathrm{rad}}$ is displayed in Fig.~\ref{vrotradco}b.
In the second equality of eq.~(\ref{eq:mdotinf}), the CO luminosity
$L'_{\mathrm{CO}}$ only involves line-of-sight velocities
within $\pm50$\,km\,s$^{-1}$
from the red curves in Fig.~\ref{co}d-g. We adopt a conversion factor
$\alpha_{\mathrm{CO}}=0.78$\,$M_{\odot}/(\mathrm{K\,km\,s^{-1}\,pc^2})$, and
implicitly assume the same brightness for the CO $1-0$ and $2-1$ lines.
Finally, $f_r$ and $f_s$ (both within the range $0.73-1$) are geometrical
factors that account for
the source inclination; $f_r$ corrects for the radial interval sampled
by the beam on the plane of the source, and $f_s$ corrects for the
projection of the circular arcs on the plane of the sky:
\begin{eqnarray}
  f_r & = & \left[ \cos^2(\theta-\theta_0) +
    \sin^2(\theta-\theta_0)\cos^2i \right]^{1/2} \nonumber \\
  f_s & = & \left[ \sin^2(\theta-\theta_0) +
    \cos^2(\theta-\theta_0)\cos^2i \right]^{1/2}  
\end{eqnarray}

Equation~(\ref{eq:mdotinf}) implicitely corrects the gas mass crossing the
circles in the inward direction ($V_{\mathrm{rad}}<0$) for that crossing
them in the outward direction ($V_{\mathrm{rad}}>0$), and can be thus
considered net inflow rates.
The values of $d\dot{M}/d\mathrm{PA}$ as a function of PA are
displayed in Fig.~\ref{vrotradco}c, where negative (positive) values indicate
inflowing (outflowing) contributions. It shows that the outflowing mass
does not cancel the inflowing mass at $r\le2''$, although a massive outflowing
clump is seen at $\mathrm{PA}\approx-90^{\circ}$ for $r=1''$.

The values of $\dot{M}_{\mathrm{inf}}$ are listed in Table~\ref{tab:rotinf}.
At $r=1.5''-1''$, $\dot{M}_{\mathrm{inf}}=16-20$\,$M_{\odot}$\,yr$^{-1}$ is
similar to our estimated nuclear SFR
($\sim16-18$\,$M_{\odot}$\,yr$^{-1}$, Sect.\,\ref{sed}), strongly suggesting
that the nuclear starburst is fed and sustained by the observed inflow.
Our $\dot{M}_{\mathrm{inf}}$ values are not corrected by the
feedback from the nuclear region, although
$\dot{M}_{\mathrm{outf}}<10$\,$M_{\odot}$\,yr$^{-1}$ \citep{per16}.
We have also estimated in Table~\ref{tab:rotinf} the inward flux of angular
momentum across the quoted circles, by including the factor
$r\,V_{\mathrm{rot}}(r,\theta)$ in the second equality of eq.~(\ref{eq:mdotinf}).
While $\dot{L}_{\mathrm{inf}}$ is negative at $r=3''$, meaning a net transfer
of angular momentum outward, its value at shorter radii does not show a
clear dependence on $r$.

%__________________________________________________ One column table
   \begin{table}
     \caption{Estimated mass inflow rates and inward
       fluxes of angular momentum across the four outer circles ($r=1''-3''$)
       depicted in Fig.~\ref{co}a-c .}
         \label{tab:rotinf}
\begin{center}
          \begin{tabular}{ccc}   
            \hline
            \noalign{\smallskip}
            $r$  & $\dot{M}_{\mathrm{inf}}$$^{\mathrm{a}}$ &
            $\dot{L}_{\mathrm{inf}}$$^{\mathrm{a}}$ \\
            $('')$ &  ($M_{\odot}$\,yr$^{-1}$) &
            ($M_{\odot}$\,pc\,km\,s$^{-1}$\,yr$^{-1}$) \\
            \noalign{\smallskip}
            \hline
            \noalign{\smallskip}
            3 & $\sim0$  & $-10^5$ \\
            2 & 11.3 & $6.2\times10^5$ \\
            1.5 & 16.0 & $8.3\times10^5$ \\
            1 & 19.9 & $6.8\times10^5$ \\
            \noalign{\smallskip}
            \hline
         \end{tabular} 
\end{center}
\begin{list}{}{}
\item[$^{\mathrm{a}}$] Both $\dot{M}_{\mathrm{inf}}$ and $\dot{L}_{\mathrm{inf}}$
  are net values.
\end{list}
   \end{table}

The timescale associated with the inflow is
$t\sim M_{\mathrm{gas}}/\dot{M}_{\mathrm{inf}}$, where $M_{\mathrm{gas}}$ is the gas
mass of the nuclear region ($4.5\times10^8$\,$M_{\odot}$, Sect.\,\ref{masses}).
This gives the time for complete nuclear gas replenishment, $t\sim23$\,Myr,
which corresponds to $\sim1$ rotation period of the nuclear bar. This timescale
is similar to the elapsed time estimated for the longest curve in
Fig.~\ref{barframe}, $t\sim24$\,Myr. Since the gas mass enclosed in
the annulus between $r=1''$ and $r=3''$ is $3.6\times10^8$\,$M_{\odot}$
(Sect.\,\ref{masses}), it gives an independent estimate of
$\dot{M}_{\mathrm{inf}}\sim15$\,$M_{\odot}$\,yr$^{-1}$. The similarity of both
estimates is encouraging, given that the former gives an ``instantaneous''
value (i.e., averaged over the time the flow crosses a radial distance
equivalent to the beam size, $\sim0.6$\,Myr at 100\,km\,s$^{-1}$) while the
latter is a value averaged over the next $\sim20$\,Myr.

Our timescale for complete nuclear gas replenishment is also similar
--to within a factor of 2-- to the equivalent timescales estimated 
for NGC~4418 \citep{gon12,sak13} and Arp\,299a \citep{fal17}, two LIRGs
with luminosities similar to ESO~320-G030 and showing also inflowing
molecular gas toward their nuclei.
It is also consistent with the expectedly short timescales of low-luminosity
  AGN duty cycles \citep{gar12}.

\subsubsection{The overall scenario}

The kinematic model derived from the CO $2-1$ data cube indicates an
efficient mechanism that drives a massive inflow along the bar. 
The gas in the arms, which is rotating faster than the bar, overruns it
with perpendicular incidence.
At and beyond the leading edge of the bar,
a negative torque is exerted by the stars that
drains angular momentum of the molecular gas \citep{gar05,hun08}, 
generating orbits of high eccentricity that make the gas flow couple to the
bar morphology. The steep change in the direction of the
velocity vectors at the leading edge of the bar suggests the presence of
a nearly radial shock front, which is offset from the bar major axis
in the forward (rotation) direction \citep{kor04}.
The gas approaching the bar in
the perpendicular direction will shock the gas flowing parallel to the
bar along its leading edge, coming from larger $r$. 
Dissipation of kinetic energy through these shocks and viscosity contribute
to drive a quasi-radial inflow in the rotating bar frame; we expect that after
just two crossings of the bar, the inflowing gas will accumulate, through a
shock that drives turbulence, around the envelope and nuclear disk, thus
feeding the nuclear starburst.
In ESO~320-G030, the inflowing gas does not stall in a ring at the ILR
of the nuclear bar, but continues all
the way toward the inner $\sim150$\,pc as evidenced by the high
concentration of warm molecular gas forming the structures probed by H$_2$O
and other species (Appendix~\ref{appa}).  
The enhanced nuclear star formation as derived from the IR luminosities of
the nuclear disk and envelope (Sect.\,\ref{sed}) strongly suggests that we are
viewing a pseudobulge in formation, namely, a proto-pseudobulge.

We obtain inflow velocities comparable in magnitude to those inferred in
NGC~1530 \citep{reg97}, and also increasing toward the center. 
With decreasing distance to the nucleus,
shocks increase the gas turbulence and the inflow becomes more disordered
and not restricted to the plane of the galaxy. Fig.~\ref{co}c shows that
velocity dispersion as measured by CO 2-1 apparently decreases just at the
nucleus. This decreasing $\Delta V$ is associated with a nuclear blue asymmetric
self-absorbed CO profile \citep{per17}, illustrating that the inflowing gas
from the SW region is also seen in front of the nucleus at small radii.
The increasing flow distortion, with gas inflowing along
orbits not contained in the galaxy plane,
is also required to account for the redshifted
velocities found in the OH 119 and 79\,$\mu$m doublets observed in
absorption (Sect.\,\ref{inflowfarir}), indicating the presence of gas with
line-of-sight velocities of $\sim100$\,km\,s$^{-1}$. 
It is plausible that this represents the effect of vertical resonances
on the gas flow that make the inflow 3D \citep{pfe90}.
The extra-planar flow of gas may also have a contribution from the
fountain effect generated by the neutral outflow \citep{caz14}, which remains
gravitationally bound to the galaxy.

   \begin{figure}
   \centering
   \includegraphics[width=7.0cm]{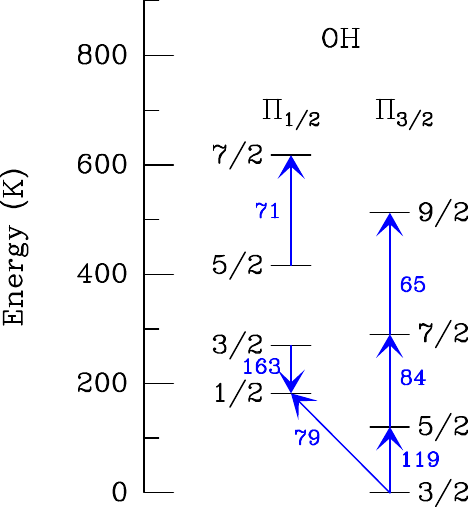}
   \caption{Energy level diagram of OH indicating the lines observed with
     {\it Herschel}/PACS (blue arrows and labels). Labels denote
     the rounded wavelengths in $\mu$m. Upward (downward) arrows indicate lines
  detected primarily in absorption (emission).
   }   
    \label{eneroh}
    \end{figure}
   \begin{figure}
   \centering
   \includegraphics[width=8.0cm]{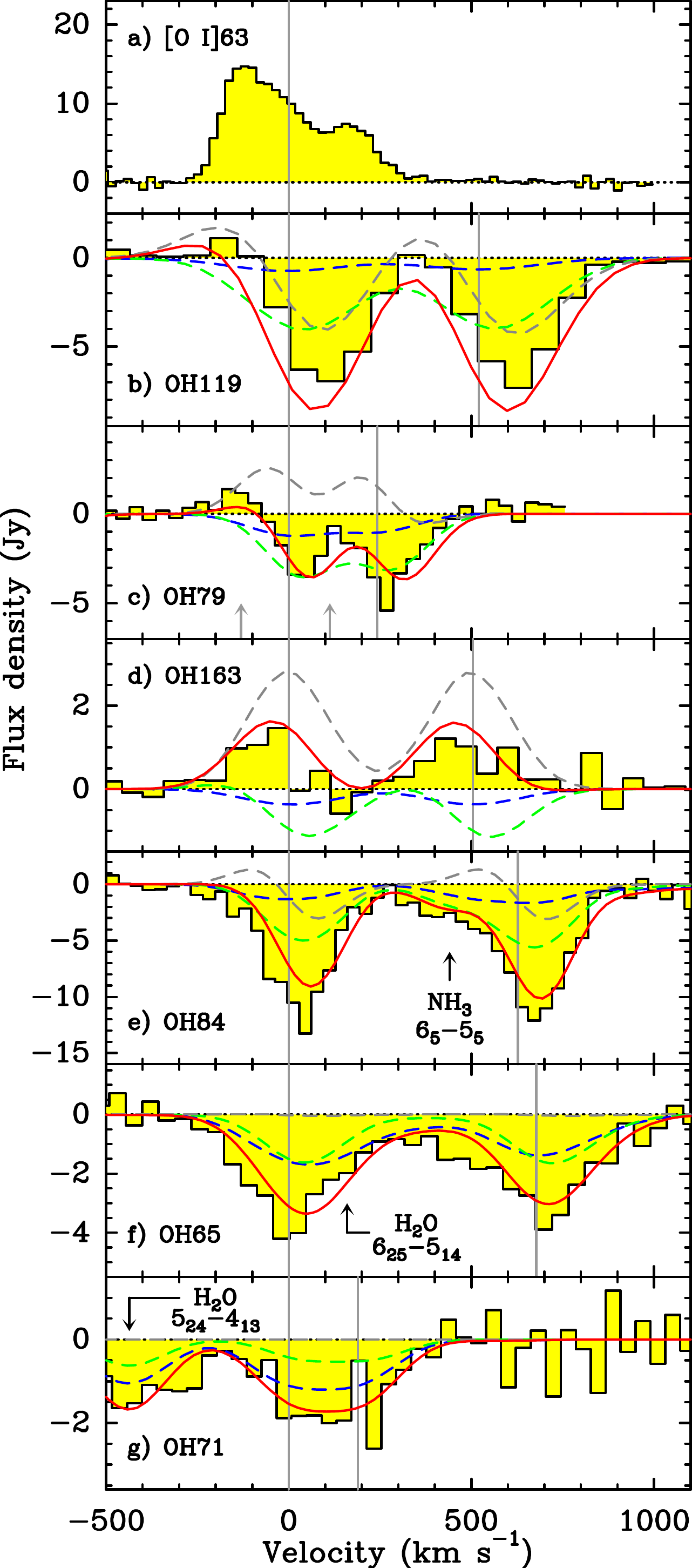}
   \caption{Inflow in ESO~320-G030 observed in the far-IR
     [O {\sc i}]\,63\,$\mu$m line and OH doublets. The gray vertical lines
     indicate the position of the two components of the OH doublets
     (blended within a single feature in the case of OH\,71\,$\mu$m
     in panel g), and
     the vertical arrows in panel c indicate the position of the
     $^{18}$OH\,79\,$\mu$m doublet components. OH 
     is modeled with the same components as derived from H$_2$O: the
     core (dashed blue lines), the nuclear disk (dashed green) and
     the envelope (dashed gray); red is total. The inflow is simulated
     in the envelope component (see Sect.\,\ref{inflowfarir}). The model
     also includes the contribution by $^{18}$OH, NH$_3$, and H$_2$O
     in panels c, e, and f-g, respectively (see Appendix~\ref{appa}).
   }   
    \label{oh}
    \end{figure}

\subsection{The inflow observed in the far-IR}
\label{inflowfarir}
  
Clear evidence of inflowing gas is also seen in the far-IR. The
[O {\sc i}]\,63\,$\mu$m line shows a blue asymmetric profile with redshifted
absorption at $\approx100$\,km\,s$^{-1}$ (Fig.~\ref{oh}a). Unlike the
case of NGC~4418 \citep{gon12}, however, the redshifted part of the
profile is seen in emission above the continuum, probably because
the continuum emission from ESO~320-G030
is less spatially concentrated than in NGC~4418.

The observed OH doublets, indicated in the energy level diagram of
  Fig.~\ref{eneroh}, show a sequence in the velocity of
the absorption as a function of the lower level energy and line optical
depth (Fig.~\ref{oh}b-g): The ground-state and optically thick
OH\,119\,$\mu$m doublet peaks at $\approx100$\,km\,s$^{-1}$;
the cross-ladder ground-state OH\,79\,$\mu$m doublet, with lower opacity,
also shows evidence for redshifted absorption but at lower velocities; the
OH\,84\,$\mu$m doublet, with $E_{\mathrm{lower}}=120$\,K, still shows
some hints of redshifted absorption, but the high-lying
OH\,65 and 71\,$\mu$m doublets, with $E_{\mathrm{lower}}=290$ and
$415$\,K, respectively, peak at central
velocities. Since the doublets progressively probe more excited
(and therefore more central) regions, the inflow dissipates its kinetic
energy when approaching the very inner regions of the nucleus
(i.e., the disk, where inflow motions are still seen, and the core).

On the other hand, the high OH column densities (see below) required to
account for the observed absorption suggest that these are produced within
the nuclear region sampled by H$_2$O. Therefore, we explore here whether
the inflow observed in the far-IR is primarily associated with the outermost
nuclear H$_2$O component, that is, the envelope, responsible
for the low-lying H$_2$O far-IR absorption
and submm emission lines. Indeed, hints of redshifted absorption are
also seen in the H$_2$O138 and  H$_2$O75 lines (Fig.~\ref{fitpacsalma}).
This inflowing region is smaller ($r\lesssim0.5''$,
Fig.~\ref{figsketch}) than that sampled by CO ($1''-2''$, Fig.~\ref{co}).

We have then applied the composite H$_2$O fiducial model to OH, but have
included a velocity field as follows: For the envelope, the gas is inflowing
with a velocity of 100\,km\,s$^{-1}$ at the outermost radius and decreasing
linearly with radius; for the disk, an inflow velocity of 60\,km\,s$^{-1}$
is adopted. No velocity field is included for the core component. Model
results are overlaid on the observed line profiles in Fig.~\ref{oh}b-e,
and are roughly consistent with the scenario that the inverse P-Cygni
OH\,119\,$\mu$m and 79\,$\mu$m profiles are driven by an inflow within
the envelope that primarily applies to the external shells ($110-150$\,pc).
Since this component is optically thin in the far-IR continuum,
it generates inverse P-Cygni profiles in OH\,119 and OH\,84\,$\mu$m
but a blue asymmetric profile in OH\,79\,$\mu$m
(gray dashed curves in Fig.~\ref{oh}). 
Toward the optically thick disk and core
components all doublets are predicted in absorption,
and the blue asymmetric profile of the OH\,163\,$\mu$m doublet is
produced by the redshifted absorption toward the disk
(green dashed curve in Fig.~\ref{oh}d).
The net modeled profiles (shown in red) resemble the observed ones although
with significant discrepancies in the OH\,119\,$\mu$m doublet.

The OH column density of the inflowing gas in the envelope is
$N_{\mathrm{OH}}=3.2\times10^{16}$\,cm$^{-2}$. To estimate the associated
mass inflow rate, we adopt a geometry consisting of two shells on opposite
sides, each with radius $R\sim130$\,pc, surface $\pi R^2$, and width
$\Delta R\sim40$\,pc, inflowing with an average
velocity of $V_{\mathrm{inf}}\sim80$\,km\,s$^{-1}$, so that
\begin{equation}
  \dot{M}_{\mathrm{inf}}\sim
\frac{2\,N_{\mathrm{OH}}\,\mu \,m_{\mathrm{H}} \,\pi R^2 \,V_{\mathrm{inf}}}{X_{\mathrm{OH}}\,\Delta R},
\end{equation}
which results in $\dot{M}_{\mathrm{inf}}\sim30$\,$M_{\odot}$\,yr$^{-1}$ for a
fiducial OH abundance of $X_{\mathrm{OH}}\sim2.5\times10^{-6}$. While this
estimate is admittedly rather uncertain, it is consistent with the scenario
of several dozens solar masses per year of gas feeding the nucleus of
ESO~320-G030, as inferred from CO.

\section{Discussion and conclusions}
\label{sec:discussion}

The combined analysis of the H$_2$O absorption and emission
lines in ESO~320-G030,
with wavelengths ranging from $58$ to $669$\,$\mu$m, and the continuum,
together with high-resolution data obtained with ALMA for the H$_2$O448 line
and the associated submm dust emission, unveils the structure of the
galactic nucleus which we suggest is evidence for the presence of
  a prominent proto-pseudobulge fed by a molecular inflow driven by a
  strong nuclear bar. The radius of the most extended region
  of the nucleus (the envelope component, $\sim200$\,pc) is in the lower range
  of measured pseudobulge sizes \citep[e.g.,][]{car99,fis08}.
  The radius will likely increase in view of the CO gas reservoir
    around the nucleus, with a mass comparable to that of the nucleus.
  Our 3D model for the H$_2$O448 line shape indicates a velocity field with
  $V_{\mathrm{rot}}/\sigma\sim1$,
  meaning that there is an increase in random motions relative to ordered gas
  motions in the nuclear disk. Stellar kinematics indicate a value of
  $V_{\mathrm{rot}}/\sigma\sim0.7$ within an aperture of $r=150$\,pc, also showing
  an increase in random motions in the nuclear region while retaining a
  memory of the rotation (A. Crespo G\'omez et al., in prep.).

The envelope has typical columns of 
$\mathrm{several}\times10^{23}$\,cm$^{-2}$ and is moderately warm
($T_{\mathrm{dust}}\approx50$\,K). With these conditions, the low-lying
H$_2$O lines at submm wavelengths ($240-400$\,$\mu$m) are efficiently pumped,
but little absorption is produced in the far-IR with only significant
absorption in the H$_2$O lines at 75 and 108$\mu$m. Nevertheless, the inferred
colums are more than enough to extinguish any line emission at short
wavelengths, and indeed
the Pa-$\alpha$ emission within $r=0.5$\,kpc primarily probes
  the western (near) side of the nuclear bar, with a morphology different from
  that of the continuum submm emission (Fig.~\ref{overplot}b).
It is within the envelope region where the CO 2-1 line becomes very broad
(Fig.~\ref{co}c), which is probably a direct consequence
of the shocks produced by the inflowing gas and may be evidence of
disordered motions that would eventually lead to a pseudobulge.

High-lying molecular absorption lines in the far-IR are produced when the
columns are so high that the far-IR continuum becomes optically thick
\citep{gon15}, and these conditions are also linked to the
emission in the H$_2$O448 line. In ESO~320-G030, a nuclear disk with
projected radius of $\approx40$\,pc attains these conditions, though still
with moderately warm dust ($T_{\mathrm{dust}}\approx55$\,K). The disk
is distorted, elongated in the direction of the bar, and highly turbulent
($\Delta V/V_{\mathrm{rot}}\approx1$);
it is thus expected to be geometrically thick \citep[e.g.,][]{caz20}.

At the center of the disk, an unresolved, extremely buried
($\tau_{100}>>1$) and very warm
\citep[$T_{\mathrm{dust}}\sim100$\,K for the far-IR photosphere; see][]{gon19}
 core component is identified 
from the absorption detected in very high-lying lines of H$_2$O, of which
the H$_2$O\,$7_{07}-6_{16}$ line at $71.95$\,$\mu$m ($E_{\mathrm{lower}}=640$\,K)
is an excellent tracer. We estimate a core radius of $R=9-16$\,pc, but
  higher angular resolution observations are
  required to better determine its size.
HCN vibrational emission 
  has been recently detected in ESO~320-G030 (N. Falstad et al., in
  preparation), additionally indicating the presence of a very warm 
  optically thick region. In such environments, trapping of IR radiation raises 
  $T_{\mathrm{dust}}$ and the mid-IR radiation density within the cocoon
  of dust, but the resulting SED does not show any enhanced mid-IR emission
  as only the photosphere is probed at mid- and even far-IR wavelengths
  \citep{gon19}.
Because of their extreme extinction, the nature of these very compact
nuclear components has been long debated;
even X-rays and mid-IR high ionization tracers from a putative AGN 
are expected to be severely attenuated.
ESO~320-G030 is undetected with the {\it Swift}/BAT all-sky survey
  observations\footnote{https://swift.gsfc.nasa.gov/results/bs105mon/}
  in the $14-195$\,keV band with a sensitivity of
  $8.4\times10^{-12}$\,erg\,s$^{-1}$\,cm$^{-2}$, which translates into a
  luminosity of $<6\times10^8$\,L$_{\odot}$.
  Assuming a $\sim5$\% contribution to the bolometric AGN luminosity
  in the quoted band, the upper limit for $L_{\mathrm{AGN}}$ is
  $\lesssim10^{10}$\,L$_{\odot}$; however, 
  absorption at $<30$\,keV is still relevant for
  $N_{\mathrm{H}}\sim10^{25}$\,cm$^{-2}$, and an AGN with
$\sim10$\% of the total galaxy luminosity is still possible.
Assuming this limiting AGN scenario, a mass accretion rate onto the black
hole of $\mathrm{BHAR}\sim0.01$\,$M_{\odot}$\,yr$^{-1}$ would be required
for a fiducial radiative efficiency of $0.1$.
This BHAR is a factor of $\sim6\times10^{-4}$ times the
estimated nuclear SFR, matching the volume-averaged
$\mathrm{BHAR/SFR}$ ratio in local bulge-dominated galaxies \citep{hec04}.
Using $M_{\mathrm{dyn}}\approx2\times10^9$\,$M_{\odot}$ for $r<r_{\mathrm{H_2O}}$ 
(Sect.\,\ref{masses}) as a proxy for the mass of the pseudobulge in formation,
and the $M_{\mathrm{BH}}/M_{\mathrm{bulge}}\sim2.5\times10^{-3}$ ratio appropriate
for small bulges \citep{kor13}, the resulting 
$M_{\mathrm{BH}}\sim5\times10^6$\,$M_{\odot}$\footnote{Using the observed
  $\Delta V\approx100$\,km\,s$^{-1}$ as a proxy for the stellar velocity
  dispersion that would result once the gas is locked onto stars, and the
  observed $M_{\mathrm{BH}}/\sigma$ correlation by \cite{tre02}, a slightly
  higher $M_{\mathrm{BH}}\sim8\times10^6$\,$M_{\odot}$ is obtained.}
would be emitting at a high level of $0.1\,L_{\mathrm{Edd}}$.
This very crude estimate
is comparable to the high Eddington ratio ($\sim0.3$) estimated for NGC~4418
if an AGN is assumed to power its compact nucleus \citep{sak13}.

The nuclear core may still be primarily powered by a starburst, with
 a limiting luminosity surface density of
 $\sigma\,T_{\mathrm{dust}}^4\sim10^{13}$\,$L_{\odot}$\,kpc$^{-2}$.
 This is close to the value theoretically expected for radiation-pressure
 supported starburst disks \citep{tho05}.
 Even in this
scenario, the core component is expected to host a growing SMBH
at the center of the galaxy, given the nuclear feeding reservoir
($M_{\mathrm{gas}}\sim10^8$\,$M_{\odot}$ within $\sim12$\,pc) and
excellent conditions for such fast growth (probably constrained by
radiation pressure on dust grains).
Statistically, Seyfert galaxies are preferentially found in barred
systems \citep[e.g.,][]{mai03}.

Excluding the luminosity of the core component, the averaged
  nuclear SFR surface density is
  $\log\Sigma_{\mathrm{SFR}}\, (\mathrm{M_{\odot}\,yr^{-1}\,kpc^{-2}})\sim2.1$,
and the averaged nuclear molecular gas surface density is
$\log\Sigma_{\mathrm{H2}}\, (\mathrm{M_{\odot}\,pc^{-2}})\sim3.6$. Thus the
nuclear region of ESO~320-G030 lies near the high end of the
Schmidt law for starburst galaxies
\citep{ken98}.

Bars within bars were long ago understood to provide an efficient way to
drive gas toward the very inner centers of galaxies \citep[e.g.,][]{shl89},
and ESO~320-G030 appears to be a prototypical example of such a system. 
The nuclear region is moderately elongated along the bar
as traced primarily by the dust continuum image at 454\,GHz,
and a massive inflow is found in the inner $\approx0.5$\,kpc of the
galaxy from the analysis of the CO 2-1 data cube.
The azimuthal velocity of the molecular gas sharply decreases across the
bar, resembling the large velocity jumps observed across optical dust lanes
associated with bars \citep[e.g.,][]{reg97}. 
The molecular inflow, with typical radial velocities of
$80-150$\,km\,s$^{-1}$, is indeed strongly associated with the nuclear bar.
Two independent estimates of the mass inflow rate from CO yield
  similar values, $\dot{M}_{\mathrm{inf}}\sim15-20$\,M$_{\odot}$\,yr$^{-1}$.
  Since these values are similar to the SFR of the nuclear region
  ($16-18$\,M$_{\odot}$\,yr$^{-1}$), we conclude that the enhanced
  nuclear starburst is fed and sustained by the observed inflow.
These inflow velocities are also observed in the ground-state OH
doublets at 119 and 79\,$\mu$m as redshifted absorption,
probing line-of-sight velocities toward the source
of far-IR continuum that indicate a complex 3D flow not
restricted to the plane of the galaxy.
The inflowing gas probed by OH appears to be more compact than that
  sampled by CO; it is associated with the envelope component and
  its kinetic energy dissipates at spatial scales of the
    nuclear disk ($<100$\,pc). The mass inflow rate inferred from OH
  is $\sim30$\,M$_{\odot}$\,yr$^{-1}$, comparable to that derived from CO.

  The timescale associated with the inflow, $\sim20$\,Myr, is also
    expected to characterize the timescale over which the current nuclear
    burst will fade, once the gas reservoir within the ILR of the
    primary bar is fully accreted onto the nucleus. This is at least
one order of magnitude shorter than typical formation timescales
of pseudobulges as inferred for circumnuclear star-forming rings
in barred galaxies \citep{kor04}, indicating
that atypically short timescale secular evolution,
extreme accumulations of gas, and plausibly fast growing SMBHs 
may characterize nuclei of galaxies with strong nuclear bars.

While the case of ESO 320-G030 is exceptional in the local universe,
how common are these nuclear gas concentrations at $z>1$,
when the cosmic accretion rate required to sustain the observed SFR
in massive main sequence
galaxies was $\gtrsim15$\,M$_{\odot}$\,yr$^{-1}$
\citep{sco17}, similar
to the value we obtain for the nuclear
region of ESO 320-G030? How does that matter accrete onto galaxies,
and how is it redistributed? What is the impact of short-range,
intense bursts of nuclear star formation on overall galaxy evolution
over cosmic time? Far-IR spectroscopy provides a unique way of
measuring the spatial structure of star formation during
prolific stages of nuclear gas accumulation,
which can be unveiled with future far-IR facilities similar to
the recently cancelled
{\it SPace Infrared telescope for Cosmology and Astrophysics}
({\it SPICA}) \citep{roe18}\footnote{During the final stages of
  preparation of this manuscript, the {\it SPICA} mission was cancelled
  by ESA prior to the scheduled Mission Selection Review; see
  \cite{abe20}.}.

\begin{acknowledgements}
EG-A is grateful for the warm hospitality of the Harvard-Smithsonian
Center for Astrophysics, where part of the present study was carried out,
and thanks Javier Goicoechea for useful conversations on the far-IR
spectrum of Sgr~B2, Alex Crespo G\'omez for sharing preliminary results on
the stellar kinematics in ESO~320-G030, and Juan Rafael
  Mart\'{\i}nez-Galarza for useful conversations on the Bayesian analysis.
  We thank the anonymous referee
  for constructive and helpful comments on the manuscript.
PACS was developed by a consortium of institutes
led by MPE (Germany) and including UVIE (Austria); KU Leuven, CSL, IMEC
(Belgium); CEA, LAM (France); MPIA (Germany); 
INAFIFSI/OAA/OAP/OAT, LENS, SISSA (Italy); IAC (Spain). This development
has been supported by the funding agencies BMVIT (Austria), ESA-PRODEX
(Belgium), CEA/CNES (France), DLR (Germany), ASI/INAF (Italy), and
CICYT/MCYT (Spain).
SPIRE was developed by a consortium of institutes led by Cardiff University
(UK) and including Univ. Lethbridge (Canada); NAOC (China); CEA, LAM (France);
IFSI, Univ. Padua (Italy); IAC (Spain); Stockholm Observatory (Sweden);
Imperial College London, RAL, UCL-MSSL, UKATC, Univ. Sussex (UK);
and Caltech, JPL, NHSC, Univ.Colorado (USA). This development has been
supported by national funding agencies:
CSA (Canada); NAOC (China); CEA, CNES, CNRS (France); ASI (Italy);
MCINN (Spain); SNSB (Sweden); STFC, UKSA (UK); and NASA (USA).
This paper makes use of the following ALMA data: ADS/JAO.ALMA\#2016.1.00263.S
and ADS/JAO.ALMA\#2013.1.00271.S.
ALMA is a partnership of ESO (representing its member states), NSF (USA) and
NINS (Japan), together with NRC (Canada) and NSC and ASIAA (Taiwan) and
KASI (Republic of Korea), in cooperation with the Republic of Chile.
The Joint ALMA Observatory is operated by ESO, AUI/NRAO and NAOJ.
Based on observations made with ESO Telescopes at the Paranal Observatory
under programme ID 086.B-0901(A).
EG-A is a Research Associate at the Harvard-Smithsonian
Center for Astrophysics. EG-A, JM-P, and FR-V thank the Spanish 
Ministerio de Econom\'{\i}a y Competitividad for support under projects
ESP2017-86582-C4-1-R and PID2019-105552RB-C41.
EG-A and HAS thank NASA grant ADAP NNX15AE56G.
MP-S acknowledges support from the Comunidad de Madrid, Spain, through
Atracci\'on de Talento Investigador Grant 2018-T1/TIC-11035 and
PID2019-105423GA-I00 (MCIU/AEI/FEDER,UE).
AA-H and SG-B acknowledge support through grant PGC2018-094671-B-I00
(MCIU/AEI/FEDER,UE). LC, AA-H, MP-S, and JM-P acknowledge support under
project No. MDM-2017-0737 Unidad de Excelencia "Mar\'{\i}a de Maeztu" -
Centro de Astrobiolog\'{\i}a (INTA-CSIC). SG-B acknowledges support from
the Spanish  MINECO  and  FEDER  funding grant AYA2016-76682-C3-2-P.
C.Y. acknowledges support from an ESO Fellowship.
LC acknowledges support  from  the Spanish Ministerio de Econom\'{\i}a y
Competitividad for support under project ESP2017-83197.
SC acknowledges financial support from the State Agency for Research of
the Spanish MCIU through the 'Center of Excellence Severo Ochoa' award
to the Instituto de Astrof\'{\i}sica de Andaluc\'{\i}a (SEV-2017-0709).
This research has made use of NASA's Astrophysics Data System (ADS)
and of GILDAS software (http://www.iram.fr/IRAMFR/GILDAS).
\end{acknowledgements}

\bibliographystyle{aa}
\bibliography{refs}

\begin{appendix} %First appendix
\section{{\it Herschel}/PACS and SPIRE observations of ESO~320-G030 and models}
\label{appa}

\begin{figure*}
   \centering
   \includegraphics[width=16.0cm]{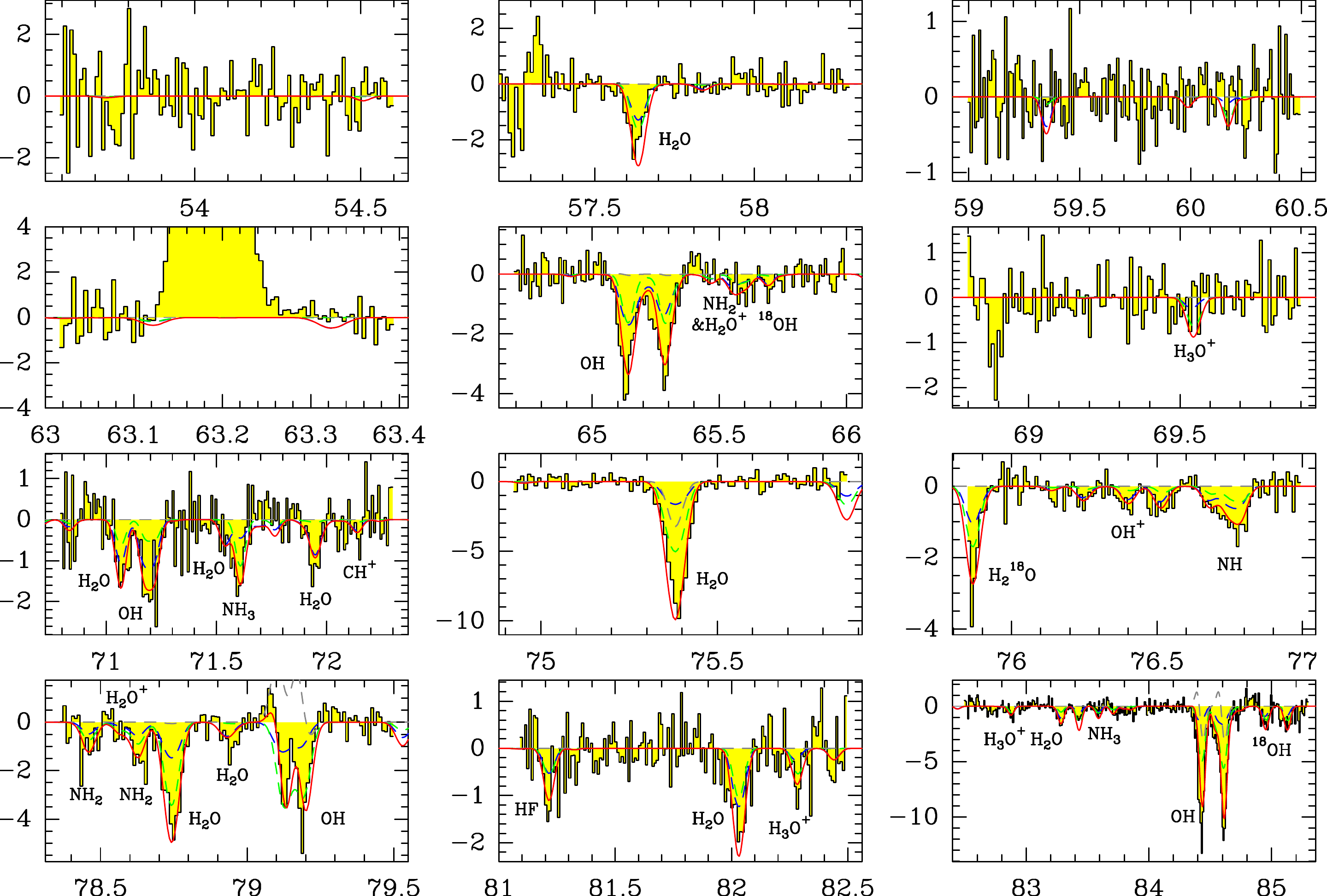}
   \includegraphics[width=16.0cm]{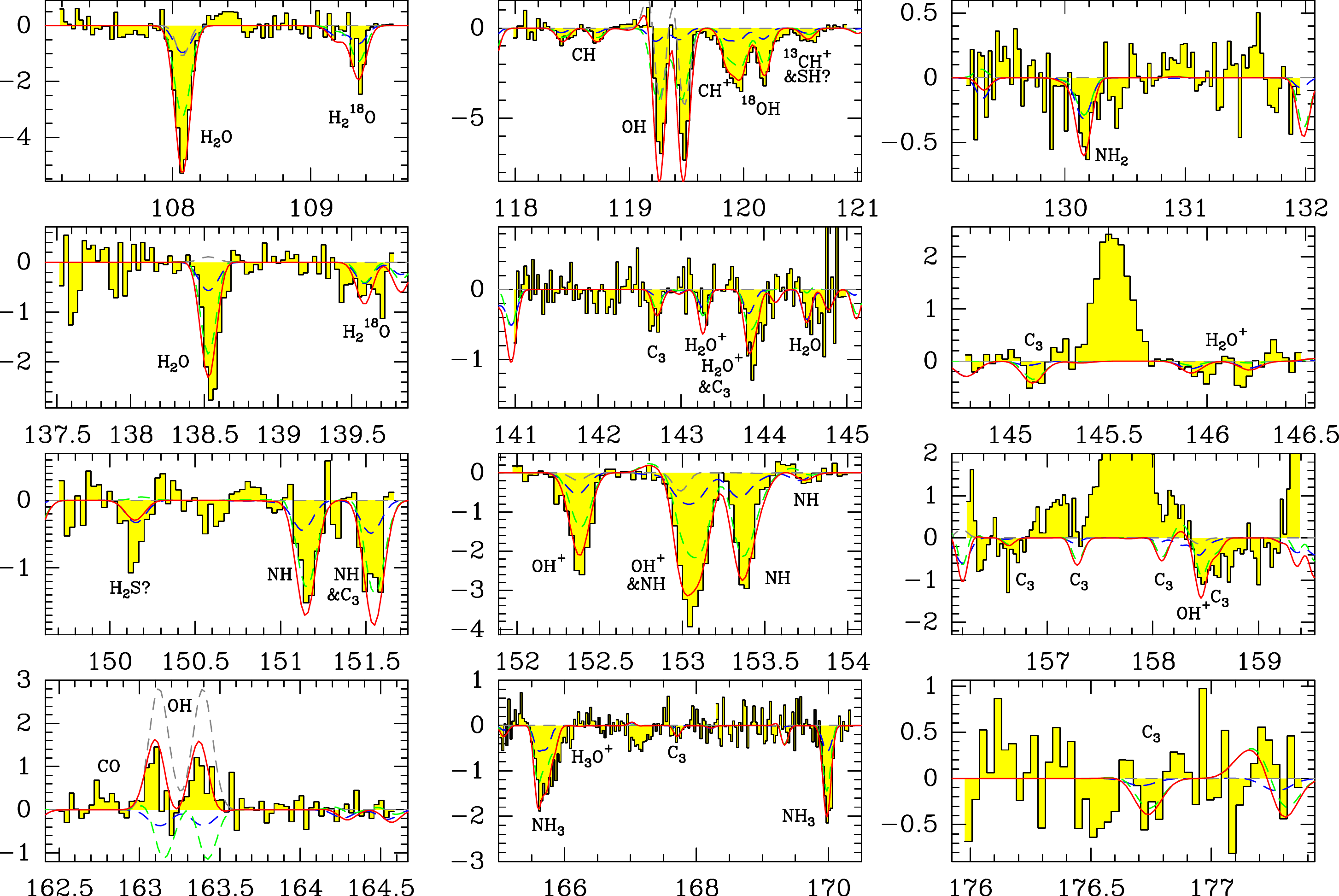}
   \caption{Overall model fit to all molecular features detected
     with {\it Herschel}/PACS in
     ESO~320-G030, including H$_2$O, H$_2^{18}$O, OH, $^{18}$OH,
     OH$^+$, H$_2$O$^+$, H$_3$O$^+$, NH, NH$_2$, NH$_3$, CH, CH$^+$, HF,
     C$_3$, and H$_2$S (for this last species, see text). The contribution
     by the core, the nuclear disk, and the envelope is shown with
     dashed blue, green, and gray; red is total. The carriers of
     the modeled features (some of them undetected) are indicated.
     The abscissa indicates rest
     wavelength in $\mu$m, and the ordinate axis is the
     continuum-subtracted flux density in Jy.
   }   
    \label{fullpacs}
    \end{figure*}

\begin{figure*}
   \centering
   \includegraphics[width=18.0cm]{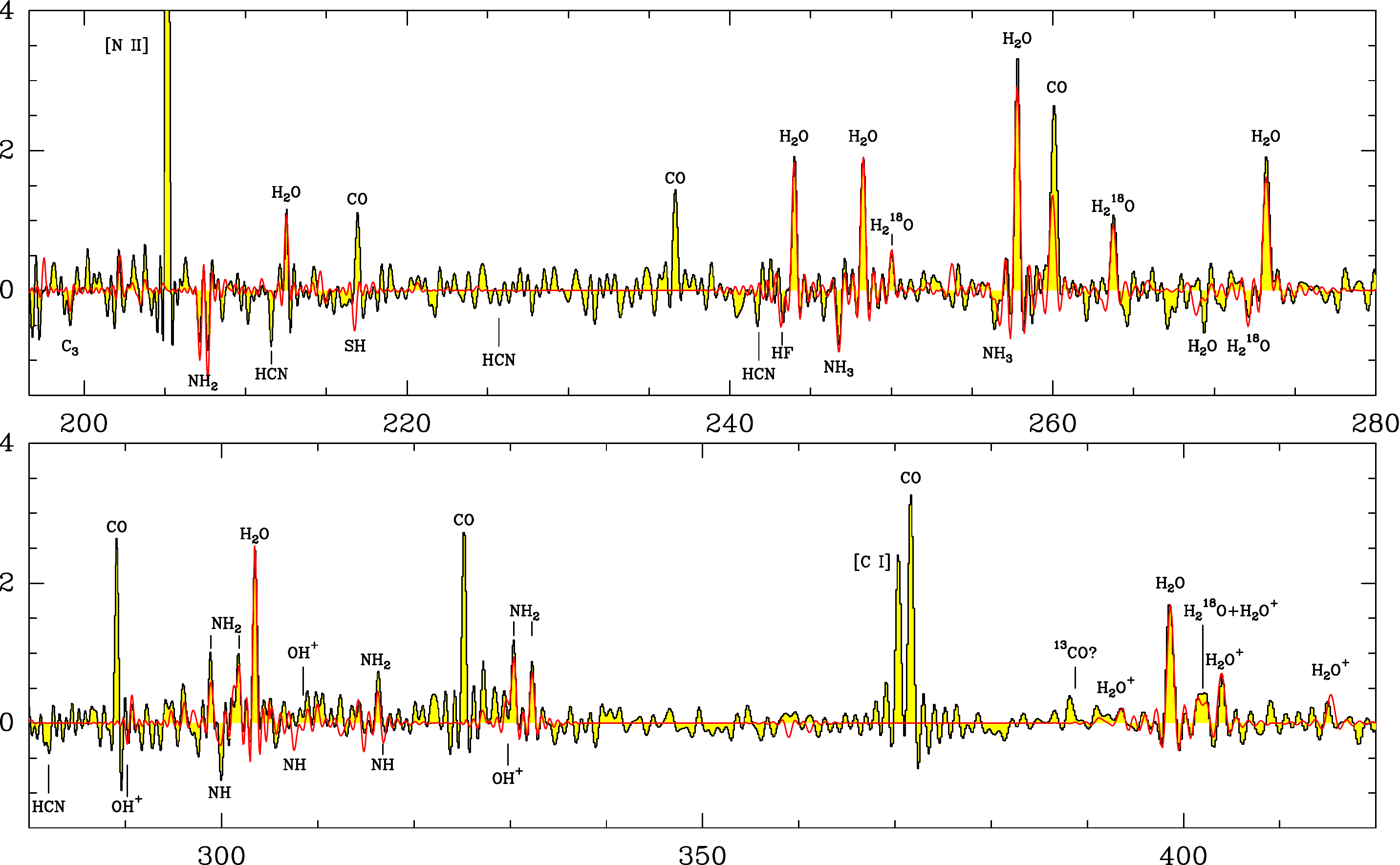}
   \caption{Comparison between the observed {\it Herschel}/SPIRE spectrum
     (up to 420\,$\mu$m, filled histograms) observed in ESO~320-G030 and the
     predictions by the multispecies model (in red).
     The modeled spectrum has been convolved to
     the appropriate sinc function. The abscissa indicates rest
     wavelength in $\mu$m, and the ordinate axis is the
     continuum-subtracted flux density in Jy.
   }   
    \label{fullspire}
   \end{figure*}

We have applied our composite model for H$_2$O to all other molecular
absorption features detected in the far-IR with {\it Herschel}/PACS.
Models were generated for H$_2$O, H$_2^{18}$O, OH, $^{18}$OH,
OH$^+$, H$_2$O$^+$, H$_3$O$^+$, NH, NH$_2$, NH$_3$, CH, CH$^+$, $^{13}$CH$^+$,
SH, HF, C$_3$, and H$_2$S, and the resulting modeled spectrum is overlaid
with the spectra of all observed PACS wavelength ranges in
Fig.~\ref{fullpacs}\footnote{The $\lambda_{\mathrm{rest}}=63.0-63.4$\,$\mu$m
  range has an Obs ID 1342212227; all other Obs ID are listed in
  Table~\ref{tab:obs}.}. Model predictions at submm wavelengths are
  also compared with the {\it Herschel}/SPIRE spectrum of ESO~320-G030
  in Fig.~\ref{fullspire}.

Models for species other than H$_2$O have the intrinsic uncertainty of
the contribution of each component to the line absorption or emission.
We have adopted the following criteria: $(i)$ Since the optically thin envelope
generates little absorption in the H$_2$O lines, it is only included when
needed, that is, for species that require it to obtain a reasonable fit in
some lines. These are OH, OH$^+$, and NH$_2$. $(ii)$ For the species H$_2^{18}$O
and $^{18}$OH, we adopt a fixed column density ratio relative to the main
isotopologues in the core and the disk. For the other species we started
by assuming the same abundance in the core and the disk, but later relaxed 
this assumption for some species to obtain a better fit to the observed lines.
$(iii)$ In addition, we allowed for some flexibility in the value of
$\tau_{100}$ for the disk and the envelope, relative to the fiducial model.
In Table~\ref{tab:res}, the envelope has a fiducial $\tau_{100}=0.22$, but
both OH and OH$^+$ are better reproduced with $\tau_{100}=0.34$. In addition,
$\tau_{100}$ of the disk is allowed to vary between the fiducial value of $1.5$
and $3$. The increase in $\tau_{100}$ has the effect of enhancing the modeled
molecular absorption at $>130$\,$\mu$m. $(iv)$ For most species
(OH, $^{18}$OH, OH$^+$, H$_3$O$^+$, NH, CH, CH$^+$, $^{13}$CH$^+$, C$_3$, and SH)
an inflow velocity of 60\,km\,s$^{-1}$ was included in the disk model to
match the position of the observed absorption features, which further
indicates that the inflowing gas is still present at galactocentric distances
of only $\sim40$\,pc.
The column densities and abundances in the very saturated core component are
obviously uncertain, and we rely mostly on the values in the disk for which
uncertainties are expected to be better than 0.4\,dex.

\underline{OH}: All observed doublets at 65.2, 71.2, 79.2, 84.3, 119.3, and
163.2\,$\mu$m are detected (see also Sect.\,\ref{inflowfarir}).
A very high OH abundance in the core is apparently required to nearly
reproduce the OH\,71 and 65\,$\mu$m absorption, but it is significantly
lower in the disk. The model overpredicts to some extent the absorption
in the OH\,119\,$\mu$m doublet. 

\underline{H$_2^{18}$O}: Absorption lines are detected at 75.9, 109.4,
and 139.6\,$\mu$m, although the first two lines are close to the edge of
the observed wavelength ranges. The lines are reproduced by assuming the same
$N(\mathrm{H_2O})/N(\mathrm{H_2^{18}O})=100-150$
ratio in the core and the disk. The resulting model reproduces
  rather well the H$_2^{18}$O submm lines at 250, 264, 272, and 402\,$\mu$m
  (Fig.~\ref{fullspire}).

\underline{$^{18}$OH}: The observed doublets at 65.7, 85, and 120\,$\mu$m
are nearly reproduced with $N(\mathrm{OH})/N(\mathrm{^{18}OH})=100-150$, a 
value similar to that found for H$_2$O. The enhancement of $^{18}$O
in the nuclear region of ESO 320-G030 is higher than in the Galactic
Center \citep[$\sim250$, see][and references therein]{wil94}, higher
than in M~82 and NGC~253 \citep{mar10}, similar to the value in Arp~220
\citep{gon12}, and lower than in Mrk~231 \citep{gon14b}.

\underline{OH$^+$}: The 3 absorption lines at $76.2-76.5$\,$\mu$m are primarily
generated in the core, while the $152.3-153.1$ and 158.4\,$\mu$m lines
at longer wavelengths are expected to be produced by the disk. 
To reproduce these features, a high OH$^+$ abundance of
$\sim(0.6-1)\times10^{-7}$
is required in both components, comparable to the value inferred in
Mrk\,231 \citep{gon18}. Additional contribution by the envelope with
a similar OH$^+$ abundance is included to better match the strong absorption
at 153\,$\mu$m. By contrast, no far-IR OH$^+$ absorption is detected
in the high spectral resolution (Fabry-P\'erot) spectrum of Sgr\,B2 taken
with the {\it Infrared Space Observatory (ISO)} \citep{pol07}.
Although the far-IR OH$^+$ absorption is prominent, the ground-state
  lines in the submm are hardly seen (Fig.~\ref{fullspire}), which is
  consistent with the model.

\underline{H$_2$O$^+$}: Absorption features are seen at 65.5
(blended with NH$_2$), 78.55, 143.3, 143.8 (blended with C$_3$), and
$145.9-146.2$\,$\mu$m, although
the latter features are shifted relative to the expected positions.
Our model assumes the same abundance in both the core and the disk,
$3\times10^{-8}$, and the 143.3\,$\mu$m is somewhat overpredicted.
Hence, a ratio $\mathrm{OH^+/H_2O^+}\sim2-3$ is inferred.
The model for H$_2$O$^+$ satisfactorily reproduces the submm
  lines at $400-420$\,$\mu$m.

\underline{H$_3$O$^+$}: The clearest evidence of H$_3$O$^+$ absorption is
found at the red edge of the NH$_3$ 166\,$\mu$m absorption, with hints
of absorption also seen at 82.3 and 82.9\,$\mu$m but no detected absorption
at 69.55\,$\mu$m. The latter constrains the H$_3$O$^+$ abundance to
$\sim3\times10^{-8}$, but the feature at 166\,$\mu$m is not fully reproduced.
It is possible that formation pumping enhances this absorption, as
favored in Arp~220 \citep{gon13}. Results are consistent with 
$\mathrm{H_2O^+/H_3O^+}\sim1$.

\underline{NH}: Strong absorption features are detected at $76.6-76.9$,
151.1, 151.5 (blended with C$_3$), $153.1$ (blended with OH$^+$), and
153.4\,$\mu$m, with additional hints of absorption at 153.7\,$\mu$m.
A high NH abundance of $\sim5\times10^{-7}$ in both components is required
to match the observed absorption. At submm wavelengths, the model
  predicts little absorption in the ground-state lines at 300, 308, and
  317\,$\mu$m. While this is consistent with the lack of
  absorption features at 308 and 317\,$\mu$m, the absorption 
  observed at 300\,$\mu$m remains underpredicted.

\underline{NH$_2$}: Absorption is detected at 65.6 (blended with H$_2$O$^+$),
$78.4-78.6$, and 130.2\,$\mu$m, which we use to estimate the NH$_2$ column
  density. NH$_2$ also has strong lines in the submm at $207-208$\,$\mu$m
  (in absorption) and $\sim300-330$\,$\mu$m (in emission), which are
  reasonably reproduced by the model after adding an envelope contribution
  to the model (with a NH$_2$ abundance of $1.5\times10^{-8}$).
The abundance of NH$_2$ in the core and disk components is about one order
of magnitude lower than NH, in contrast with the abundance ratio in Sgr B2
where NH$_2$ is much more abundant than NH \citep{goi04}. This suggests
that there is an additional source of ionization in ESO~320-G030, probably
due to cosmic rays.

\underline{NH$_3$}: Absorption features are detected at 71.6, 83.4,
$83.6-84.0$, 84.5 (blended with OH), 165.7, and 170\,$\mu$m,
which are reproduced with a high NH$_3$ abundance of $\sim10^{-6}$.
  The model also approximately accounts for the observed ground-state
  para-NH$_3$ absorption at 256.6\,$\mu$m ($2_1^--1_1^+$). We then
infer $\mathrm{NH_3/NH_2}\sim20$,
a ratio similar to the value in Sgr B2 \citep{goi04}. The nitrogen
chemistry in ESO~320-G030 appears to be the result of a combination of
shock chemistry and high ionization rates.

\underline{CH}: The doublet $N,J=3,7/2\leftarrow 2,5/2$ with
$E_{\mathrm{lower}}=105$\,K is detected at $118.4-118.7$\,$\mu$m.
The model, with a CH abundance of $\sim5\times10^{-8}$, is consistent
with the lack of detection of CH at 203-204\,$\mu$m.
Our derived abundance is a factor of $\approx2.5$ higher
than the value derived in dark molecular clouds \citep{mat86} and
diffuse clouds \citep{she08}. 
Formation of CH from CH$^+$, which is very abundant (see below),
may be favored \citep{wel06}.

\underline{CH$^+$}: Clear absorption is observed at 119.8\,$\mu$m, adjacent
to the redshift component of the $^{18}$OH doublet at 120\,$\mu$m.
While this absorption is in principle attributable to both the
CH$^+$\,$3-2$ line and to $^{17}$OH ground-state absorption,
the latter species is not expected
to contribute significantly because the other component of the doublet at
$\approx119.62$\,$\mu$m is not detected \citep[see also][]{fis10}. On the
other hand, the CH$^+$\,$5-4$ line at 72.3\,$\mu$m is not detected.
To account for the CH$^+$\,119.8\,$\mu$m absorption, 
a very high abundance of $\sim2\times10^{-7}$ is required, which is
  consistent with the lack of detection of the ground-state line
  at 359\,$\mu$m.
A similarly high CH$^+$ abundance has been inferred by \cite{nag13}
toward the Orion Bar, but only within a narrow $A_V-$range where
reaction of C$^+$ with vibrationally excited H$_2$ can overcome the
high activation barrier of the formation reaction
$\mathrm{C^++H_2\rightarrow CH^++H_2}$. The much higher implied 
column densities of CH$^+$ in the nucleus of ESO~320-G030 may indicate 
the additional combined effect of widespread dissipation of turbulence,
shocks, and ionization by cosmic rays. A column density ratio
$N(\mathrm{CH})/N(\mathrm{CH^+})\sim0.7-2$ is found in the Magellanic
Clouds \citep{wel06}, while this ratio is $\sim0.25$ in ESO~320-G030.

\underline{$^{13}$CH$^+$}: A broad absorption feature is detected at
120.55\,$\mu$m, which could be associated with either $^{13}$CH$^+$\,$3-2$
and/or to SH (see below). To check if it can be reproduced with only
$^{13}$CH$^+$, a model with a fixed abundance ratio
$\mathrm{^{13}CH^+/CH^+}=0.05$ is used, appropriate for the central regions
of starburst galaxies \citep{tan19}. The resulting modeled $^{13}$CH$^+$\,$3-2$
absorption accounts for approximately half of the observed 120.55\,$\mu$m
feature. It is possible that the $^{13}$CH$^+$ abundance in ESO~320-G030 is even
higher than the adopted value, as in the very center of NGC~4945 \citep{tan19}.

\underline{SH}: We have attempted to fill in the remaining 120.55\,$\mu$m
absorption by including a model for SH; the transition that may contribute to 
the observed absorption feature is $^2\Pi_{3/2}\,J=9/2 \leftarrow 7/2$
($E_{\mathrm{lower}}\approx160$\,K). A constraint on the SH model is that
  it generates ground-state absorption at 217\,$\mu$m, close to the
  CO $12-11$ line. From the observed CO SLED, we expect little contamination
  by SH to CO $12-11$, implying an upper limit to the SH abundance of
  $\sim2\times10^{-8}$. The 120.55\,$\mu$m absorption is then still
  underpredicted. The quoted SH abundance is a factor of $\approx1.5$ higher
  than the highest SH abundance inferred in diffuse clouds \citep{neu12,neu15},
  where SH only accounts for $<<1$\% of the gas-phase sulfur chemistry.

\underline{HF}: A single far-IR feature is observed at 81.2\,$\mu$m. We fix the
HF abundance in both the core and the disk to the gas-phase fluorine
abundance \citep{sno07,ind13},
and the observed HF\,$3-2$ line is approximately reproduced.
The model is also consistent with the apparent ground-state absorption 
at 243\,$\mu$m. An undepleted chemistry is strongly suggested by these results.

\underline{C$_3$}: Weak absorption features coincident with lines of the
$\nu_2$ band of C$_3$ are observed at 142.7, 143.8 (blended with H$_2$O$^+$),
145.1 and, imprinted on a wing emission in the [C {\sc ii}]157\,$\mu$m
line, at 157.3 and 158.1\,$\mu$m. We thus favor the detection of
C$_3$ in ESO~320-G030. The observed absorption lines are dominated by
the disk, for which a very high abundance of $\mathrm{several}\times10^{-7}$
is required. For comparison, \cite{cer00} and \cite{moo10} infer an abundance
of C$_3$ relative to H$_2$ in the galactic sources Sgr B2 and W31C of
$(1-5)\times10^{-8}$. The modeled spectrum predicts C$_3$ absorption features
at 167.7 and 176.7\,$\mu$m which are within the S/N of the observed spectra.

\underline{H$_2$S?}: A relatively strong feature is detected at
150.15\,$\mu$m, matching the expected position of the
H$_2$S\,$4_{32}-3_{21}$ line ($E_{\mathrm{lower}}\approx155$\,K).
However, a similar or deeper absorption
would be expected in the $3_{21}-2_{12}$ line at 144.78\,$\mu$m, but it is
not detected. The 151.15\,$\mu$m feature is not detected in the 
{\it (ISO)} spectrum of Sgr\,B2 \citep{pol07}. The possible carrier of this
absorption is considered unknown.

We have also indicated in Fig.~\ref{fullspire} the position of the
  HCN $12-11$ to $16-15$ lines (except the $13-12$ line that is blended
  with CO at 260\,$\mu$m). Apparent absorption features are detected
  at the wavelengths of the $12-11$ (282\,$\mu$m), $14-13$ (242\,$\mu$m),
  and $16-15$ (212\,$\mu$m) lines, but not at the position of the
  $15-14$ transition at 226\,$\mu$m. While this does not allow us
  to unambiguously associate the quoted spectral features to HCN, no
  alternative, reliable carriers have been found.

%__________________________________________________ Two columns table
   \begin{table*}
     \caption{Column density ratios and abundances $X$ of species $Y$ included
       in the overall fit of the {\it Herschel}/PACS spectrum of ESO 320-G030}
         \label{tab:fullpacs}
\begin{center}
          \begin{tabular}{l|cc|cc}   
            \hline
            \noalign{\smallskip}
  & \multicolumn{2}{c|}{CORE} & \multicolumn{2}{c}{DISK}   \\  
            Species $Y$   &  $N(\mathrm{H_2O})/N(Y)$ & $X$$^{\mathrm{a}}$ &
            $N(\mathrm{H_2O})/N(Y)$ & $X$$^{\mathrm{a}}$  \\
            \noalign{\smallskip}
            \hline
            \noalign{\smallskip}
H$_2$O  & 1 & $4\times10^{-5}$  & 1 & $1\times10^{-5}$ \\
H$_2^{18}$O  & 100 & $4\times10^{-7}$  & 100 & $1\times10^{-7}$ \\
OH  & 2 & $2\times10^{-5}$  & 1 & $5\times10^{-6}$ \\
$^{18}$OH  & 100 & $2\times10^{-7}$  & 100 & $5\times10^{-8}$ \\
OH$^+$  & 280 & $1\times10^{-7}$  & 140 & $6\times10^{-8}$ \\
H$_2$O$^+$ & 1200 & $3\times10^{-8}$  & 350 & $3\times10^{-8}$ \\
H$_3$O$^+$ & 1200 & $3\times10^{-8}$  & 350 & $3\times10^{-8}$ \\
NH  & 75  & $5\times10^{-7}$  & 20 & $4\times10^{-7}$ \\
NH$_2$  & 620  & $6\times10^{-8}$  & 175  & $7\times10^{-8}$ \\
NH$_3$  & 38  & $1\times10^{-6}$  & 10  & $1\times10^{-6}$ \\
CH  & 740  & $5\times10^{-8}$  & 210 & $4\times10^{-8}$ \\
CH$^+$  & 180  & $2\times10^{-7}$  & 50 & $2\times10^{-7}$ \\
$^{13}$CH$^+$  & 3600  & $1\times10^{-8}$  & 1050 & $1\times10^{-8}$ \\
SH  & 1800  & $2\times10^{-8}$  & 520 & $2\times10^{-8}$ \\
HF  & 1800  & $2\times10^{-8}$  & 520 & $2\times10^{-8}$ \\
C$_3$  & 180  & $2\times10^{-7}$  & 30 & $3\times10^{-7}$ \\
            \noalign{\smallskip}
            \hline
         \end{tabular} 
\end{center}
\begin{list}{}{}
\item[$^{\mathrm{a}}$] Abundances $X$ of species $Y$ relative to H nuclei
  are estimated from the column density $N(Y)$ and the continuum optical
  depth at 100\,$\mu$m ($N_{\mathrm{H}}=1.3\times10^{24}\,\tau_{100}$).
\end{list}
   \end{table*}
%__________________________________________________ 

In summary, high enhancements in the abundance and column density of
light hydrides are observed in the nuclear region of ESO~320-G030.
In relation with Sgr~B2, the prototypical high-mass star
forming region in our galaxy with
a nucleus optically thick in the far-IR, qualitative differences are seen in
the absorption due to excited OH$^+$, CH$^+$, and also NH. These highly
reactive species are widely observed in diffuse clouds through absorption
from the ground-state level, but not in dense
regions through absorption from rotational levels above the ground-state.
Since the X-ray emission from the nucleus of ESO~320-G030 is weak
\citep{per11}, the source of molecular ionization is likely to be cosmic
rays \citep[see also the case of Mrk~231 in][]{gon18}.
In addition, an undepleted chemistry (i.e., no grain mantles) generated by
shocks and warm dust is strongly suggested.

\section{2D likelihood distributions of the free parameters}
\label{appb}

\begin{figure*}
   \centering
   \includegraphics[width=18.0cm]{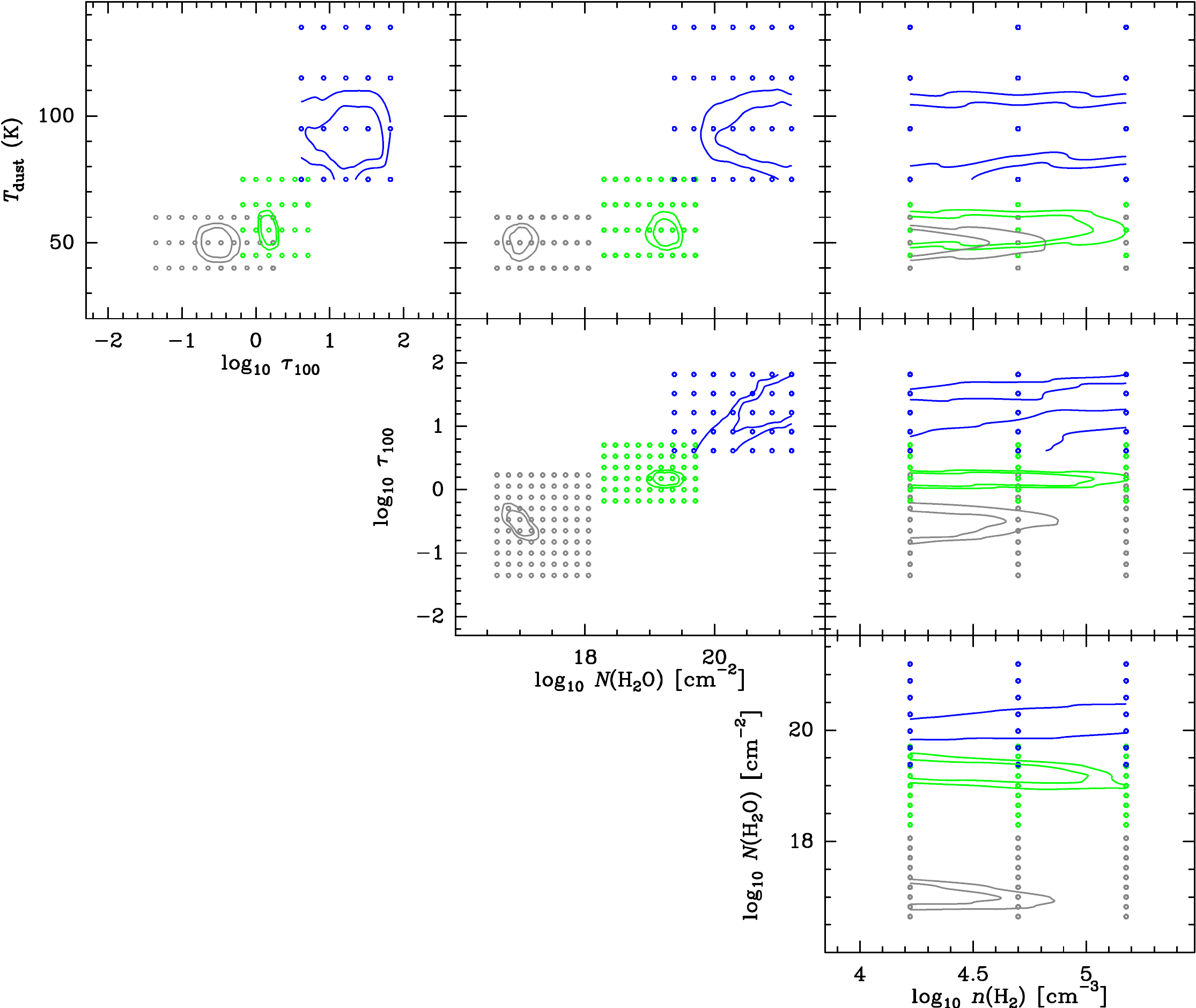}
   \caption{2D marginalized posterior distributions of the
     free physical parameters of each component ($T_{\mathrm{dust}}$, $\tau_{100}$,
     $N_{\mathrm{H2O}}$, $n_{\mathrm{H2}}$) included in our fits to the
     H$_2$O fluxes and continuum flux densities (Section~\ref{bayesian}).
     Each panel displays contours at 25\% and 50\%
     of the peak likelihood in the parameter-parameter space
     for each of the three components. Blue, green, and gray colors
     correspond to the core, the nuclear disk, and the envelope, respectively.
   }
    \label{2dmarg}
   \end{figure*}

The posterior distribution of eq.~(\ref{prob2}) is marginalized
  over to produce the 2D likelihood distributions of the
  free physical parameters, as shown in Fig.~\ref{2dmarg}. This enables
  an evaluation of the degeneracies among these parameters.

We find two main degeneracies: First, $\tau_{100}$ is degenerate with
  $N_{\mathrm{H2O}}$ in the core component. The extinction in this component is
  important even in the submm, so that an increase in $\tau_{100}$ reduces the
  width of the external shell responsible for the line absorption. As a
  consequence, the required H$_2$O column density increases to maintain the
  same value in the photosphere that can be traced. Second, the opposite effect 
  is to some extent found in the envelope, where an increase in $\tau_{100}$
  is accompanied by a decrease in $N_{\mathrm{H2O}}$. In this component,
  extinction by dust is negligible, and any increase in $\tau_{100}$ involves
  a stronger radiation field that is responsible for the H$_2$O excitation,
  thereby reducing to some extent the value $N_{\mathrm{H2O}}$ required to
  explain the observed line fluxes. Nevertheless, the values of $\tau_{100}$
  and $N_{\mathrm{H2O}}$ in the envelope are still well constrained by the data.

\end{appendix}

\end{document}